\documentclass[prd,nofootinbib,aps,twocolumn,showpacs,preprintnumbers,amsmath,amssymb,floatfix]{revtex4-1}
\usepackage{graphicx}
\usepackage{xspace}
\usepackage{booktabs}
\newcommand{\be}{\begin{equation}}
\newcommand{\ee}{\end{equation}}
\newcommand{\benn}{\begin{equation*}}
\newcommand{\eenn}{\end{equation*}}
\newcommand{\bea}{\begin{equation}\begin{aligned}}
\newcommand{\eea}{\end{aligned}\end{equation}}
\newcommand{\beann}{\begin{equation*}\begin{aligned}}
\newcommand{\eeann}{\end{aligned}\end{equation*}}
\newcommand{\bse}{\begin{subequations}}
\newcommand{\ese}{\end{subequations}}
\newcommand{\Order}{{\cal O}}
\newcommand{\ie}{i.e.}

\newcommand{\PQ}{PQ}
\newcommand{\eq}{\ensuremath{{\text{eq}}}}
\newcommand{\Reheating}{\mathrm{R}}
\newcommand{\TP}{\mathrm{TP}}

\newcommand{\BR}{\mathrm{BR}}
\newcommand{\Hypercharge}{\mathrm{Y}}
\newcommand{\Weak}{\mathrm{L}}
\newcommand{\Color}{\mathrm{c}}
\newcommand{\LSP}{\mathrm{LSP}}
\newcommand{\LOSP}{\mathrm{LOSP}}
\newcommand{\dr}{\mathrm{dr}}
\newcommand{\alphaS}{\alpha_\mathrm{s}}
\newcommand{\gS}{g_\mathrm{s}}
\newcommand{\X}{X}
\newcommand{\EWIP}{\mathrm{EWIP}}
\newcommand{\Hzero}{{H_0}}
\newcommand{\seconds}{\mathrm{s}}
\newcommand{\km}{\mathrm{km}}
\newcommand{\Mpc}{\mathrm{Mpc}}

\newcommand{\eV}{\mathrm{eV}}
\newcommand{\meV}{\mathrm{meV}}
\newcommand{\mueV}{\mu\mathrm{eV}}

\newcommand{\MeV}{\mathrm{MeV}}
\newcommand{\GeV}{\mathrm{GeV}}
\newcommand{\TeV}{\mathrm{TeV}}
\newcommand{\axion}{\ensuremath{a}}
\newcommand{\saxion}{\ensuremath{\sigma}\xspace}
\newcommand{\axino}{\ensuremath{\tilde{a}}\xspace}
\newcommand{\gravitino}{\ensuremath{\widetilde{G}}\xspace}
\newcommand{\ax}{\ensuremath{_{a}}}

\newcommand{\neutralino}{\ensuremath{\widetilde{\chi}^{0}_{1}}\xspace}
\newcommand{\slepton}{\ensuremath{\widetilde{{l}}}_{1}}
\newcommand{\sneutrino}{\ensuremath{\widetilde{\nu}_{1}}}

\newcommand{\lepton}{\ensuremath{{l}}}
\newcommand{\gluino}{\ensuremath{\tilde{g}}\xspace}
\newcommand{\quark}{\ensuremath{{{q}}}}
\newcommand{\antiquark}{\ensuremath{{\bar{{q}}}}}
\newcommand{\maxion}{\ensuremath{m_{a}}}
\newcommand{\msaxion}{\ensuremath{m_\saxion}\xspace}
\newcommand{\maxino}{\ensuremath{m_{\tilde{a}}}\xspace}
\newcommand{\mgravitino}{\ensuremath{m_{\widetilde{G}}}\xspace}
\newcommand{\mgluino}{\ensuremath{m_{\tilde{g}}}\xspace}

\newcommand{\mgravsax}{\ensuremath{m_{\widetilde{G},\saxion}}\xspace}
\newcommand{\mLOSP}{m_{\LOSP}}
\newcommand{\MRone}{M_{\mathrm{R}1}}
\newcommand{\monetwo}{m_{1/2}}

\newcommand{\mneutralino}{m_{\neutralino}}
\newcommand{\mslepton}{m_{\slepton}}
\newcommand{\msneutrino}{m_{\sneutrino}}

\newcommand{\MGUT}{m_{\mathrm{GUT}}}
\newcommand{\tausaxion}{\ensuremath{\tau_\saxion}\xspace}
\newcommand{\TD}{T_{\mathrm{D}}}
\newcommand{\TDLOSP}{T_{\mathrm{D}}^{\LOSP}}
\newcommand{\TR}{T_{\Reheating}}
\newcommand{\TL}{T_{\mathrm{low}}}
\newcommand{\Tafter}{T_{\mathrm{after}}}
\newcommand{\MPlanck}{\ensuremath{M_{\mathrm{P}}}}
\newcommand{\mPlanck}{\ensuremath{m_{\mathrm{P}}}}
\newcommand{\gstarS}{g_{*S}}
\newcommand{\gstar}{g_{*}}
\newcommand{\fax}{\ensuremath{f_{\mathrm{PQ}}}}
\newcommand{\vax}{\ensuremath{v_{\mathrm{PQ}}}}
\newcommand{\Yprimordial}{Y_{\mathrm{p}}}

\newcommand{\Ysaxeq}{Y_{\saxion}^{\mathrm{eq}}}
\newcommand{\YaxeqTP}{Y_{\axion}^{\mathrm{eq/TP}}}

\newcommand{\YsaxeqTP}{Y_{\saxion}^{\mathrm{eq/TP}}}

\newcommand{\YgravitinoTP}{Y_{\gravitino}^{\mathrm{TP}}}
\newcommand{\YaxinoTP}{Y_{\axino}^{\mathrm{TP}}}
\newcommand{\Yaxinoeq}{Y_{\axino}^{\mathrm{eq}}}
\newcommand{\YaxinoeqTP}{Y_{\axino}^{\mathrm{eq/TP}}}
\newcommand{\YLOSP}{Y_{\LOSP}}

\newcommand{\OmegaaxinoeqTP}{\Omega_{\axino}^{\mathrm{eq/TP}}}

\newcommand{\OmegaDM}{\Omega_{\mathrm{CDM}}}

\newcommand{\OmegaGravitino}{\Omega_{\gravitino}}
\newcommand{\OmegaAxino}{\Omega_{\axino}}
\newcommand{\OmegaAxionMIS}{\Omega_{a}^\mathrm{MIS}}
\newcommand{\neff}{\ensuremath{N_\text{eff}}}
\newcommand{\rhorad}{\rho_\mathrm{rad}}
\newcommand{\rhoMSSMrad}{\rho_\mathrm{rad}}
\newcommand{\rhodr}{\rho_\mathrm{dr}}

\newcommand{\gstarMSSM}{g_{*}}
\begin{document}
% __________________________________________________________________
%
% ___ Preprint Numbers ___________________________________________________
%
\preprint{arXiv:1302.2143}
\preprint{MPP-2013-30}
%
%
% ___ Preamble ______________________________________________________
%
% __________________________________________________________________
\title{Dark radiation and dark matter in supersymmetric axion models\\
with high reheating temperature}
% __________________________________________________________________
\author{Peter Graf}
%\email{graf@mpp.mpg.de}
\affiliation{Max-Planck-Institut f\"ur Physik, 
F\"ohringer Ring 6,
D--80805 Munich, Germany}
% __________________________________________________________________
\author{Frank Daniel Steffen}
%\email{steffen@mpp.mpg.de}
\affiliation{Max-Planck-Institut f\"ur Physik, 
F\"ohringer Ring 6,
D--80805 Munich, Germany}
% __________________________________________________________________
%
% ___ Abstract _________________________________________________________
%
\begin{abstract}
  Recent studies of
  the cosmic microwave background, large scale structure,
  and big bang nucleosynthesis (BBN)
  show trends towards extra radiation.
  Within the framework of supersymmetric hadronic axion models,
  we explore two high-reheating-temperature scenarios 
  that can explain consistently  
  extra radiation 
  and cold dark matter (CDM),
  with the latter residing either in gravitinos or in axions.
  In the gravitino CDM case,  
  axions from decays of thermal saxions
  provide extra radiation already prior to BBN 
  and decays of 
  axinos with a cosmologically required TeV-scale mass
  can produce extra entropy.
  In the axion CDM case,  
  cosmological constraints
  are respected with light eV-scale axinos and weak-scale gravitinos  
  that decay into axions and axinos.
  These decays lead to late extra radiation 
  which can coexist with 
  the early contributions from saxion decays.
  Recent results of the Planck satellite probe
  extra radiation at late times  
  and thereby both scenarios.
  Further tests are the  
  searches for axions at ADMX and
  for supersymmetric particles at the LHC.
\end{abstract}
\pacs{14.80.Va, 11.30.Pb, 98.80.Cq, 98.80.Es}
%
%% 11.30.Pb 	Supersymmetry
%% 14.80.Va     Axions
%% 98.80.Cq     Particle-theory and field-theory models of the early Universe
%% 98.80.Es 	Observational cosmology (including Hubble constant, distance scale, cosmological constant, early Universe, etc) 
%
%\keywords{}
\maketitle
%
%
%______________________________________________
\section{Introduction}
\label{Sec:Intro}
%______________________________________________

Recent cosmological studies show trends towards
a radiation content of the Universe at the onset of 
big bang nucleosynthesis (BBN) and much later that exceeds
expectations for standard three active neutrino species.
The obtained limits 
on non-standard contributions $\Delta\neff$
to the effective number of light neutrino species $\neff$ 
are still consistent with the standard value $\neff\simeq 3$
at the $1$--$2\sigma$ level.
However, BBN likelihood analyses based on recent studies
of the mass fraction $\Yprimordial$ of primordial 
helium~\cite{Izotov:2010ca,Aver:2010wq} find posterior maxima of
$\Delta\neff\simeq 0.7$--$0.8$~\cite{Izotov:2010ca,Hamann:2010bk,Graf:2012hb} 
and precision cosmology studies
of the cosmic microwave background (CMB) 
and large scale structure (LSS) means of 
$\Delta\neff\simeq 0.8$--$1.8$~\cite{Komatsu:2010fb,Hamann:2010pw,Hinshaw:2012fq}
prior to the announcement of the new Planck results~\cite{Ade:2013zuv}.
While the BBN studies are limited by systematic errors
(see e.g.~\cite{Aver:2010wq}),
the Planck satellite mission has recently 
probed $\neff$ at the CMB decoupling epoch 
-- as expected~\cite{Perotto:2006rj,Hamann:2007sb} --
with an unprecedented sensitivity of 
$\Delta\neff\simeq 0.26$
at the $1\sigma$ level.
In fact, the Planck results point to favored values of
$\Delta\neff\simeq 0.25$--$0.6$ 
and upper  limits 
of $\Delta\neff\lesssim 1$ at the $2\sigma$ level~\cite{Ade:2013zuv}.
In particular, with the above $\Delta\neff$ values,
a tension between Planck data
and direct measurements of the Hubble constant $\Hzero$~\cite{Riess:2011yx}
is relieved that is present in the base $\Lambda$CDM model 
that does not allow for the possibility of $\Delta\neff>0$.
Indeed, new astrophysical data sets on $\Hzero$
seem crucial to clarify whether there is extra radiation
pointing to new physics or a Hubble constant that is considerably below
the current values from direct measurements.

Various explanations for $\Delta\neff\sim 1$ 
have been explored in the literature
invoking, e.g., light sterile neutrinos~\cite{Hamann:2010bk,Hamann:2011ge},
other light species~\cite{Nakayama:2010vs,Boehm:2012gr},
neutrino asymmetries~\cite{Pastor:2008ti,Mangano:2010ei}, 
or decays of heavy particles~\cite{Chun:1995hc,Chun:2000jr,Ichikawa:2007jv,Kawasaki:2007mk,Fischler:2010xz,Hasenkamp:2011em,Hooper:2011aj,Choi:2012zn,Graf:2012hb,Cicoli:2012aq,Higaki:2012ar,GonzalezGarcia:2012yq,Hasenkamp:2012ii,Bae:2013qr,Jeong:2013axf}.
Here 
we study two classes of supersymmetric (SUSY) hadronic axion models
which describe consistently extra radiation and cold dark matter (CDM)
for a high reheating temperature after inflation 
of up to $\TR\sim 10^9\,\GeV$ or $10^{11}\,\GeV$.
In the considered (R-parity-conserving) models, 
it may thereby be possible to generate
the baryon asymmetry, 
e.g., via thermal leptogenesis 
with hierarchical heavy Majorana neutrinos~\cite{Buchmuller:2005eh}.
Moreover, SUSY axion models are compelling 
since both the strong CP problem
and the hierarchy problem are solved simultaneously. 
These models come with new fields including
the axion $a$, the saxion $\sigma$, the axino~$\axino$,
and the gravitino~$\gravitino$, which
can play important cosmological roles
depending on their masses, the Peccei--Quinn (PQ) scale $\fax$, 
and the reheating temperature $\TR$.

As the pseudo-Nambu-Goldstone boson 
associated with the U(1)$_\mathrm{PQ}$ symmetry 
broken spontaneously at $\fax$~\cite{Sikivie:2006ni,Kim:2008hd},
the axion has interactions suppressed by $\fax$ 
and a mass of 
$\maxion
\simeq 
6~\meV 
(10^{9}\,\GeV/\fax)$.
With laboratory, astrophysical, 
and cosmological studies~\cite{Raffelt:2006cw,Beringer}
pointing to
$\fax \gtrsim 6 \times 10^8~\GeV$,
the axion is predicted to be 
an extremely weakly interacting particle (EWIP) 
with a tiny mass of $\maxion\lesssim 10~\meV$.
In SUSY settings, the saxion and the axino appear respectively as
the scalar and the fermionic partner of the axion.
They are EWIPs as well with masses $\msaxion$ and $\maxino$
that depend on details of the model and of SUSY breaking.
For example, one expects the saxion mass $\msaxion$ to be
of the order of the gravitino mass $\mgravitino$
in gravity-mediated SUSY breaking. 
As the gauge field associated with local SUSY transformations,
the gravitino is another EWIP with
interactions suppressed by the (reduced) Planck scale
$\MPlanck=2.4\times 10^{18}\,\GeV$
and a mass that depends on the SUSY breaking scale.
While we do not assume a specific SUSY breaking model, 
$\msaxion=\mgravitino$ is used in the main part of this work.
Other than that, $\mgravitino$ (together with $\msaxion$)
and $\maxino$ are treated as free parameters set in a way
to evade cosmological constraints.
Model building aspects of the considered mass hierarchies
will be considered elsewhere. 

In the first of the two classes that we consider, 
the gravitino is 
the lightest supersymmetric particle (LSP)
that provides CDM.
Here decays of thermal saxions into axions can provide 
$\Delta\neff\sim 0.5$ prior to BBN~\cite{Chun:1995hc,Chang:1996ih,Kawasaki:2007mk,Graf:2012hb,Bae:2013qr};
see also~\cite{Chang:1996ih,Asaka:1998ns,Ichikawa:2007jv,Kawasaki:2007mk,Bae:2013qr,Jeong:2013axf} 
for extra radiation 
from late decays of non-thermal saxions.
In the second class, a very light axino is the LSP, 
the gravitino the next-to-LSP (NLSP) 
and CDM resides in axions from the misalignment mechanism.
Again, it is possible to have 
$\Delta\neff\sim 0.5$ from decays of thermal saxions into axions
already prior to BBN.
However, now there can be an additional contribution 
of $\Delta\neff\sim 0.5$ but only
well after BBN from gravitino decays 
into the axion and the axino~\cite{Ichikawa:2007jv,Hasenkamp:2011em}.
For both classes, we show updated $\Delta\neff$ contours 
that point to new limits on $\TR$
accounting for the recent Planck results on $\Delta\neff$~\cite{Ade:2013zuv}.
Moreover,
we devote particular attention to cosmological viability
and to the interplay with present and potential future
insights from SUSY searches at the LHC.

Some points by which our present study goes beyond directly
related existing studies~\cite{Hasenkamp:2011em,Graf:2012hb}
are the following.
Decays are treated
beyond the sudden-decay approximation.
In the $\gravitino$ LSP case,
the resulting $\Delta\neff$ contours are confronted explicitly
with the $\TR$ limit imposed by a gravitino density $\OmegaGravitino$
that cannot exceed the dark matter density $\OmegaDM$.
Here cosmological constraints require 
$\maxino\gtrsim 2~\TeV$ such that axinos 
decay prior to the 
decoupling of the lightest ordinary sparticle (LOSP),
which denotes the lightest sparticle 
within the minimal supersymmetric standard model (MSSM).
Axinos can then provide a sizable fraction 
of the total energy density of the Universe when decaying
and thereby produce entropy~\cite{Choi:2008zq,Hasenkamp:2010if,Hasenkamp:2011xh}.
This is included in our calculations, 
as is the gravitino density $\OmegaGravitino^{\axino\to a\gravitino}$
from rare axino decays into axions and gravitinos.
Here we apply an updated result for the axino abundance 
produced thermally in the early Universe, 
which we obtain by including 
quartic axino-squark-antisquark-gluino interactions~\cite{Strumia:2010aa}
omitted in an earlier calculation~\cite{Brandenburg:2004du}.
In the $\axino$ LSP case with the $\gravitino$ NLSP,
we present $\Delta\neff$ contours that account for
both decays, $\gravitino\to a\axino$ and $\saxion\to a a$, explicitly.
Moreover, our treatment includes 
contributions of the gravitino-spin-3/2 components
and of electroweak processes  
to the thermally produced gravitino yield. 
In both of the considered LSP cases, 
we account systematically for the possibility 
that saxion decays into gluon pairs can have a sizable branching ratio
and can thereby produce significant amounts of entropy.

The remainder of this paper is organized as follows.
In the next section we discuss the observational hints
towards extra radiation beyond the SM and possible scenarios
in light of the recent Planck results.
Section~\ref{Sec:HighTR} is devoted to general aspects
of the considered SUSY hadronic axion models in
high-$\TR$ scenarios, which apply to the
two explored LSP cases.
This section contains our updated result for the 
primordial abundance of thermally produced axinos. 
The gravitino CDM and the axion CDM scenarios are presented
in Sects.~\ref{Sec:GravitinoCDM} and~\ref{Sec:AxionCDM} respectively.
Here we consider the corresponding contributions to $\OmegaDM$,
$\Delta\neff$, and entropy, provide resulting $\TR$ limits,
and address the testability of these scenarios.
We summarize our conclusions in Sect.~\ref{Sec:Conclusion}.
Appendix~\ref{Sec:AxinoTP} provides details on our updated calculation 
of the thermally produced axino abundance, where 
hard thermal loop (HTL) 
resummation~\cite{Braaten:1989mz,Braaten:1991dd} is used 
to treat screening effects of the primordial plasma
as in Ref.~\cite{Brandenburg:2004du}.
In Appendix~\ref{Sec:AnalyticApprox} 
approximate expressions for 
the numerical results obtained 
in Sects.~\ref{Sec:GravitinoCDM} and~\ref{Sec:AxionCDM}
are given
that allow for a qualitative understanding of those results.
While $\msaxion=\mgravitino$
is assumed throughout the main part of this work,
we briefly describe the changes 
that occur for $\msaxion\neq\mgravitino$
in Appendix~\ref{Sec:MassDiffSaxionGravitino}.

%______________________________________________
\section{Extra radiation}
\label{Sec:ExtraRadiation}
%______________________________________________

One of our key motivations for the studies presented in this work
is the trend towards extra radiation 
inferred from current cosmological investigations
as summarized briefly in the Introduction.
In this section we 
expand slightly on the description of 
the current situation and outline different possible perspectives
accounting for the new Planck results
on $\Delta\neff$.

The standard model (SM) predictions 
of the total relativistic energy density,
\be
\rho_\text{rad}^{\mathrm{tot}}(T)
=
\left[ 1+\frac{7}{8}\neff\left(\frac{T_\nu}{T}\right)^4\right] 
\rho_\gamma(T) ,
\label{Eq:neff_def}
\ee
are given by $\neff=3$ and $T_\nu=T$ 
at $T\sim 1~\MeV$ (before neutrino decoupling and $e^+e^-$ annihilation)
and by $\neff=3.046$ and $T_\nu=(4/11)^{1/3}T$ 
after neutrino decoupling.
Here $\rho_\gamma$ is the photon energy density and
$T_{(\nu)}$ the temperature of photons (neutrinos). 
The effective number of light neutrino species $\neff$
increases slightly due to residual neutrino heating 
by $e^+e^-$ annihilation~\cite{Mangano:2005cc}.

There are various ways to probe $\neff$  
and thereby non-standard contributions $\Delta\neff$ 
to which we refer as extra radiation.
At the epoch of BBN, 
a speed-up of the Hubble expansion rate caused by $\Delta\neff>0$
leads to a more efficient $^{4}$He output than in standard BBN.
Observationally inferred limits 
on the primordial $^{4}$He mass fraction $\Yprimordial$ 
can thus be translated into $\Delta\neff$ limits.
Much later, at the epoch of CMB decoupling, $\Delta\neff>0$
affects the time of radiation-matter equality, 
leads to a less efficient early integrated Sachs--Wolfe effect, 
and reduces the scale of the sound horizon. 
This affects the CMB power spectrum 
by increasing the height of the first peak 
and by shifting the peak positions 
towards higher multipole momenta.
Moreover, free-streaming of the relativistic populations 
associated with $\Delta\neff>0$ suppresses power on small scales
and thereby affects the matter power spectrum 
inferred from studies of the LSS.
Based on those observables, numerous studies of BBN, CMB, and LSS
have explored limits and favored values for 
$\Delta\neff$~\cite{Hamann:2007pi,Reid:2009nq,Komatsu:2010fb,Hamann:2010pw,GonzalezGarcia:2010un,Hinshaw:2012fq,Ade:2013zuv} 
with the outcome outlined in the Introduction.

To motivate the $\Delta\neff$ values considered in our study, 
we quote representative current constraints on $\Delta\neff$
imposed by BBN and precision cosmology 
in Table~\ref{Tab:Neffconstrains}.
%
%-------------------------------------
%
\begin{table}[t]
\centering
\caption{{Constraints on $\Delta\neff$ from BBN and precision cosmology. 
The first two lines give the posterior maximum (p.m.) 
and the minimal 99.7\% credible interval imposed by BBN
as obtained in Ref.~\cite{Graf:2012hb} 
using the indicated data sets
and the prior $\Delta\neff\geq 0$.
The third line lists the mean and the 95\% CL upper limit on $\Delta\neff$ 
from the precision cosmology study~\cite{Hamann:2010pw} 
based on CMB data, 
the Sloan Digital Sky Survey (SDSS) data-release 7 halo power spectrum (HPS),
and data from the Hubble Space Telescope (HST).
The last two lines provide the mean and the 95\% CL upper limit 
on $\Delta\neff$ $(=\neff-3.046)$
as obtained by the Planck collaboration~\cite{Ade:2013zuv}
when combining Planck CMB data with
WMAP polarization data (WP), data from high-$l$ experiments (highL),
and data on baryon acoustic oscillations (BAO).
The values in the last line emerge
when results of Ref.~\cite{Riess:2011yx}
on a direct measurement of the 
Hubble constant $\Hzero$ are taken into account.
}}
\label{Tab:Neffconstrains}
\begin{ruledtabular}
\begin{tabular}{@{\extracolsep{\fill}}lcc}
Data
&  p.m./mean
&  upper limit\\
\noalign{\smallskip}
\hline
\noalign{\smallskip}
$Y_\text{p}^\text{IT}$~\cite{Izotov:2010ca} 
+ $[\text{D/H}]_\text{p}$~\cite{MNR:MNR13921}
& 0.76
& $<1.97~(3\sigma)$ \\
$Y_\text{p}^\text{Av}$~\cite{Aver:2010wq} 
+ $[\text{D/H}]_\text{p}$~\cite{MNR:MNR13921}
& 0.77
& $<3.53~(3\sigma)$ \\
CMB + HPS + HST~\cite{Hamann:2010pw}
& 1.73
& $<3.59~(2\sigma)$ \\
Planck+WP+highL+BAO~\cite{Ade:2013zuv}
& 0.25
& $<0.79~(2\sigma)$ \\
Planck+WP+highL+$\Hzero$+BAO~\cite{Ade:2013zuv}
& 0.47
& $<0.95~(2\sigma)$ \\
\end{tabular}
\end{ruledtabular}
\end{table}
%
%------------------------------------
%
The first two lines have been obtained in 
a BBN-likelihood analysis~\cite{Graf:2012hb}
based on the recent $\Yprimordial$ studies 
of Izotov and Thuan~\cite{Izotov:2010ca} 
and of Aver \textit{et al.}~\cite{Aver:2010wq}.
Those studies report primordial $^{4}$He abundances 
of $Y^\text{IT}_\text{p} = 0.2565\pm0.001(\text{stat.})\pm0.005(\text{syst.})$ 
and $Y^\text{Av}_\text{p} = 0.2561\pm0.0108$, respectively,
with errors referring to 68\% intervals.
Moreover, a primordial D abundance 
of $\log[\text{D/H}]_\text{p}=-4.56\pm0.04$~\cite{MNR:MNR13921}
and a free-neutron lifetime 
of $\tau_{\mathrm{n}}=880.1\pm1.1~\seconds$~\cite{Beringer} 
have been used
in the determination of the listed posterior maxima (p.m.) and 
the $3\sigma$ upper limits.
The third line gives the mean and the 95\% confidence level (CL) 
upper limit on $\Delta\neff$ obtained in the precision cosmology study
of Ref.~\cite{Hamann:2010pw} based on CMB data, 
the Sloan Digital Sky Survey (SDSS) data-release 7 halo power spectrum (HPS),
and data from the Hubble Space Telescope (HST).
Compatibility with $\Delta\neff=0$ is found  
at the $1$--$2\sigma$ level
in both the BBN and that precision cosmology study.
While a more decisive compatibility test 
seems to be difficult for BBN investigations
due to significant systematic uncertainties
(see e.g.~\cite{Aver:2010wq}), 
the new results of the Planck satellite mission
have improved the $\Delta\neff$ accuracy 
of precision cosmology investigations substantially~\cite{Ade:2013zuv}.
Even with the improved accuracy, 
compatibility with $\Delta\neff=0$ is found
to hold still at the $1$--$2\sigma$ level.
In the last two lines of Table~\ref{Tab:Neffconstrains}
we provide the mean and the 95\% CL upper limit 
on $\Delta\neff$ $(=\neff-3.046)$
obtained by the Planck collaboration~\cite{Ade:2013zuv}
when combining CMB data from Planck with
WMAP polarization data (WP), data from high-$l$ experiments (highL),
and data on baryon acoustic oscillations (BAO).
The values in the last line emerge 
with a Gaussian prior on $\Hzero$
based on the direct measurement 
of the Hubble constant of Ref.~\cite{Riess:2011yx}.

The Planck results quoted in Table~\ref{Tab:Neffconstrains}, 
still allow for (or even favor) a relatively small amount
of extra radiation, e.g., from saxion decays and/or gravitino decays.
With the current BBN limits, the following scenarios are possible:
(i)~this small amount was already present at the onset of BBN 
with no additional contribution after BBN,
(ii)~this small amount was generated only well after BBN, or
(iii)~part of this small amount was generated already prior to BBN
and the remaining part well after BBN.

We will see below that composition~(i) is the only one that
can be realized in the considered gravitino LSP case,
whereas the alternative axino LSP case 
allows for all three compositions.
Contours of $\Delta\neff=0.25$, $0.47$, $0.79$, and $0.95$
will be explored in the respective parameter regions
corresponding to the means and the $2\sigma$ upper limits
obtained by the Planck collaboration~\cite{Ade:2013zuv} 
as quoted in the last two lines of Table~\ref{Tab:Neffconstrains}.%
\footnote{Accidentally, $\Delta\neff=0.79$ nearly coincides 
with the posterior maxima from the BBN analysis of~\cite{Graf:2012hb}
quoted in Table~\ref{Tab:Neffconstrains}. 
Thus, the respective contours allow us to infer also 
parameter regions in which one finds 
the $\Delta\neff$ value favored by BBN studies.}

%______________________________________________
\section{High-Reheating-Temperature Scenarios}
\label{Sec:HighTR}
%______________________________________________

Throughout this work it is assumed 
that inflation has governed the earliest moments of the Universe,
as suggested by its flatness, isotropy, and homogeneity.
Accordingly, any initial EWIP population was diluted away 
by the exponential expansion during the slow-roll phase
of the inflaton field.
A radiation-dominated epoch with an initial temperature of $\TR$ 
emerged from the subsequent reheating phase
in which inflaton decays repopulate the Universe.%
\footnote{Inflaton decays into EWIPs may have been efficient.
However, we do not include such contributions
since there are inflation models 
in which this production mechanism can be 
negligible~\cite{Asaka:2006bv,Endo:2007sz}.}
While inflation models may point to $\TR$ well above $10^{10}\,\GeV$,
we limit our studies to the case $\TR<\fax$ 
in which no \PQ\ symmetry restoration takes place after inflation.
Focussing on high-reheating temperature scenarios with $\TR>10^7\,\GeV$,
axions, saxions, axinos, and gravitinos can be produced efficiently 
in thermal scattering of MSSM fields in the hot plasma. 
Depending on the PQ scale $\fax$ and on $\TR$,
even scenarios in which the fields of the axion supermultiplet 
were in thermal equilibrium are conceivable.

For the axion and the saxion, our estimate for 
the decoupling temperature reads~\cite{Graf:2012hb}
\be
\TD^{\axion,\saxion}
\approx 
1.4\times 10^9\,\GeV \left( \frac{\fax}{10^{11}\,\GeV} \right) ^{2}.
\label{Eq:TDSAxion}
\ee
Following the approach of Ref.~\cite{Graf:2012hb}
and using our results for thermal axino production 
presented below and in Appendix~\ref{Sec:AxinoTP},
we estimate the axino decoupling temperature as
\be
\TD^{\axino}
\approx 
5.2\times 10^8\,\GeV 
\left(\frac{\fax}{10^{11}\,\GeV}\right)^{2}.
\label{Eq:TDAxino}
\ee

In cosmological scenarios with $\TR>\TD^{\axino}$
(or even $\TR>\TD^{\axion,\saxion}$), axinos (together with axions/saxions)
were in thermal equilibrium before decoupling as a relativistic species
provided $\maxino\ll\TD^{\axino}$ (and $\msaxion\ll\TD^{\saxion}$).
Then the yield of those thermal relic axions/saxions and axinos
after decoupling is given respectively by
\be
Y_{\axion,\saxion}^\eq 
= \frac{n^\eq_{\axion,\saxion}}{s}\approx1.2\times10^{-3}
\label{Eq:YAxionSaxionEq}
\ee
and
\be
Y_{\axino}^\eq 
= \frac{n^\eq_{\axino}}{s}\approx1.8\times10^{-3}.
\label{Eq:YAxinoEq}
\ee
Here $n_{j}^{(\eq)}$ denotes the corresponding (equilibrium)
number density of species $j$ and $s$ the entropy density.
For the latter, we use $s(T)=2\pi^2 \gstarS T^3/45$ 
with an effective number
of relativistic degrees of freedom of $\gstarS(\TD)\simeq 232.5$
that accounts for the MSSM and the axion multiplet fields, 
which can all be considered as relativistic 
at $\TD$ for $m_{\saxion,\axino}\ll\TD$.

In scenarios with $\TR<\TD^{\axion,\saxion,\axino}$, 
the axion multiplet fields
can still be  thermally produced (TP)
via scattering of colored (s)particles in the primordial plasma.
The resulting yields are given by~\cite{Graf:2012hb}
\begin{equation}
Y_{\axion,\saxion}^\TP
= 
1.33\times 10^{-3} g_s^6 
\ln \! \left(\frac{1.01}{g_s}\right) \!\!
\left( \frac{10^{11}\,\GeV}{\fax} \right)^{\!\!2} \!\!
\left( \frac{\TR}{10^{8}\,\GeV} \right)\!\!
\label{Eq:SAxionYieldTP}
\end{equation}
and, as derived by updating the result of Ref.~\cite{Brandenburg:2004du}
in Appendix~\ref{Sec:AxinoTP}, by
\begin{equation}
Y_{\axino}^\TP
= 
1.98\times 10^{-3} g_s^6 
\ln \! \left(\frac{1.27}{g_s}\right) \!\!
\left( \frac{10^{11}\,\GeV}{\fax} \right)^{\!\!2} \!\!
\left( \frac{\TR}{10^{8}\,\GeV} \right) \!.
\label{Eq:AxinoYieldTP}
\end{equation}
Here the strong gauge coupling is understood
to be evaluated at $\TR$, i.e.,
$g_s\equiv g_s(\TR)=\sqrt{4\pi\alpha_s(\TR)}$,
which we calculate
according to its 1-loop renormalization group running within the MSSM
from $\alpha_s(m_\mathrm{Z})=0.1176$ at the Z-boson mass
$m_\mathrm{Z}=91.1876~\GeV$.
 
Note that our focus is on hadronic 
or KSVZ axion models~\cite{Kim:1979if,Shifman:1979if} 
in a SUSY setting, with $N_Q=1$ 
heavy KSVZ (s)quark multiplets $Q_L$ and $\bar{Q}_R$.
After integrating out the KSVZ fields, 
we obtain the effective Lagrangian~\cite{Graf:2012hb}
\begin{align}
\mathcal{L}_\text{\PQ}^\text{int} 
= 
&\frac{\alpha_s}{8\pi\fax} 
\bigg[ 
\saxion 
\left( G^{b\, \mu\nu}G^b_{\mu\nu} 
- 2D^b D^b
- 2i\bar{\tilde{g}}^b_M\gamma^\mu D_\mu\tilde{g}_M^b \right) 
\nonumber\\
&\quad\,\,\,\,
+a
\left( G^{b\,\mu\nu}\widetilde{G}^b_{\mu\nu} 
+ 2\bar{\tilde{g}}_M^b\gamma^\mu\gamma^5D_\mu\tilde{g}_M^b \right) 
\nonumber\\ 
&\quad\,\,\,\,
- i\bar{\axino}_M \frac{[\gamma^\mu,\gamma^\nu]}{2}\gamma^5\tilde{g}_M^b G^b_{\mu\nu} 
+2\bar{\axino}_M D^b\tilde{g}_M^b \bigg]
\ ,
\label{Eq:eff_lag}
\end{align}
where 
$b$ is a color index,
$D_\mu$ the corresponding color-gauge covariant derivative,
${G}^b_{\mu\nu}$ the gluon-field-strength tensor, 
$\widetilde{G}^b_{\mu\nu}=\epsilon_{\mu\nu\rho\sigma}G^{b\,\rho\sigma}/2$ its dual,
$\tilde{g}^b$ the gluino field, and
$D^b=-g_s \sum_{\tilde{q}} \tilde{q}_i^* T_{ij}^b \tilde{q}_j$
with a sum over all squark fields $\tilde{q}$ 
and the SU(3)$_c$ generators $T_{ij}^b$
in their fundamental representation;
the subscript $M$ indicates 4-component Majorana spinors.%
\footnote{Slightly different expressions 
for $\mathcal{L}_\text{\PQ}^\text{int}$ 
can be found in~\cite{Strumia:2010aa,Choi:2011yf}. 
We use the space-time metric 
$g_{\mu\nu}=g^{\mu\nu}=\mathrm{diag}(+1,-1,-1,-1)$ 
and other conventions and notations of Ref.~\cite{Dreiner:2008tw}
and, except for a different sign of the Levi-Civita tensor 
$\epsilon^{0123}=+1$, of Ref.~\cite{Drees:2004jm}.}
In the considered framework, 
the Lagrangian~\eqref{Eq:eff_lag}
describes the relevant saxion/axion/axino interactions 
even in a conceivable very hot early stage 
of the primordial plasma with temperatures $T$ not too far below $\fax$.%
\footnote{We do not consider scenarios 
with a radiation-dominated epoch with $T$ 
above the masses of the heavy KSVZ (s)quarks $m_{Q,\tilde{Q}}$ 
such as those considered in Ref.~\cite{Bae:2011jb}.} 
Based on~\eqref{Eq:eff_lag} 
the presented results~\eqref{Eq:TDSAxion}, \eqref{Eq:TDAxino},
\eqref{Eq:SAxionYieldTP}, and~\eqref{Eq:AxinoYieldTP}
are obtained.
In particular, 
as outlined in more detail 
in Appendix~\ref{Sec:AxinoTP},
our result for the thermally produced 
axino yield~\eqref{Eq:AxinoYieldTP} accounts 
for the second term in the third line of~\eqref{Eq:eff_lag} 
that describes the quartic axino-squark-antisquark-gluino 
interaction~\cite{Strumia:2010aa}, 
whereas the corresponding result 
of Ref.~\cite{Brandenburg:2004du}
was based on only the first term in that line.

Gravitinos with mass values of $\mgravitino\gtrsim 1~\GeV$,
which are the ones considered in this work, have never been
in thermal equilibrium with the primordial plasma.
Nevertheless, they can be produced efficiently 
in thermal scattering of MSSM fields in the hot plasma.
Derived in a gauge-invariant treatment, 
the resulting thermally produced gravitino yield 
reads~\cite{Bolz:2000fu,Pradler:2006qh,Pradler:2006hh}
\begin{equation}
Y_{\gravitino}^{\TP}
=
\sum_{i=1}^3y_i\, g_i^2
\left(1+\frac{M^2_{i}}{3\mgravitino^2}\right) 
\ln\left(\frac{k_i}{g_i}\right)
        \left(\frac{\TR}{10^{8}\,\GeV} \right)
        \ ,
\label{Eq:YgravitinoTP}
\end{equation}
with $y_i$, the gauge couplings $g_i$, the gaugino mass parameters
$M_i$, and $k_i$ as given in Table~\ref{Tab:Constants}.
%
% ______________________________________________________
\begin{table}[t]
  \caption{Assignments of the index $i$, the gauge coupling $g_i$, and 
    the gaugino mass parameter $M_i$, to the gauge groups
    U(1)$_\Hypercharge$, SU(2)$_\Weak$, and SU(3)$_\Color$,
    and the constants $k_i$, $y_i$, and $\omega_i$.}
  \label{Tab:Constants}
\begin{center}
\renewcommand{\arraystretch}{1.25}
\begin{ruledtabular}
\begin{tabular*}{3.25in}{@{\extracolsep\fill}cccccccc}
gauge group         & $i$ & $g_i$ & $M_i$  &  $k_i$ &  $(y_i/10^{-14})$ & $\omega_i$ 
\\ \hline
U(1)$_\Hypercharge$ & 1 & $g'$    & $M_1$  & 1.266  & 0.653 & 0.018 
\\
SU(2)$_\Weak$       & 2 & $g$     & $M_2$   & 1.312  & 1.604 & 0.044 
\\
SU(3)$_\Color$ & 3 & $g_\mathrm{s}$ & $M_3$ & 1.271  & 4.276 & 0.117 
\end{tabular*}
\end{ruledtabular}
\end{center}
\end{table}
% __________________________________________________________________
%
%
Here $M_i$ and $g_i$ are understood to be evaluated at $\TR$.

In the following we consider universal gaugino masses,
$\monetwo=M_i(\MGUT)$, 
at the grand unification scale 
$\MGUT\simeq 2\times 10^{16}\,\GeV$.
We do not specify a SUSY model. 
Nevertheless, we use certain pairs of $\monetwo$ and 
the weak-scale gluino mass $\mgluino$
keeping in mind that these values are related
via renormalization group evolution.
In particular, we will associate
$\mgluino\simeq 1$, $1.25$, and $1.5~\TeV$ 
with
$\monetwo=400$, $500$, and $600~\GeV$, respectively.
Computing the renormalization group evolution with 
the spectrum generator {\tt SPHENO}~\cite{Porod:2003um,Porod:2011nf},
these relations are obtained within the Constrained MSSM (CMSSM)
with a universal scalar mass parameter of $m_0=1.7~\TeV$,
the trilinear coupling $A_0=0$, a positive higgsino mass parameter, 
$\mu>0$, and a mixing angle in the Higgs sector of $\tan\beta=10$.
The above combinations are still allowed 
by current SUSY searches at the LHC
but are well within reach of the ongoing experiments; 
see e.g.\ Ref.~\cite{ATLAS:2012us}. 

Note that the field-theoretical 
methods~\cite{Braaten:1989mz,Braaten:1991dd} applied in the derivations 
of~\eqref{Eq:SAxionYieldTP}, \eqref{Eq:AxinoYieldTP}, 
and \eqref{Eq:YgravitinoTP} require weak couplings $g_i\ll 1$ 
and thus $T \gg 10^6~\GeV$.%
\footnote{The methods developed 
in Refs.~\cite{Braaten:1989mz,Braaten:1991dd}
are compelling since they allow for a gauge-invariant treatment
of plasma screening effects in calculations of 
thermal EWIP production~\cite{Bolz:2000fu,Brandenburg:2004du,Pradler:2006qh,Pradler:2006hh,Graf:2010tv,Graf:2012hb}.
For alternative approaches, 
see~\cite{Moroi:1993mb,Bolz:1998ek,Covi:2001nw,Rychkov:2007uq,Strumia:2010aa,Choi:2011yf}.}
Moreover, in those derivations, a hot thermal plasma 
consisting of the particle content of the MSSM
is considered in the high-temperature limit.
In fact, it is assumed that 
radiation governs the energy density of the Universe 
as long as thermal production of the respective EWIP is efficient,
\ie, for $T$ down to at least $T\sim 0.01~\TR$.
This is assumed in this work also.
However, we will encounter
situations with significant entropy production at smaller temperatures
generated by decays of by then non-relativistic
saxions and/or axinos from thermal processes.
Then this can dilute the yield 
of a stable or long-lived EWIP from thermal processes 
in the earliest epoch correspondingly
with dilution factors of $\Delta>1$:
\begin{equation}
Y^{\eq/\TP}_{\EWIP}
\to
\frac{1}{\Delta}
Y^{\eq/\TP}_{\EWIP}.
\label{Eq:YXoverDelta}
\end{equation}
Abundances of decoupled species that emerge from decays
of thermally produced EWIPs prior to the entropy producing event 
are equally affected.

In high-reheating temperature scenarios, the LOSP usually
freezes-out as a weakly interacting massive particle (WIMP) 
at a decoupling temperature of
$\TD^{\LOSP}\simeq\mLOSP/25$ 
with an abundance
$Y_{\LOSP}$ that can be determined by solving the corresponding
Boltzmann equations. In the case of entropy production after LOSP 
decoupling, this abundance will be diluted
\begin{equation}
Y_{\LOSP}\to
\frac{1}{\Delta}
Y_{\LOSP}
\label{Eq:YLOSPoverDelta}
\end{equation}
as well~\cite{Buchmuller:2006tt,Pradler:2006hh,Hasenkamp:2010if}.
However, in situations in which the entropy producing event
ends well before LOSP decoupling, $Y_{\LOSP}$ is not affected.
Here we assume in both cases that LOSP decoupling
takes place in a radiation-dominated epoch.
This is justified in the settings considered below
where the contribution of long-lived non-relativistic species 
to the total energy density 
(that enters the Friedmann equation)
is negligible during LOSP freeze-out.

In high-reheating temperature scenarios, 
thermal leptogenesis with hierarchical heavy Majorana neutrinos
can explain the baryon asymmetry of the Universe~\cite{Buchmuller:2005eh}.
Without late-time entropy production, 
$\MRone\sim\TR$ of at least about $10^9\,\GeV$
is then required to generate the observed baryon asymmetry $\eta$,
where $\MRone$ denotes the mass 
of the lightest among the heavy right-handed Majorana neutrinos.
With late-time entropy production, 
a baryon asymmetry generated prior to the entropy-producing events
must have been larger by the associated dilution factor $\Delta$.
In the framework of thermal leptogenesis, this can be realized
for up to $\Delta\sim 10^4$
with $\MRone\sim\TR\sim10^{13}\,\GeV$,
as can be seen in Fig.~7(a)
of Ref.~\cite{Buchmuller:2002rq} and in Fig.~2 of
Ref.~\cite{Buchmuller:2002jk};
see also~\cite{Pradler:2006hh,Hasenkamp:2010if}.
In fact, with a dilution factor of $\Delta$, 
the required minimum temperature 
for successful leptogenesis has to be larger by
that factor: 
\begin{equation}
\TR\gtrsim 10^9\,\GeV
\quad\to\quad
\frac{1}{\Delta}\,\TR\gtrsim 10^9\,\GeV .
\label{Eq:TRminTLG}
\end{equation}
Together with~\eqref{Eq:YXoverDelta} 
and~\eqref{Eq:YLOSPoverDelta},
this motivates us to carefully calculate $\Delta$
and to monitor the results 
for the two scenarios discussed in the following.

%______________________________________________
\section{Gravitino CDM Case}
\label{Sec:GravitinoCDM}
%______________________________________________
 
In this section we look at the R-parity conserving SUSY scenario 
in which a gravitino with mass $\mgravitino\gtrsim 1~\GeV$
is the stable LSP whose thermally produced density parameter
\begin{equation}
\OmegaGravitino^{\TP} h^2 
= 
\mgravitino Y_{\gravitino}^{\TP}(T_0)s(T_0)h^2/\rho_c
\label{Eq:Omegah2Gravitino}
\end{equation}
provides a substantial part of the CDM density $\OmegaDM h^2$,
where $T_{0}=0.235~\meV$ is the present photon temperature, 
$h$ the Hubble constant  
in units of $100~\km\,\Mpc^{-1}\seconds^{-1}$, 
and $\rho_{c}/[s(T_0)h^2]=3.6~\eV$.
Motivated by the recent finding of the Planck collaboration of~\cite{Ade:2013zuv}
\be
\OmegaDM h^2 = 0.1187\pm 0.0017\ (1\sigma)
\label{Eq:OmegaCDM}
\ee
obtained  
from the Planck+WP+highL+BAO data set
for the base $\Lambda$CDM model,%
\footnote{Settings beyond the base $\Lambda$CDM model
and $\Delta\neff$ contours obtained from the Planck+WP+highL+$\Hzero$+BAO data
are explored in this work.
Nevertheless, for our studies, 
we consider the upper limit~\eqref{Eq:OmegaCDMLimit} 
to be sufficiently precise.}
we will consider a nominal $3\sigma$ upper limit of
\be
\OmegaDM h^2 \leq 0.124.
\label{Eq:OmegaCDMLimit}
\ee

In the gravitino LSP case, all heavier sparticles including the LOSP 
and the axino are unstable. In turn,
each LOSP and each axino present in the Universe after
LOSP decoupling will decay directly or via a cascade
into one gravitino.
Depending on $\YLOSP$, the contribution to $\OmegaGravitino$
from decays of thermal relic LOSPs can be small as
will be discussed below in more detail.
This is different for long-lived axinos 
that decay at temperatures below a fiducial
$\TL\ll\TDLOSP$. 
For settings 
with $\OmegaGravitino^{\TP}\sim\OmegaDM$ and $\fax<10^{12}\,\GeV$,
their contribution
\be
\OmegaGravitino^{\axino\to\gravitino \X}h^{2}
=\mgravitino\YaxinoeqTP(\TL)s(T_0)h^2/\rho_c
\label{Eq:OmegaGravitinoAxinoDecay}
\ee
exceeds~\eqref{Eq:OmegaCDMLimit} by many orders of magnitude. 
This can be immediately seen 
when comparing~\eqref{Eq:YAxinoEq} 
and~\eqref{Eq:AxinoYieldTP} 
with~\eqref{Eq:YgravitinoTP}.
To avoid this excess,
we focus in this section
on $\gravitino$ LSP scenarios in which 
axinos decay dominantly into gluons and gluinos 
well before LOSP decoupling
with a rate that
can be derived from the effective Lagrangian~\eqref{Eq:eff_lag},
\be
\Gamma_{\axino} \simeq \Gamma_{\axino \to g \gluino} 
= \frac{\alpha_s^2\maxino^3}{16\pi^3 \fax^2} \left( 1-\frac{\mgluino^2}{\maxino^2} \right)^3.
\label{Eq:axino_lifetime}
\ee
While the gluinos will be brought into chemical thermal equilibrium 
when emitted prior to LOSP decoupling,
gravitinos from the rare axino decay $\axino\to a\gravitino$ 
will still contribute to the gravitino density
\begin{equation}
\OmegaGravitino^{\axino\to a\gravitino}h^2 
= 
\mgravitino \BR(\axino\to a\gravitino)\YaxinoeqTP(\TL) s(T_0)h^2/\rho_c
\label{Eq:Omega2GravitinoAxinoDecay}
\end{equation}
even when axinos decay well before LOSP decoupling,
i.e., at temperatures below the fiducial $\TL$ but above $\TDLOSP$.
The corresponding partial decay width~\cite{Chun:1993vz,Kim:1994ub}
\be
\Gamma_{\axino\to a\gravitino} 
\simeq
\frac{\maxino^5}{96\pi\MPlanck^2\mgravitino^2}
\label{Eq:GammaAxinoGravitino}
\ee
governs the branching ratio of that rare decay%
\footnote{Additional decays of the axino, e.g., 
into a neutralino LOSP or another LOSP candidate 
are possible in the considered scenarios.
The corresponding partial decay width is suppressed 
by a factor of $\Order(\alpha^{2}/\alpha_{s}^{2})$ 
with respect to $\Gamma_{\axino \to g \gluino}$ 
when $\maxino$ is well above $\mgluino$, 
where $\alpha$ denotes the fine-structure constant;
cf.\ Eq.~(4) in Ref.~\cite{Choi:2008zq}.
Their contribution to $\Gamma_{\axino}$
can then be neglected.} 
\be
\BR(\axino\to a\gravitino) 
\simeq 
\frac{\Gamma_{\axino\to a\gravitino}}{\Gamma_{\axino \to g \gluino}}
\simeq
\frac{\pi^2 \fax^2}{6 \alpha_s^2 \MPlanck^2}
\frac{\maxino^2}{\mgravitino^2}
\left(\! 1-\frac{\mgluino^2}{\maxino^2} \right)^{\!\!-3}
\!\!\!\!
,
\label{Eq:BRAxinoGravitino}
\ee
where $\MPlanck=\mPlanck/\sqrt{8\pi}=2.44\times 10^{18}\,\GeV$ 
is the reduced Planck scale
and the limit $\maxino\gg\mgravitino$ is considered.
For example, we find a small branching ratio of
$\BR(\axino\to a\gravitino)\lesssim 10^{-5}$ for $\mgravitino\gtrsim 1~\GeV$,
$\fax\lesssim 10^{11}\,\GeV$, 
and $\maxino\lesssim 6~\TeV$ well above $\mgluino\sim 1~\TeV$.
For large $\YaxinoeqTP\sim 10^{-3}$, 
$\OmegaGravitino^{\axino\to a\gravitino} h^2$ 
can still contribute significantly to the CDM density.
Accordingly, we will consider contours of 
$\OmegaGravitino^{\TP} h^2+\OmegaGravitino^{\axino\to a\gravitino} h^2=0.124$
in this section.

In the $\gravitino$ LSP scenarios considered in this section,
axions from decays of thermal saxions prior to BBN
are the only significant contribution to $\Delta\neff$, 
as already mentioned in 
Sects.~\ref{Sec:Intro} and~\ref{Sec:ExtraRadiation}.
The Lagrangian that allows for the relevant $\saxion\to a a$
decay reads~\cite{Chun:1995hc}
\begin{align}
\mathcal{L}_\text{\PQ}^\text{kin} 
= &
\left(1+\frac{\sqrt{2}x}{\vax}\saxion \right)
\label{Eq:axion_saxion}\\
&
\times
\left[ 
\frac{1}{2}\partial^{\mu} a \partial_\mu a
+\frac{1}{2}\partial^{\mu} \saxion \partial_\mu \saxion
+ i\bar{\axino}\gamma^{\mu}\partial_{\mu}\axino
\right] 
+ \dots
\nonumber
\end{align}
and the associated decay rate
\be
\Gamma_{\saxion\to aa} 
= \frac{x^2 \msaxion^3}{32\pi \fax^2},
\label{Eq:SaxionAxiAxi}
\ee
where  $x=\sum_i q_i^3v_i^2/\vax^2$
depends on the axion model
with $q_i$ denoting the charges 
and $v_i$ the vacuum expectation values 
of the fundamental 
PQ fields~\cite{Chun:1995hc}. 
For example, $x=1$ in a KSVZ axion model 
with just one \PQ\ scalar (with $q=1$ and $v=\vax$)
and $x\ll 1$ in such a model with two \PQ\ scalars 
with $q_1=-q_2=1$ and similar vacuum expectation values, 
$v_1\simeq v_2\simeq\vax/\sqrt{2}$.
The two scales $\vax = \sqrt{\sum_i v_i^2q_i^2}$ and $\fax$ 
are related via $\fax = \sqrt{2}\vax$~\cite{Graf:2012hb}.

For $\msaxion\gtrsim 1~\GeV$,
the saxion decay into two gluons, $\saxion\to g g$,
can become a competing decay mode 
towards small values of $x$.
The associated rate reads
\be
\Gamma_{\saxion\to gg} 
= 
\frac{\alphaS^2\msaxion^3}{16\pi^3\fax^2},
\label{Eq:SaxionGluGlu}
\ee
and is derived from~\eqref{Eq:eff_lag}.
The saxion decay into photons is subdominant
whenever the $\saxion\to g g$ decay is kinematically viable,
i.e., for $\msaxion$ above the threshold to form hadrons.
Saxion decays into gluinos or axinos
are kinematically not possible 
in the $\gravitino$ LSP case 
with $\msaxion=\mgravitino$.
Accordingly, the lifetime of the saxion 
and the branching ratio of its decays into axions
and into gluons are well described by
\begin{eqnarray}
&&\!\!\!\!\!\!\!\!\!\!
\tausaxion
=
\frac{1}{\Gamma_\saxion}
\simeq 
\frac{1}{\Gamma_{\saxion\to aa} 
+\Gamma_{\saxion\to gg}}
=
\frac{32\pi\fax^2}{\msaxion^3[x^2+2(\alphaS/\pi)^2]},
\label{Eq:LifetimeSaxion}
\\
&&\!\!\!\!\!\!\!\!\!\! 
\BR(\saxion\to a a)
\simeq
\frac{x^{2}}{x^2+2(\alphaS/\pi)^2},
\label{Eq:BRsaxionTwoAxions}
\\
&&\!\!\!\!\!\!\!\!\!\! 
\BR(\saxion\to gg) 
\simeq
1-\BR(\saxion\to a a)
=\!
\left[1+\frac{x^2}{2(\alphaS/\pi)^{2}}\right]^{\!-1}
\!\!\!\!\!\!\!,
\label{Eq:BRsaxionTwoGluons}
\end{eqnarray}
respectively, with $\alphaS\equiv\alphaS(\msaxion)$.
For example, for $x\gtrsim 0.2$ and $\msaxion\gtrsim 10~\GeV$, 
one finds $\BR(\saxion\to a a)\gtrsim 0.9$
so that $\tausaxion$ is governed by the decay into axions.
Towards smaller $x$ and/or $\msaxion$, the saxion decay
into gluon pairs becomes important with effects discussed below.

When decaying, 
both the axino and the saxion are non-relativistic. 
Accordingly, we encounter two types of decays
of non-relativistic particles: 
(i)~decays into axions and gravitinos
which are by then decoupled from the thermal plasma
and thereby inert relativistic species
and
(ii)~decays into relativistic species that are 
rapidly thermalized and thereby associated with entropy production.
We can indeed face simultaneously situations 
studied previously for the generic cases
of out-of-equilibrium decays of non-relativistic particles 
into inert radiation~\cite{Scherrer:1987rrPart2}
and into thermalizing radiation 
that produce entropy~\cite{Scherrer:1987rrPart1}.

Let us now calculate the contribution to $\Delta\neff$ 
of the energy density of relativistic axions $\rho_{\axion}$ 
from thermal processes in the earliest moments of the Universe,
from axino decays, 
and -- most importantly -- from late decays of thermal saxions,
\be
\Delta\neff(T) 
= 
\frac{120}{7\pi^2 T_\nu^4}\, \rho_{\axion}(T),
\label{Eq:DeltaNeffAxion}
\ee
and the relic density of gravitinos from thermal production 
and from decays of thermal axinos,
$\OmegaGravitino^{\TP}+\OmegaGravitino^{\axino\to a\gravitino}$.
By taking into account the possibility 
of entropy production in both axino and saxion decays,
we generalize and refine 
our related previous study~\cite{Graf:2012hb}.
Moreover, our numerical results are now obtained 
beyond the sudden decay approximation. 
Nevertheless, we will return to that approximation
to derive expressions that allow 
for a qualitative understanding of the behavior
of our numerical solutions in Appendix~\ref{Sec:AnalyticApprox}.

In the epoch when thermal processes involving EWIPs are no longer efficient
and when axinos and saxions from such processes are non-relativistic,
the time evolution of the energy densities 
of axinos, saxions, and relativistic axions 
is described by the following Boltzmann equations
\begin{align}
\dot{\rho}_{\axino} + 3H\rho_{\axino} 
&= -\Gamma_{\axino}\rho_{\axino}, 
\label{Eq:axinoEvol1}\\
\dot{\rho}_{\saxion} + 3H\rho_{\saxion} 
&= -\Gamma_{\saxion}\rho_{\saxion}, 
\label{Eq:saxEvol1}\\
\dot{\rho}_{\axion} + 4H\rho_{\axion} 
&\simeq \BR(\saxion\to aa)\Gamma_{\saxion}\rho_{\saxion}
\nonumber\\
&\quad +\BR(\axino\to a\gravitino)\Gamma_{\axino}\rho_{\axino}/2,
\label{Eq:axionEvol1}
\end{align}
with the Hubble expansion rate $H \equiv \dot{R}/R$, 
the dot indicating derivation with respect to cosmic time $t$,
and the second term on the right-hand side of~\eqref{Eq:axionEvol1} 
providing a valid approximation for $\maxino\gg\mgravitino$.
The time evolution of entropy $S$ is given by
\begin{align}
S^{1/3}\dot{S} 
&\simeq 
R^4\left(\frac{2\pi^2}{45}\gstarS\right)^{1/3} 
\Big\{ [1-\BR(\axino\to a\gravitino)]\Gamma_{\axino}\rho_{\axino}
\nonumber\\
&\quad\quad\quad\quad\quad\quad
+[1-\BR(\saxion\to aa)]\Gamma_{\saxion}\rho_{\saxion} 
\Big\}, 
\label{Eq:EntropyEvol1}
\end{align}
and the one of the cosmic scale factor $R$ 
by the Friedmann equation for a flat Universe
\be
H^2 \simeq \frac{8\pi}{3\mPlanck^2}
\left( 
\rho_{\axino}+\rho_{\saxion}+\rho_{\axion}+\rhoMSSMrad 
\right)
\label{Eq:Friedmann1}
\ee
with the Planck mass
$\mPlanck = 1.22\times10^{19}\,\GeV$
and the energy density of the thermal MSSM radiation background 
\be
\rhoMSSMrad 
\equiv
\frac{\pi^{2}}{30}\gstarMSSM T^{4}
=
\frac{3}{4}
\frac{\gstarMSSM}{\gstarS}
\left( \frac{45}{2\pi^2 \gstarS} \right)^{1/3} 
\frac{S^{4/3}}{R^4},
\label{Eq:RadiationfromEntropy}
\ee
where $\gstarMSSM$ is the effective number of relativistic 
degrees of freedom within the MSSM only, 
i.e., without the axion multiplet and the gravitino.
In this section, $\gstarMSSM=\gstarS$
holds for the interval over which we integrate
the Boltzmann equations. 
(This will be different in Sect.~\ref{Sec:AxionCDM}.)

Equations~\eqref{Eq:axinoEvol1}--\eqref{Eq:Friedmann1} 
form a closed set of differential equations that we solve numerically. 
We begin our computation at $t_i=1.6\times10^{-13}\,\seconds$
corresponding to $T_i=1~\TeV$ with $R(t_i)=1~\GeV^{-1}$
and end at $t_f=0.7~\seconds$ corresponding to $T_f\simeq 1~\MeV$. 
For the initial values of the energy densities, we use
\begin{align}
\rho_{\axino}(t_i) 
&=
\maxino \YaxinoeqTP s(T_i), 
\\
\rho_{\saxion}(t_i) 
&=
\msaxion \YsaxeqTP s(T_i), 
\\
\rho_{\axion}(t_i) 
&=
\langle p_{\axion,i}^\text{th}\rangle\YaxeqTP s(T_i),
\label{Eq:Initialaxions}
\end{align}
where the average thermal axion momentum is 
$\langle p_{\axion,i}^\text{th}\rangle=2.701\,T_{\axion,i}$ 
and $T_{\axion,i} = [\gstarS(T_i)/228.75]^{1/3}T_i$.
Note that saxions can be treated as a non-relativistic species 
throughout the time interval $[t_i,\,t_f]$ although a
saxion, e.g., with $\msaxion=100~\GeV$ will be relativistic
at an initial temperature of $T_{i}=1~\TeV$.
At times at which saxions are relativistic,
their contribution $\rho_{\saxion}$ on the right-hand side 
of the Friedmann equation~\eqref{Eq:Friedmann1} is negligible.
Whenever their contribution becomes sizable,
they are non-relativistic, 
which justifies the simplified treatment.

With the initial entropy $S(t_i) = s(T_i)R(t_i)^3$
and after numerical integration, we obtain the dilution factor 
\be
\Delta = \frac{S(t_f)}{S(t_i)}.
\label{Eq:Delta}
\ee
As described already in the previous section,
this factor quantifies the dilution due to entropy release
which affects the yield of species not in thermal equilibrium
such as $Y_{\gravitino}^{\TP}$ and thereby~\eqref{Eq:Omegah2Gravitino}.
The relic gravitino density from axino 
decays~\eqref{Eq:Omega2GravitinoAxinoDecay}
is affected by this dilution as well.
In fact, since $\rho_{\gravitino}^{\TP}$ and $\rho_{\gravitino}^{\axino\to a\gravitino}$
can be safely neglected in \eqref{Eq:Friedmann1}
at the considered times and since the gravitino is stable 
in the case considered here, 
it is not necessary to include the Boltzmann equation 
for the gravitino in the described calculation.
While gravitinos from $\axino\to a\gravitino$ decays 
may still be relativistic
at the onset of BBN for $\mgravitino\ll\maxino$, 
their contribution to $\Delta\neff$
is negligible in the considered parameter regions.
This holds equally for the contribution 
of the relativistic axions emitted in those decays.
In fact, the terms $\propto\BR(\axino\to a\gravitino)$ 
in~\eqref{Eq:axionEvol1} and~\eqref{Eq:EntropyEvol1}
can be set to zero as they do not affect the presented results.

\begin{figure*}[th]
\begin{center}
\includegraphics[width=.455\textwidth,clip=true]{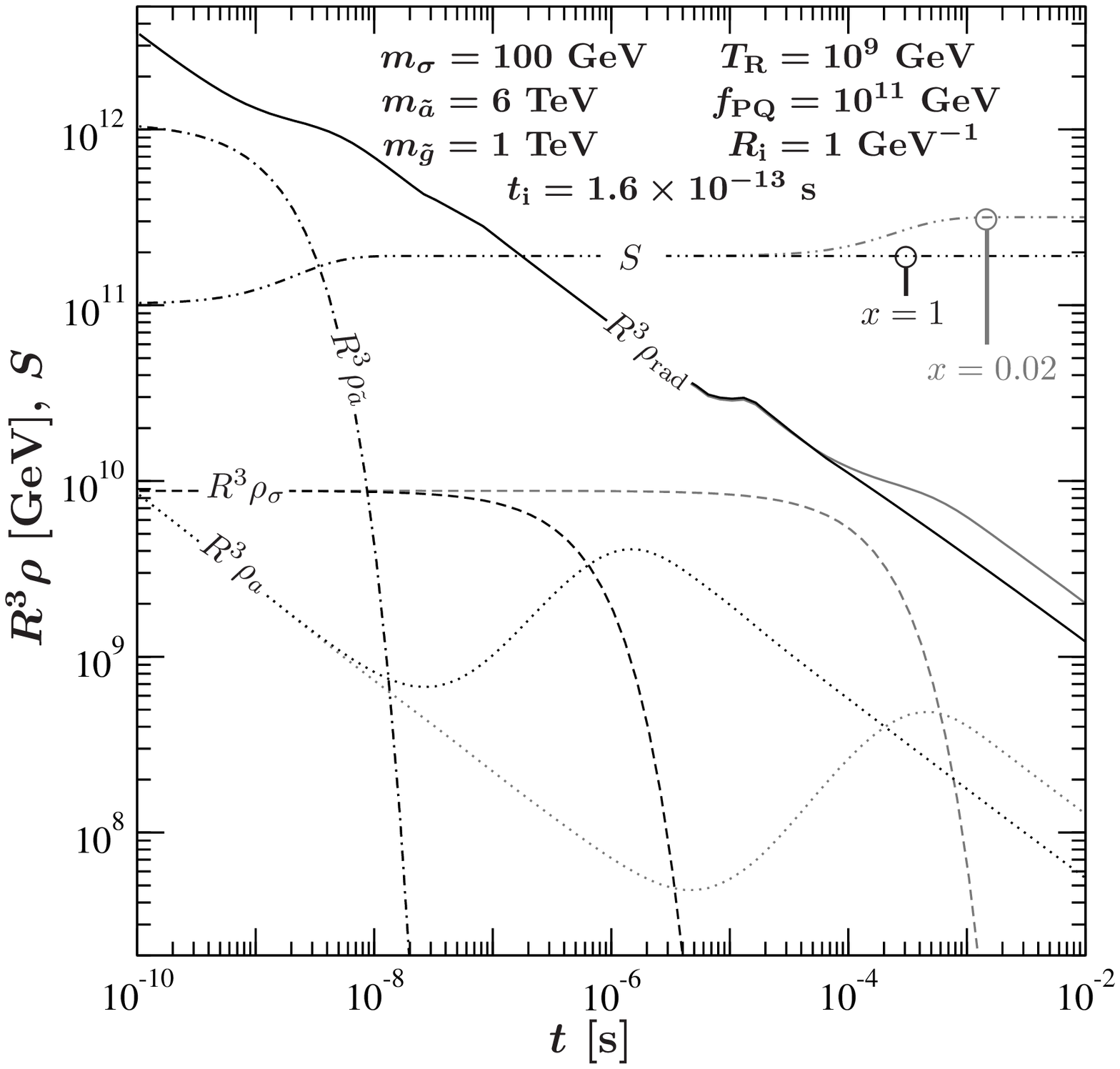}
\hskip 0.75cm
\includegraphics[width=.46\textwidth,clip=true]{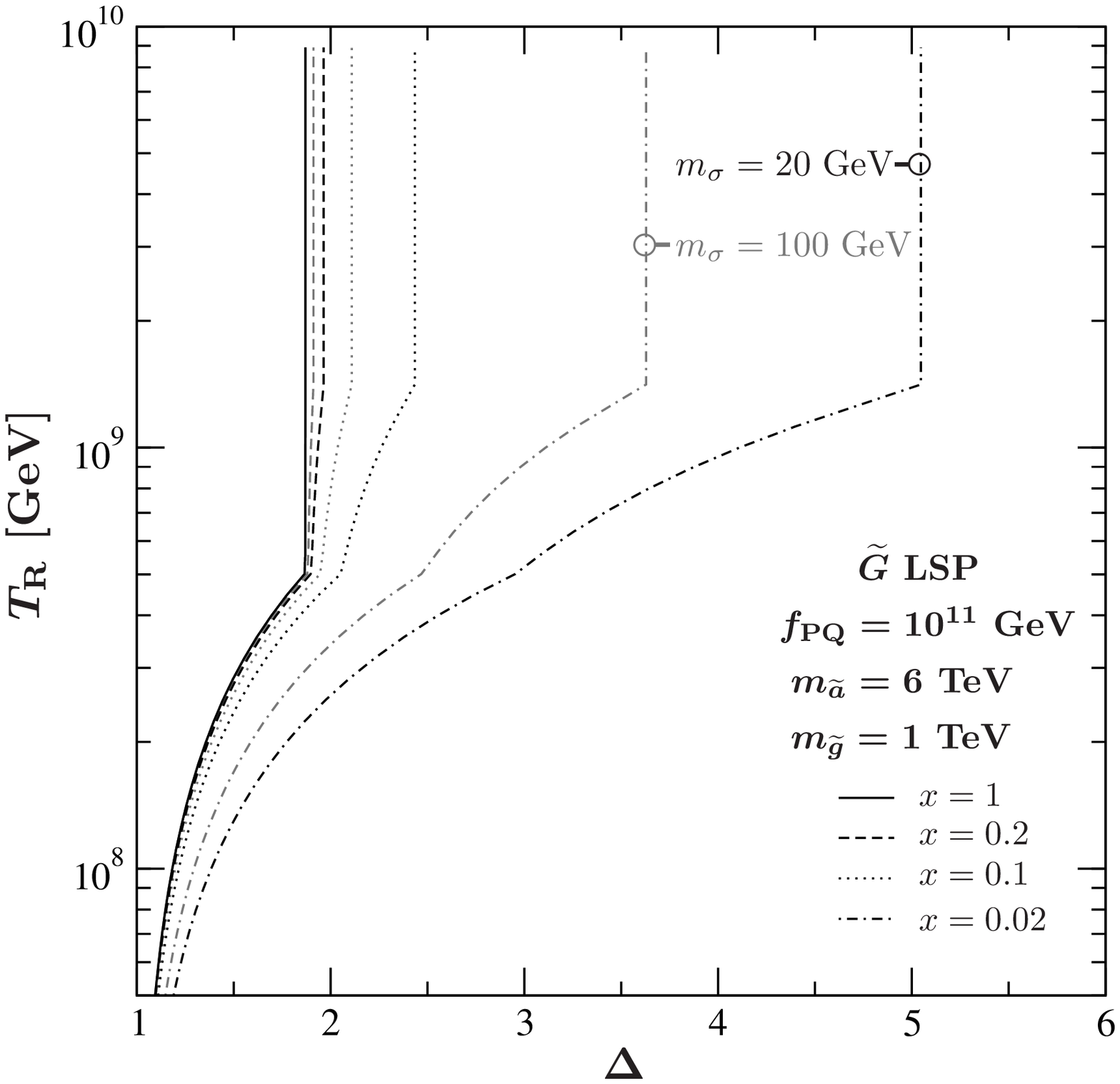} 
\vskip -0.5cm
\makebox[.46\textwidth][l]{\textbf{(a)}}\hfill
\makebox[.46\textwidth][l]{\textbf{(b)}}\hfill
\vskip 0.5cm
\caption{\small 
{(a)} 
Time evolution of the energy per comoving volume, $R^{3}\rho$,
of axinos (dash-dotted), saxions (dashed), axions (dotted) 
and other radiation (solid) and of entropy $S$ (dash-double-dotted).
Here $\msaxion=100~\GeV$, $\maxino=6~\TeV$, $\mgluino=1~\TeV$, 
$\TR=10^9\,\GeV$, and $\fax=10^{11}\,\GeV$. 
The initial value of the scale factor is set to $R_i=1~\GeV^{-1}$ 
at the temperature $T_{i}=1~\TeV$ corresponding to
a time of $t_{i}=1.6\times 10^{-13}\,\seconds$.
Black (gray) lines refer to the case with $x=1$ ($0.02$).
{(b)} The dilution factor $\Delta$ 
as a function of the reheating temperature $\TR$
for $x=1$, 0.2, 0.1, and 0.02 shown 
by the solid, dashed, dotted, and dash-dotted lines, respectively.
Black (gray) lines are obtained with $\msaxion=20~(100)~\GeV$,
whereas all other parameter are as in panel~(a).
} 
\label{Fig:GDMEvoDelta}
\end{center}
\end{figure*}

Results of our numerical integration are illustrated in Fig.~\ref{Fig:GDMEvoDelta}
for $\maxino=6~\TeV$, $\mgluino=1~\TeV$, and $\fax=10^{11}\,\GeV$.
For this setting, $\Tafter\simeq 10~\GeV$ 
is the temperature at the end of the axino-decay epoch, 
at which $\Gamma_{\axino}\simeq 3H$ is satisfied.
Therefore, a realistic LOSP with $m_\text{LOSP}\lesssim 250~\GeV$ 
is compatible with the requirement $\Tafter>\TD^{\LOSP}(\simeq \mLOSP/25)$
that is crucial as discussed at the beginning of this section.
As mentioned in Sect.~\ref{Sec:HighTR}, 
the considered gluino mass is still compatible
with limits from SUSY searches
at the LHC~\cite{ATLAS:2012us}.

Figure~\ref{Fig:GDMEvoDelta}~(a) shows 
the time evolution of the energy per comoving volume, $R^{3}\rho$,
of axinos (dash-dotted), saxions (dashed), axions (dotted) 
and other radiation (solid) and of entropy $S$ (dash-double-dotted)
for  $\msaxion=100~\GeV$ and $\TR=10^9\,\GeV$,
where black and gray lines are obtained with $x=1$ and $0.02$, respectively.
In both cases there is extra radiation in the form of axions. 
Considering the dotted lines prior to saxion decay, 
one can see clearly that only very minor contributions
reside in axions from thermal processes
or from decays of thermal axinos.
Thus, the extra radiation $\Delta\neff$ resides basically
fully in axions from saxion decays, as can be seen by
the rise of the dotted lines that results from those decays.
For $x=1$, one can see that entropy 
with a dilution factor of $\Delta\sim 2$ 
is produced in axino decays only. 
This is different for $x=0.02$ 
where additional significant late contributions to $S$ 
and thereby to $\Delta$ emerge from saxion decays into gluons. 
Considering $R^{3}\rhorad$, one sees that it decreases slower 
during the entropy producing event(s), 
whereas other dips of that solid line 
result from changes in the effective number 
of relativistic degrees of freedom.
In contrast to the axino, which is required to decay prior to LOSP freeze-out,
entropy released in late saxion decays dilutes $Y_{\LOSP}$
in addition to, e.g., $Y_{\gravitino}^{\TP}$,
$Y_{\gravitino}^{\axino\to a\gravitino}\equiv\BR(\axino\to a\gravitino)\YaxinoeqTP(\TL)$,
or the baryon asymmetry.

In general, towards small $x$, both the saxion lifetime $\tausaxion$ 
and $\BR(\saxion\to gg)$
increase which leads to larger values of $\Delta$.
This effect becomes even more pronounced towards smaller $\msaxion$
as long as the decay $\saxion\to gg$ is not kinematically suppressed.
Figure~\ref{Fig:GDMEvoDelta}(b) illustrates this behavior, 
which also becomes manifest in 
the approximations~\eqref{Eq:DeltaSaxionSmall}
and~\eqref{Eq:DeltaSaxionLarge} 
obtained in Appendix~\ref{Sec:AnalyticApprox}.
This panel of Fig.~\ref{Fig:GDMEvoDelta} shows the dilution factor $\Delta$ 
as a function of the reheating temperature $\TR$.
Here black and gray lines refer to $\msaxion=20$ and $100~\GeV$, respectively,
and are presented for $x=1$ (solid), 0.2 (dashed), 0.1 (dotted), and 0.02 (dash-dotted).
The $\TR$ dependence of $\Delta$ results 
from the one of $\YaxinoeqTP$ and of $\YsaxeqTP$;
cf.~\eqref{Eq:DeltaAxinoSmall}--\eqref{Eq:DeltaSaxionLarge}
in Appendix~\ref{Sec:AnalyticApprox}.
The kink in the $\Delta$ contour that is visible already for $x=1$ 
indicates the $\TR$ value that coincides with
the decoupling temperature of axinos $\TD^{\axino}$
given in~\eqref{Eq:TDAxino}.
In cosmological scenarios with $\TR>\TD^{\axino}$, 
$\Yaxinoeq$ applies which is independent of $\TR$.
The other kinks at larger $\TR$ that appear for $x\ll1$ 
indicate the corresponding $\TR$ value above which $\TR>\TD^{\saxion}$,
where the latter is given in~\eqref{Eq:TDSAxion}.
With a more careful treatment that includes axino/saxion
disappearance reactions when calculating the thermally produced yields
for $\TR$ near the respective decoupling temperatures,
these kinks will disappear. 
Expecting smoother curves that are close the shown ones, 
we leave such a treatment for future work.

\begin{figure*}[t]
\begin{center}
\includegraphics[width=.46\textwidth,clip=true]{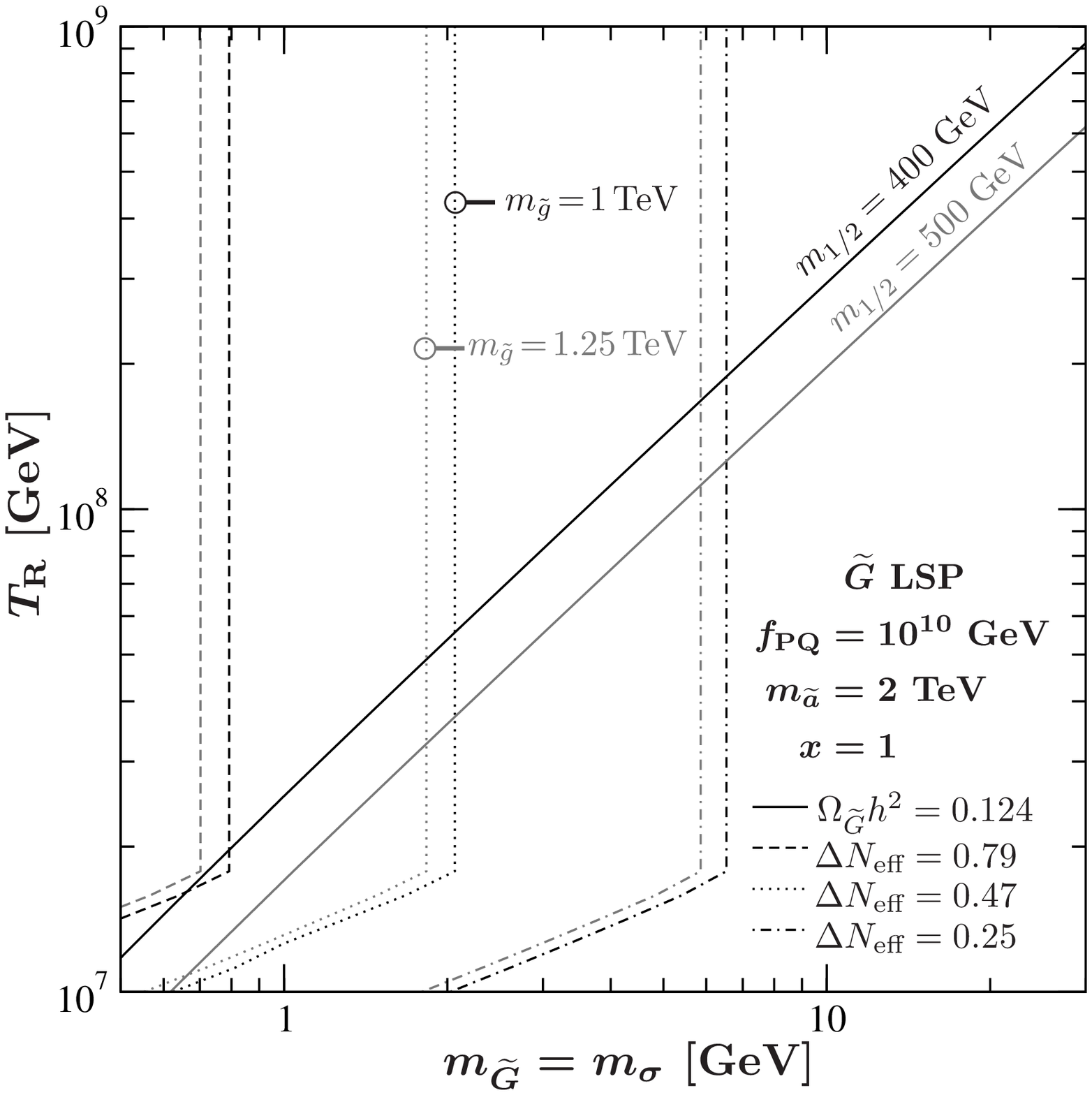} 
\hskip 0.75cm
\includegraphics[width=.468\textwidth,clip=true]{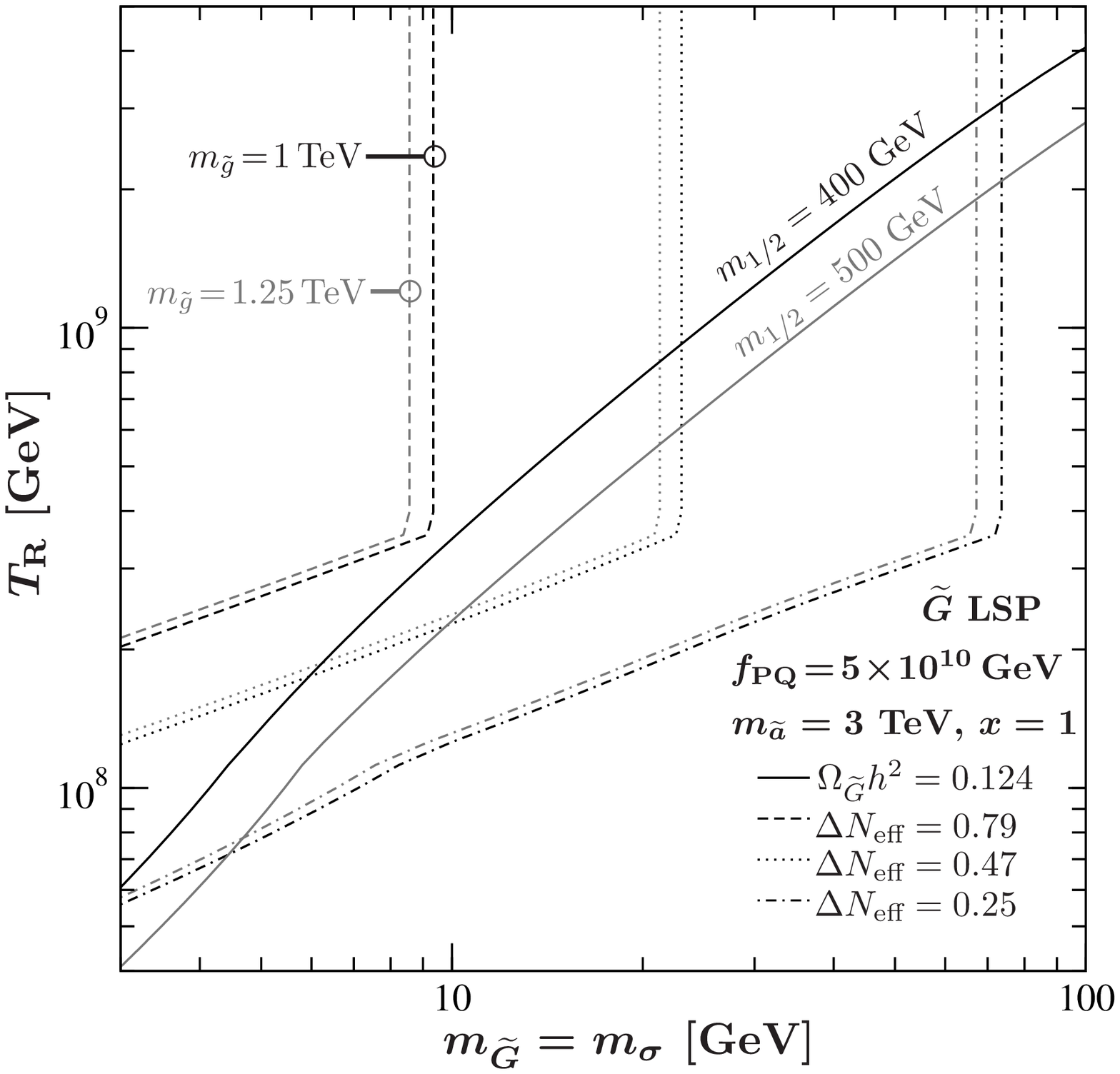} 
\vskip -0.5cm
\makebox[.46\textwidth][l]{\textbf{(a)}}\hfill
\makebox[.46\textwidth][l]{\textbf{(b)}}\hfill
\vskip 0.75cm
\includegraphics[width=.46\textwidth,clip=true]{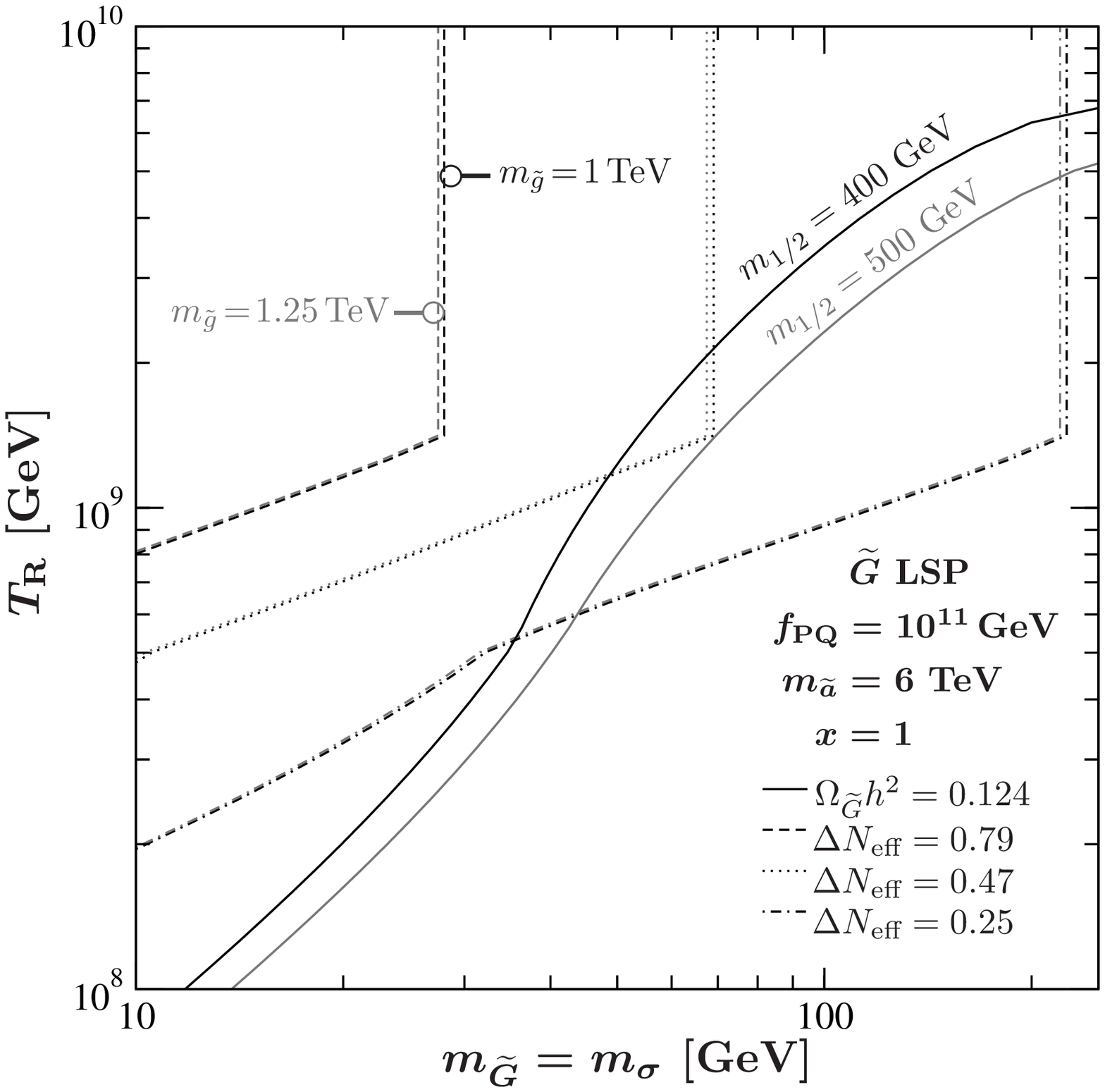} 
\hskip 0.75cm
\includegraphics[width=.475\textwidth,clip=true]{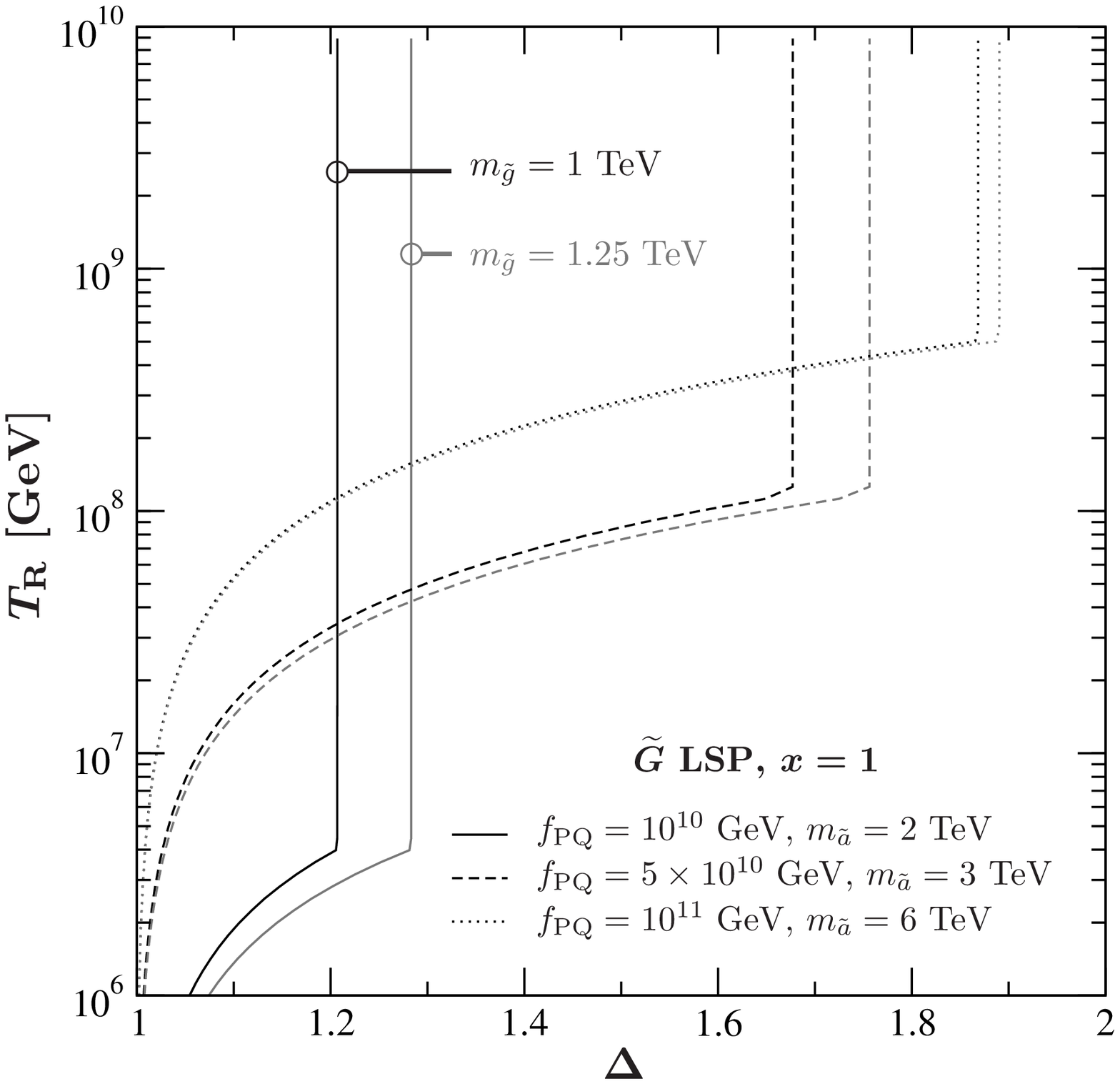} 
\vskip -0.5cm
\makebox[.46\textwidth][l]{\textbf{(c)}}\hfill
\makebox[.46\textwidth][l]{\textbf{(d)}}\hfill
\vskip 0.5cm
\caption{\small 
{(a)--(c)}~Contours of $\Delta\neff=0.25$ (dash-dotted), 
$0.47$ (dotted), and $0.79$ (dashed) 
provided by axions from decays of thermal saxions 
and of 
$\OmegaGravitino^{\TP} h^2
+\OmegaGravitino^{\axino\to a\gravitino} h^2
=0.124$ (solid)
in the $\mgravitino$--$\TR$ parameter plane
for gravitino LSP scenarios with $\msaxion=\mgravitino$ and $x=1$.
Black (gray) curves are obtained with 
$\monetwo=400~(500)~\GeV$ and $\mgluino=1~(1.25)~\TeV$. 
The PQ scale and the axino mass are set to 
(a)~$\fax=10^{10}\,\GeV$ and $\maxino=2~\TeV$,
(b)~$\fax=5\times 10^{10}\,\GeV$ and $\maxino=3~\TeV$,
and (c)~$\fax = 10^{11}\,\GeV$ and $\maxino= 6~\TeV$,
respectively.
Regions above the solid lines are
disfavored
by a gravitino density parameter 
that exceeds $\OmegaDM$ at the 3$\sigma$ level.
{(d)}~The dilution factor $\Delta$  
 as a function of the reheating temperature $\TR$ for $x=1$. 
The black (gray) solid, dashed, and dotted lines 
are obtained with $\mgluino=1~(1.25)~\TeV$
for the $f_a$ and $\maxino$ combinations 
considered in panels (a), (b), and (c), respectively.} 
\label{Fig:GCDMx1}
\end{center}
\end{figure*}

\begin{figure*}[t]
\begin{center}
\includegraphics[width=.46\textwidth,clip=true]{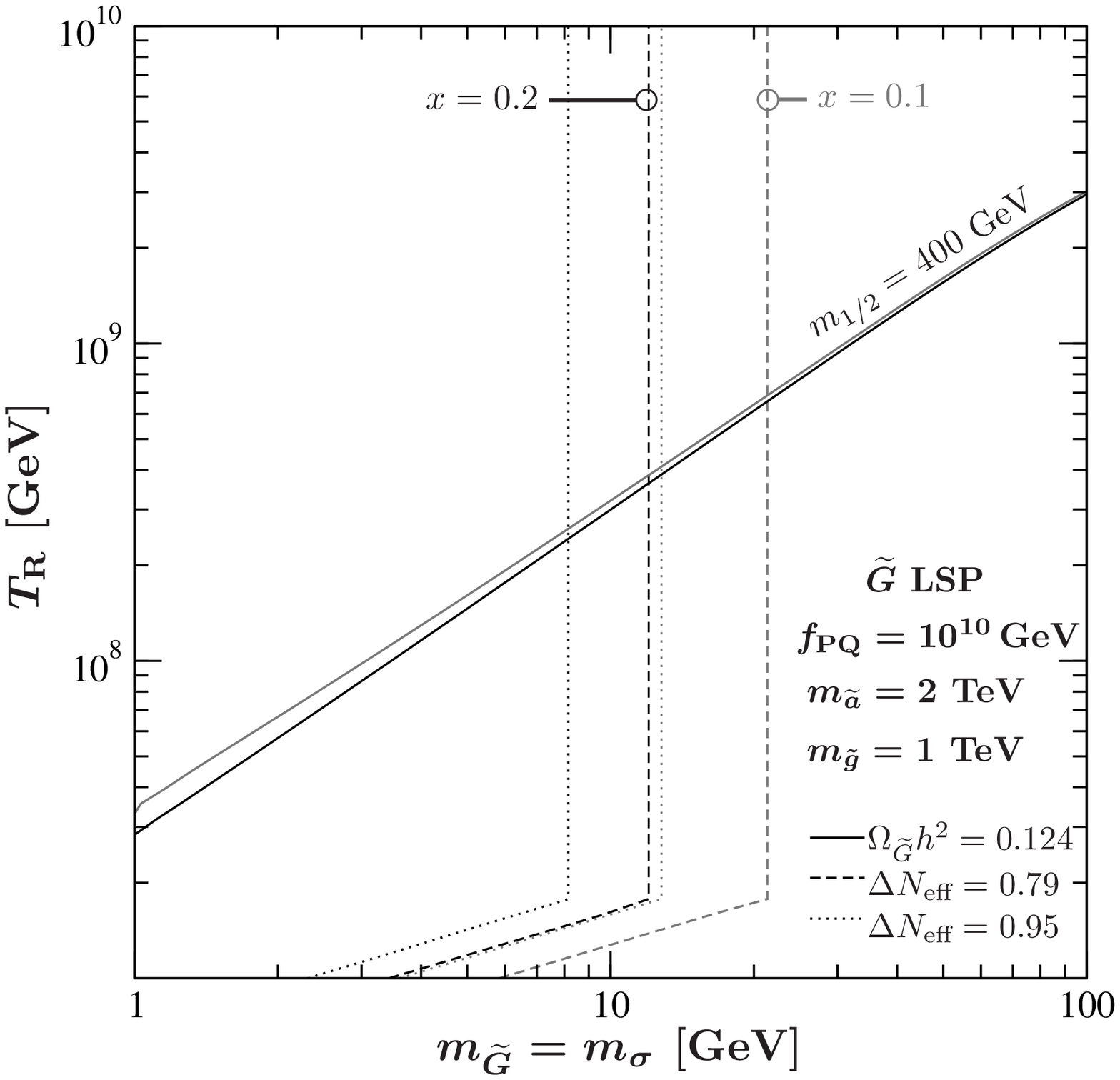} 
\hskip 0.75cm
\includegraphics[width=.45\textwidth,clip=true]{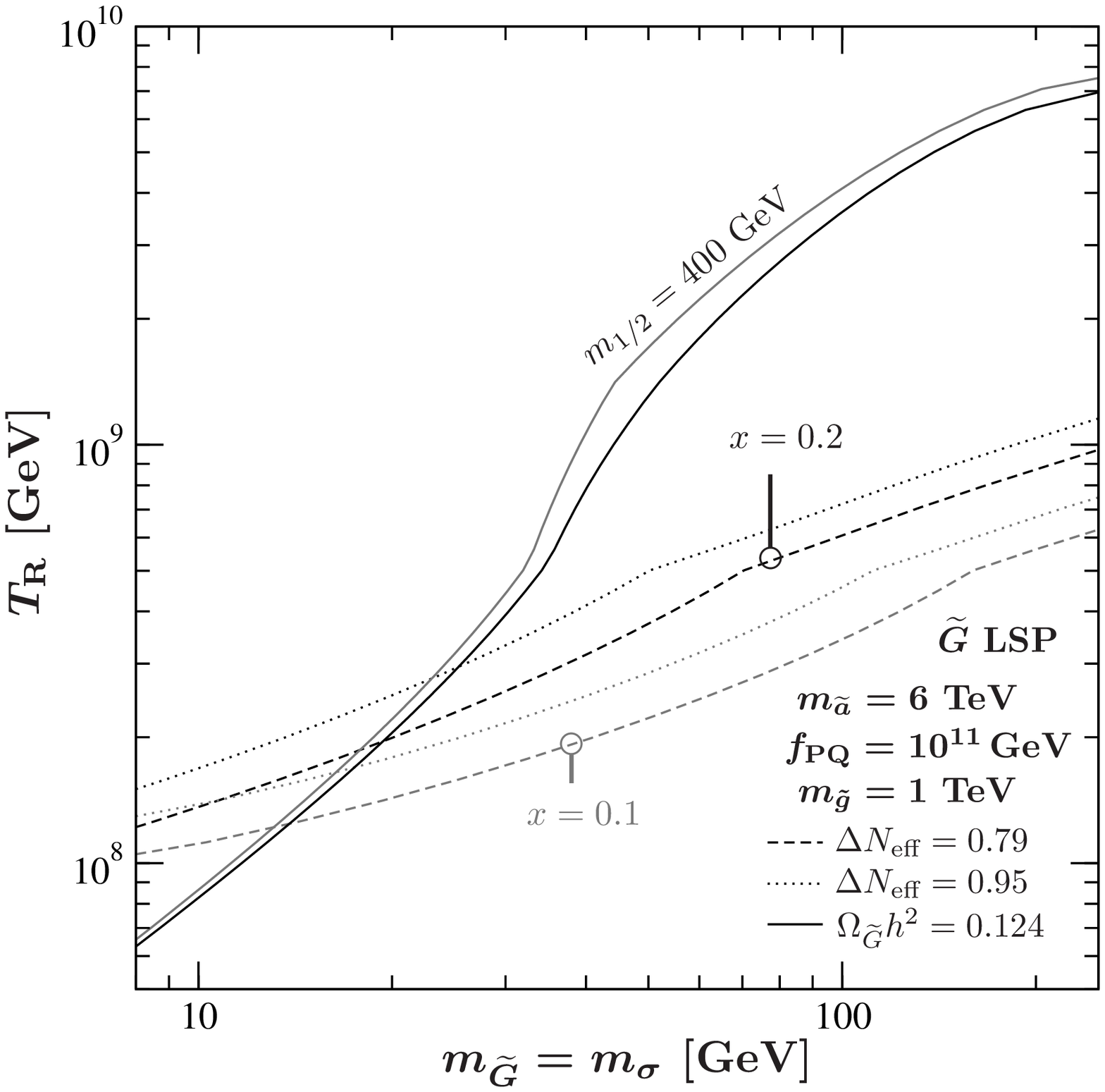} 
\vskip -0.5cm
\makebox[.46\textwidth][l]{\textbf{(a)}}\hfill
\makebox[.46\textwidth][l]{\textbf{(b)}}\hfill
\vskip 0.5cm
\caption{\small
Contours of $\Delta\neff=0.79$ (dashed) and $0.95$ (dotted)
provided by axions from decays of thermal saxions 
and of 
$\OmegaGravitino^{\TP} h^2
+\OmegaGravitino^{\axino\to a\gravitino} h^2
=0.124$ (solid)
in the $\mgravitino$--$\TR$ parameter plane
for gravitino LSP scenarios with $\msaxion=\mgravitino$,
$\monetwo=400~\GeV$, and $\mgluino=1\,\TeV$.
Black (gray) curves are obtained with $x=0.2$ ($0.1$).
The considered values of the PQ scale and the axino mass are
respectively 
(a)~$\fax=10^{10}\,\GeV$ and $\maxino=2~\TeV$
and 
(b)~$\fax = 10^{11}\,\GeV$ and $\maxino= 6~\TeV$.
Regions above the solid lines are
disfavored by a gravitino density parameter 
that exceeds $\OmegaDM$ at the 3$\sigma$ level.
In the regions above and to the left of the dashed (dotted) lines,
$\Delta\neff$ exceeds the  $2\sigma$ upper limit 
inferred from the Planck+WP+highL+BAO
(Planck+WP+highL+$\Hzero$+BAO) data set~\cite{Ade:2013zuv}.
} 
\label{Fig:GDMxplot}
\end{center}
\end{figure*}

Let us now explore systematically the amount of extra radiation 
released by saxion decays
and regions in which the constraint 
$\OmegaGravitino^{\TP}+\OmegaGravitino^{\axino\to a\gravitino}\leq\OmegaDM$
is respected.
Results for $x=1$ are presented in Fig.~\ref{Fig:GCDMx1} 
and for $x=0.1$ and $0.2$ in Fig.~\ref{Fig:GDMxplot}.
In both figures, we consider $\msaxion=\mgravitino$ and 
$\monetwo=M_i(\MGUT)$.
As already discussed in Sect.~\ref{Sec:ExtraRadiation}, 
there are hints towards the existence of extra radiation.
These hints could be an indication for the existence of axions from saxion decay.
We investigate this possibility for $\fax = 10^{10}$, $5\times10^{10}$, and $10^{11}\,\GeV$. 
For each of these values, 
$\maxino$ and $\mgluino$ are chosen such 
that the axino decay can take place 
before the freeze-out of a not too massive LOSP.
We report the considered combinations in Table~\ref{Tab:Tafter}
together with $\Tafter$ at which $\Gamma_{\axino}=3 H$
and the mass of a LOSP $\mLOSP^{\max}$
for which its decoupling temperature satisfies 
$\TD^{\LOSP}\simeq\mLOSP/25=\Tafter$.
This table shows explicitly that the viability 
of these gravitino LSP scenarios 
requires the axino to be quite heavy
and the LOSP to be relatively light.
%
% ______________________________________________________
\begin{table}[t]
 \caption{The temperature $\Tafter$ at which $\Gamma_{\axino}\simeq 3 H$
for different combinations of the PQ scale $\fax$, the axino mass $\maxino$,
and the gluino mass $\mgluino$ together with the LOSP mass $\mLOSP^{\max}$
for which $\TD^{\LOSP}\simeq\mLOSP/25\simeq\Tafter$.}
\label{Tab:Tafter}
\begin{center}
\renewcommand{\arraystretch}{1.25}
\begin{ruledtabular}
\begin{tabular*}{3.25in}{@{\extracolsep{\fill}}ccccc}
$\fax$       & $\maxino$ & $\mgluino$ & $\Tafter$  &  $\mLOSP^{\max}$
\\
$[\GeV]$	& [TeV]	    & [TeV]		& [GeV]	& [GeV]
\\ \hline
$10^{10}$ 	& 2 		    & 1 (1.25)     & 13 (9)  & 325 (225)
\\
$5\times 10^{10}$ & 3	    & 1 (1.25)     & \hphantom{0}6 (5)  & 150 (135)
\\
$10^{11}$ 	& 6 		    & 1 (1.25)     &  10 (9)  & 250 (235)
\end{tabular*}
\end{ruledtabular}
\end{center}
\end{table}
% __________________________________________________________________

Figures~\ref{Fig:GCDMx1}(a)--(c) show the amount of extra radiation $\Delta\neff$ 
provided by axions from decays of thermal saxions  for $x=1$ 
together with the upper limit on $\TR$ imposed 
by 
$\OmegaGravitino^{\TP}h^2+\OmegaGravitino^{\axino\to a\gravitino}h^2\leq 0.124$ 
at the $3\sigma$ level. 
The solid black (gray) contours indicate 
$\OmegaGravitino^{\TP}h^2+\OmegaGravitino^{\axino\to a\gravitino}h^2=0.124$  
for $m_{1/2}=400~(500)~\GeV$. The high $\TR$ regions above these contours 
are disfavored by overly efficient gravitino production.
The dashed, dotted, and dash-dotted contours indicate respectively
$\Delta\neff=0.79$, 0.47, and 0.25
and thereby the Planck+WP+highL+BAO $2\sigma$ upper limit,
the Planck+WP+highL+$\Hzero$+BAO mean,
and the Planck+WP+highL+BAO mean~\cite{Ade:2013zuv};
cf.\ Table~\ref{Tab:Neffconstrains}.
Here black (gray) contours are obtained with $\mgluino=1~(1.25)~\TeV$,
which is  compatible with $m_{1/2}=400~(500)~\GeV$ used to evaluate 
$\OmegaGravitino^{\TP}$.
The $\TR$ dependence of the $\Delta\neff$ contours disappears 
for cosmological scenarios with $\TR>\TD^{\saxion}$.
The difference between the black and gray $\Delta\neff$ contours 
for a fixed $\Delta\neff$ 
results from the dependence of the dilution factor $\Delta$ on $\mgluino$.
The corresponding dilution factors $\Delta$ can be read 
from Fig.~\ref{Fig:GCDMx1}(d).
The enhanced kinematical suppression of axino decays for a heavier gluino
leads to a longer axino lifetime and thereby to a larger~$\Delta$,
which also can be seen in~\eqref{Eq:DeltaAxinoSmall} 
of Appendix~\ref{Sec:AnalyticApprox}.
This dilutes $\YsaxeqTP$ or $\rho_{\axion}$ more strongly
and thus reduces $\Delta\neff$ correspondingly 
at a given combination of $\msaxion$ and $\TR$;
cf.~\eqref{Eq:DeltaNapprox1} and~\eqref{Eq:DeltaNapprox2} 
in Appendix~\ref{Sec:AnalyticApprox}.
Moreover, in Figs.~\ref{Fig:GCDMx1}(b) and (c), 
one can see slight kinks in the $\Delta\neff$ contours at $\TR$ values 
below $\TD^{\saxion}$. 
Those kinks appear at the same $\TR$ values 
as the kinks in Fig.~\ref{Fig:GCDMx1}(d)
and indicate the point above which $\TR>\TD^{\axino}$.
The dilution factors obtained for $\mgluino=1$ and $1.25~\TeV$
are also included in the calculation of the 
$\OmegaGravitino^{\TP}+\OmegaGravitino^{\axino\to a\gravitino}$ contours 
for $m_{1/2}=400$ and $500~\GeV$, respectively.
Indeed, the slight kinks in the 
$\OmegaGravitino^{\TP}+\OmegaGravitino^{\axino\to a\gravitino}$ contours 
visible in the panels~(b) and (c)
result from the $\Delta$ behavior shown in panel~(d).
Despite the larger $\Delta$ for larger $\mgluino$,
the $\TR$ limit imposed 
by $\OmegaGravitino^{\TP}+\OmegaGravitino^{\axino\to a\gravitino}\leq\OmegaDM$
is still more restrictive for larger $m_{1/2}$ 
due to the $M_i$ dependence of~\eqref{Eq:YgravitinoTP}.
While $\OmegaGravitino^{\TP}$ governs this limit towards 
$\fax\sim 10^{10}\,\GeV$ and $\maxino\sim 2~\TeV$
for the considered range $\mgravitino>0.5~\GeV$, 
$\OmegaGravitino^{\axino\to a\gravitino}$
becomes more relevant, e.g., 
for $\fax\sim 10^{11}\,\GeV$ and $\maxino=6~\TeV$
towards small $\mgravitino$ below $100~\GeV$; 
cf.\ Fig.~\ref{Fig:GCDMx1}(c).
%
%--------------------------------------------------
%--------------------------------------------------

As one can see from Figs.~\ref{Fig:GCDMx1}(a)--(c), 
axions from saxion decay can contribute to the amount of extra radiation.
However, for the considered $x=1$ case, values of $\Delta\neff\simeq 0.8$ 
are almost completely disfavored by the 
$\OmegaGravitino^{\TP}+\OmegaGravitino^{\axino\to a\gravitino}\leq\OmegaDM$
constraint if $\mgluino=1~\TeV$ and $\monetwo=400~\GeV$. 
In fact, if SUSY searches at the LHC point to 
minimum $\mgluino$ and $\monetwo$ values
of respectively $1.25~\TeV$ and $500~\GeV$, 
the 
$\OmegaGravitino^{\TP}+\OmegaGravitino^{\axino\to a\gravitino}\leq\OmegaDM$ constraint will clearly 
disfavor $\Delta\neff\simeq 0.8$ and the BBN-inferred posterior maxima 
$\Delta\neff=0.76$ and $0.77$ given in Table~\ref{Tab:Neffconstrains}.
Still axions from decays of thermal saxions
can then provide a viable explanation of, e.g., 
$\Delta\neff\lesssim 0.5$.
This includes the means obtained 
by the Planck collaboration~\cite{Ade:2013zuv}
as quoted in Table~\ref{Tab:Neffconstrains}.

To explore the simultaneous viability of successful leptogenesis
and an explanation of, e.g., $\Delta\neff\sim 0.25 - 0.47$
by axions from decays of thermal saxions,
one has to consider the minimum $\TR$ value together with
the dilution factors shown in Fig.~\ref{Fig:GCDMx1}(d)
as described in~\eqref{Eq:TRminTLG}.
Indeed, if the minimum $\TR$ is $10^{9}~\GeV$ without the 
entropy producing axino decays, it will become almost twice
as large in the scenarios with $\fax\gtrsim 5\times 10^{10}\,\GeV$.
Accordingly, as can be seen in Figs.~\ref{Fig:GCDMx1}(b) and~(c),
experimental insights on $\mgluino$ and $\monetwo$
will decide on such a simultaneous viability for $x=1$.
For the lower $\fax$ value considered 
in Fig.~\ref{Fig:GCDMx1}(a),
that simultaneous viability is excluded 
already with $\mgluino\simeq 1~\TeV$ and $\monetwo\simeq 400~\GeV$.

The described pictures changes considerably if $x\ll 1$.
This is shown for $x=0.2$ (black) and $0.1$ (gray)
in Figs.~\ref{Fig:GDMxplot}(a) and (b).
%
%
%--------------------------------------------------
%--------------------------------------------------
%
Here the dashed and dotted lines indicate 
$\Delta\neff=0.79$ and $0.95$, respectively.
The solid lines show 
$\OmegaGravitino^{\TP} h^2+\OmegaGravitino^{\axino\to a\gravitino} h^2=0.124$ contours.
In both panels, $\mgluino=1~\TeV$ and $m_{1/2}=400~\GeV$. 
For both (a)~$\fax=10^{10}\,\GeV$ and (b)~$10^{11}\,\GeV$,
one finds that the amount of extra radiation $\Delta\neff$
that is compatible with the 
$\OmegaGravitino^{\TP}+\OmegaGravitino^{\axino\to a\gravitino}\leq\OmegaDM$
constraint is now significantly larger
than in the corresponding $x=1$ cases.
For example, the posterior maxima 
inferred from BBN analyses, $\Delta\neff=0.79$, or $\Delta\neff=0.95$
can be easily explained by axions from thermal saxions
in the part of the $\mgravitino$--$\TR$ parameter plane
in which 
$\OmegaGravitino^{\TP}+\OmegaGravitino^{\axino\to a\gravitino}\leq\OmegaDM$.
Moreover, one sees in both panels 
that the $2\sigma$ upper limit 
from the Planck+WP+highL(+$\Hzero$)+BAO data set
translates into an upper limit on $\TR$ 
that can be significantly more restrictive than the one 
from 
$\OmegaGravitino^{\TP}+\OmegaGravitino^{\axino\to a\gravitino}\leq\OmegaDM$.

Towards smaller $x$ values in the range $0.1\lesssim x <1$,
$\Delta\neff$ increases considerably
because of the later decay of the saxion;
cf.~\eqref{Eq:LifetimeSaxion}.
At the same time, there is
also a growing branching ratio~\eqref{Eq:BRsaxionTwoGluons} 
of the entropy producing saxion decays into gluons.
By comparing the 
$\OmegaGravitino^{\TP} h^2
+\OmegaGravitino^{\axino\to a\gravitino} h^2
=0.124$ 
contours 
shown for $\monetwo=400~\GeV$ and $\fax=10^{10}\,\GeV$ 
in Fig.~\ref{Fig:GCDMx1}(a)
with the ones in Fig.~\ref{Fig:GDMxplot}(a),
one can however see that there is only a minor additional dilution 
for $x=0.1$ and $0.2$ due to $\saxion\to gg$ decays.
Also for $\fax=10^{11}\,\GeV$, the additional dilution
from saxion decays stays rather modest at those $x$ values.
This can be seen explicitly in Fig.~\ref{Fig:GDMEvoDelta}(b).
The additional kinks on the solid lines in Fig.~\ref{Fig:GDMxplot}(b)
that appear for $x=0.1$ at $\TR=\TD^{\saxion}$ can still be understood 
as a manifestation of this.
For even smaller $x$ values below $0.1$,
the dilution from saxion decays can become substantial,
as shown in Fig.~\ref{Fig:GDMEvoDelta}.
Together with the decreasing branching ratio~\eqref{Eq:BRsaxionTwoAxions},
this then leads to a reduction of $\Delta\neff$,
which can also be seen in~\eqref{Eq:DeltaNapprox2} 
of Appendix~\ref{Sec:AnalyticApprox}. 
In fact, we find the maximum viable $\Delta\neff$ values for $x\sim 0.1$.

For $0.1\lesssim x \ll 1$, 
a simultaneous viability of successful leptogenesis 
working a minimum $\TR\sim 10^{9}\,\GeV$ and
of a sizable $\Delta\neff$ provided 
by axions from decays of thermal saxions
can now be found towards $\fax\sim10^{10}\,\GeV$.
As can be inferred from Fig.~\ref{Fig:GDMxplot}(a), 
where $\Delta$ is close to 1, 
$\TR\sim 10^{9}\,\GeV$ together with $\Delta\neff\sim 0.7$
is in the allowed region 
when $x\sim 0.1$, $\mgravsax\sim 30~\GeV$ and $\fax=10^{10}\,\GeV$.
On the other hand,
towards larger $\fax\sim10^{11}\,\GeV$, 
the larger $\Delta\neff$ values obtained for $0.1\lesssim x \ll 1$
together with the 2$\sigma$ upper limits 
from the Planck collaboration~\cite{Ade:2013zuv}
impose new more restrictive $\TR$ limits
that can disfavor such a simultaneous viability.
This can be seen explicitly in Fig.~\ref{Fig:GDMxplot}(b)
for the shown $\mgravsax$ range.
Even for $\mgravsax\sim 200~\GeV$ and $x=0.2$,
that simultaneous viability is not possible 
since $\Delta$ is close to 2 for $\TR\gtrsim 10^9\,\GeV$.
A minimum of $\TR\sim 2\times 10^9\,\GeV$
will then be required for a leptogenesis scenario
working otherwise (i.e.\ for $\Delta=1$) 
at a minimum of $\TR\sim 10^{9}\,\GeV$.

At this point, it should be stressed that contributions to 
the saxion energy density can reside in coherent oscillations 
of the saxion field.
This can give additional and even dominating contributions 
to $\rho_\axion$ and thereby to $\Delta\neff$~\cite{Chang:1996ih,Asaka:1998ns,Ichikawa:2007jv,Kawasaki:2007mk,Bae:2013qr,Jeong:2013axf}.
However, these contributions depend on the initial misalignment 
of the saxion field $\sigma_i$. 
In fact, for the considered values of $\msaxion$ and $\fax$, 
the contribution of this non-thermal source is negligible 
if $\sigma_i \sim \fax$, as often assumed in the literature.

As mentioned at the beginning of this section,
we focus here on scenarios in which 
$\OmegaGravitino^{\TP}+\OmegaGravitino^{\axino\to a\gravitino}$
provides the dominant part of $\OmegaDM$.
In Figs.~\ref{Fig:GCDMx1} and~\ref{Fig:GDMxplot}, 
this holds in the region close to the solid lines
for the respective gaugino masses.
Moving away from those solid lines towards smaller $\TR$,
there is room for additional contributions to $\OmegaDM$
when assuming the considered gaugino masses.

There can be a contribution to $\OmegaDM$
from coherent oscillations of the axion field
after it acquires a mass due to instanton effects 
at $T\lesssim 1~\GeV$.
The resulting axion relic density from this misalignment mechanism 
depends on the initial misalignment angle $-\pi<\theta_i\leq\pi$ 
and $\fax$~\cite{Sikivie:2006ni,Kim:2008hd,Beltran:2006sq}
\be
\Omega\ax^\text{MIS}h^2
\sim 
0.15\,\xi\,f(\theta_i^2)\,\theta_i^2 
\left(\frac{\fax}{10^{12}\,\GeV}\right)^{7/6},
\label{Eq:OmegaAxionMIS}
\ee
where $\xi=\Order(1)$ parametrizes theoretical uncertainties
related, e.g., to details of the quark--hadron transition and
of the $T$ dependence of $\maxion$. 
Moreover, $f(\theta_i^2)$ is the anharmonicity factor
which satisfies $f(\theta_i^2)\to 1$ for small $\theta_i^{2}\to 0$
and becomes sizable towards large $\theta_{i}^{2}\to\pi^{2}$
\cite{Turner:1985si,Lyth:1991ub,Visinelli:2009zm}.
Here the possibility of late time entropy production 
is not included that can lead to a dilution of $\Omega\ax^\text{MIS}h^2$.
However, already for $\Delta=1$ and $\theta_{i}^{2}\sim 1$,
$\Omega\ax^\text{MIS}$ is only a minor fraction of $\OmegaDM$
for $\fax\lesssim 10^{11}\,\GeV$ considered in this section.
In fact, with the possibility of $\theta_{i}^{2}\ll 1$,
$\Omega\ax^\text{MIS}$ can be negligible and then
does not tighten the $\TR$ limits indicated by
the solid lines in Figs.~\ref{Fig:GCDMx1} and~\ref{Fig:GDMxplot}.

The contribution from decays of the LOSP into the gravitino LSP,
\begin{equation}
\OmegaGravitino^{\LOSP\to\gravitino\X} h^2 
= 
\mgravitino Y_{\LOSP} s(T_0)h^2/\rho_c,
\label{Eq:OmegaLOSPDecay}
\end{equation}
depends strongly on the LOSP type, its mass and couplings,
and other details of the considered point in the SUSY parameter space.
For the case in which the lightest neutralino $\neutralino$ is the LOSP, 
the yield after freeze-out can be sizable~\cite{Feng:2004mt,Kawasaki:2008qe},
\begin{equation}
Y_{\neutralino\,\LOSP}
\sim 
(1-4)\times 10^{-12}
\left(\frac{\mneutralino}{100~\GeV}\right),
\label{Eq:YLOSPNeutralino}
\end{equation}
and can thus imply $\TR$ constraints 
that are significantly more restrictive 
than those shown in Figs.~\ref{Fig:GCDMx1} and~\ref{Fig:GDMxplot}.
This holds even
with $Y_{\LOSP}$-diluting entropy production in saxion decays 
leading to the maximum contribution of $\Delta^{\saxion\to g g}\sim 2.5$
the total dilution factor $\Delta$ seen in Fig.~\ref{Fig:GDMEvoDelta}(b).
In contrast, for a charged slepton LOSP $\slepton$ 
or a sneutrino LOSP $\sneutrino$
respecting the upper limits on $\mLOSP$ 
given in Table~\ref{Tab:Tafter}, 
$Y_{\LOSP}$ is relatively 
small~\cite{Asaka:2000zh,Feng:2004mt,Kawasaki:2008qe,Freitas:2011fx},
\begin{eqnarray}
Y_{\slepton\,\LOSP}
&\lesssim& 
(0.7-2)\times 10^{-13}
\left(\frac{\mslepton}{100~\GeV}\right),
\label{Eq:YLOSPSlepton}\\
Y_{\sneutrino\,\LOSP}
&\sim& 
2\times 10^{-14}
\left(\frac{\msneutrino}{100~\GeV}\right),
\label{Eq:YLOSPSneutrino}
\end{eqnarray}
and basically negligible 
already without a possible dilution~\eqref{Eq:YLOSPoverDelta}.
Then the $\TR$ limits imposed by 
$\OmegaGravitino^{\TP}
+\OmegaGravitino^{\axino\to a\gravitino}
+\OmegaGravitino^{\LOSP\to\gravitino\X}
\leq\OmegaDM$
are very similar to the ones indicated by
the solid lines in Figs.~\ref{Fig:GCDMx1} and~\ref{Fig:GDMxplot}.

In the considered gravitino LSP scenarios with a LOSP being the NLSP,
the LOSP has a long lifetime before decaying into the gravitino.
Often such decays are found to take place during and after BBN.
For the $\neutralino$ LOSP, decays such as 
$\neutralino\to\gravitino\quark\antiquark$~\cite{Feng:2004mt}
can then reprocess the primordial light elements
via electromagnetic and hadronic energy injection.
Thereby, the observationally inferred primordial abundances
of those elements translate into upper limits on $\YLOSP$ 
that depend on the lifetime of the LOSP $\tau_{\LOSP}$.
Towards small values of $\tau_{\LOSP}$ 
which occur towards smaller values of $\mgravitino$,
the $\YLOSP$ limits become weaker and disappear.
For a $\neutralino$ LOSP 
with mass $m_{\neutralino}\leq\mLOSP^{\max}$
and the latter given in Table~\ref{Tab:Tafter}, 
BBN constraints exclude $\mgravitino\gtrsim 1~\GeV$
and thereby most of the interesting parameter regions 
considered above~\cite{Feng:2004mt,Kawasaki:2008qe}.

For the charged slepton and sneutrino LOSP cases, 
hadronic energy injection requires 4-body decays such as
$\slepton\to\gravitino\lepton\quark\antiquark$~\cite{Steffen:2006hw} or
$\sneutrino\to\gravitino\nu\quark\antiquark$~\cite{Kanzaki:2006hm}
and is thereby less efficient.
However, a long-lived charged slepton can form bound states 
with the primordial nuclei and thereby catalyze, e.g., the
primordial production of lithium-6 substantially.
This catalyzed BBN (CBBN) then imposes the upper limit
$\tau_{\slepton}\lesssim 5\times 10^{3}\,\seconds$~\cite{Pospelov:2006sc}.
Together with an upper limit on the slepton mass
$m_{\slepton}\leq\mLOSP^{\max}\lesssim 300~\GeV$
imposed by axino cosmology,
this translates into the constraint 
$\mgravitino\lesssim 4~\GeV$~\cite{Steffen:2006wx,Pospelov:2008ta}, 
which again disfavors most of the interesting parameter regions 
considered above.
Moreover, the current lower limit from searches 
for long-lived charged sleptons at the LHC, 
$m_{\slepton}\gtrsim 300~\GeV$~\cite{Chatrchyan:2012sp,ATLAS-CONF-2012-075},
is already in conflict with most of the $\mLOSP^{\max}$
values listed in Table~\ref{Tab:Tafter}.
In fact, the production of 
a long-lived charged slepton LOSP
could leave clear signatures at the LHC
and allow for a precise measurement of its mass.
For example, with $m_{\slepton}\sim 400~(500)~\GeV$,
axino-imposed constraints become difficult to evade 
and the CBBN limit on $\tau_{\slepton}$ implies 
$\mgravitino\lesssim 6~(10)~\GeV$~\cite{Steffen:2006wx,Pospelov:2008ta}.%
\footnote{Scenarios in which long-lived staus 
have an exceptionally small thermal relic abundance
well below~\eqref{Eq:YLOSPSlepton}
have been found in which CBBN limits 
can be evaded~\cite{Ratz:2008qh,Pradler:2008qc,Lindert:2011td}.
However, these scenarios often require a relatively light stau mass of
$m_{\tilde{\tau}_{1}}\lesssim 200~\GeV$ 
which is in conflict with the mentioned limit from LHC searches~\cite{Chatrchyan:2012sp,ATLAS-CONF-2012-075}.}
Such a discovery will thus not be compatible 
with the $\Delta\neff$ explanation  
via decays of thermal saxions for $\TR\gtrsim 10^{9}\,\GeV$.
It may instead point to smaller $\TR<10^8\,\GeV$, 
smaller $\fax\lesssim 10^{10}\,\GeV$, and larger $\maxino$
or to the axion CDM scenarios with an eV-scale axino LSP
and the gravitino NLSP considered in the next section.

For the sneutrino LOSP case, 
the presented high-$\TR$ explanations 
of additional radiation via decays of thermal saxions
are still viable.
The BBN constraints 
imposed by hadronic and electromagnetic energy release
become relevant only 
for large $m_{\sneutrino}\gtrsim 500~\GeV$~\cite{Kanzaki:2006hm,Kawasaki:2008qe}.
At smaller $m_{\sneutrino}\leq\mLOSP^{\max}$,
the only bound on the gravitino mass 
then results from the hierarchy
$\mgravitino<m_{\sneutrino}$
assumed in this section.
In comparison to the $\slepton$ LOSP, 
it will be much more challenging 
to identify a sneutrino $\sneutrino$ as the LOSP
and to measure its mass at the 
LHC~\cite{Covi:2007xj,Ellis:2008as,Katz:2009qx,Figy:2010hu}.
Such a measurement
will allow us to test the presented scenarios in two ways:
(i)~by confronting $m_{\sneutrino}$ with 
the upper limit $\mLOSP^{\max}$ imposed by the axino
and (ii)~by exploring the maximum $\TR$ values
for the maximum viable mass 
of the gravitino LSP
which is then $\mgravitino=m_{\sneutrino}$.

%______________________________________________
\section{Axion CDM Case}
\label{Sec:AxionCDM}
%______________________________________________

In this section we consider SUSY scenarios
in which the axion with a mass of $\maxion\sim 6~\mueV$
provides the CDM density $\OmegaDM h^{2}$
via the misalignment mechanism.
The associated relic density $\Omega\ax^\text{MIS}h^2$
that resides in coherent oscillations of the axion field
is given by~\eqref{Eq:OmegaAxionMIS}
in the absence of late-time entropy production.
With entropy production 
after the QCD phase transition, $T\ll 1~\GeV$,
the corresponding dilution factor $\Delta$ 
has to be taken into account
that reduces the density parameter by a factor of $1/\Delta$. 
Accordingly,
$\Omega\ax^\text{MIS}h^2=\OmegaDM h^{2}$ 
holds with $\fax=10^{12}\,\GeV$, e.g., 
for $f(\theta_{i}^{2})\theta_{i}^{2}\sim 1$ and $\Delta\sim 1$
or equally for $f(\theta_{i}^{2})\theta_{i}^{2}\sim 10$ and $\Delta\sim 10$.
In fact, 
also in the situations with a sizable $\Delta\sim 30$ encountered below, 
the CDM density can be explained fully by the axion condensate 
provided $\theta_{i}^{2}$ and 
the associated anharmonicity factor $f(\theta_{i}^{2})$
are sufficiently large to compensate for the dilution.

In a setting with $\OmegaDM$ provided by the axion condensate,
the LSP is no longer required to be a CDM particle.
In turn, the LSP can be a very light particle such as an
axino with $\maxino\lesssim 37~\eV$,
which is the scenario considered in this section.
Such a light axino can still be produced thermally when $\TR<\TD^{\axino}$
or decouple as a thermal relic when $\TR>\TD^{\axino}$.
The resulting population can contribute to hot dark matter (HDM).
In fact, the upper limit of $\maxino\lesssim 37~\eV$
is inferred from LSS constraints on HDM contributions 
in mixed models with CDM~\cite{Freitas:2011fx}.
When relativistic, the axino population from thermal processes 
contributes a small amount of
$(\Delta\neff)_{\axino}^{\mathrm{eq/TP}}\lesssim 0.017$~\cite{Freitas:2011fx}
to dark radiation, 
which is not included in our calculations below.

In our considerations
the gravitino is the NLSP that is lighter than the LOSP, 
i.e., than the lightest sparticle in the MSSM.
Thereby, the explored scenarios are not subject 
to the restrictive upper limits on $\TR$ 
imposed by BBN constraints
on hadronic/electromagnetic energy injection
in late decays of gravitinos 
into MSSM particles~\cite{Olive:1984bi,Asaka:2000ew}.
In the R-parity conserving settings considered in this section,
gravitinos can decay into axions and axinos only.
The gravitino lifetime $\tau_{\gravitino}$
is then governed by the associated decay 
rate~\cite{Olive:1984bi,Asaka:2000ew}
\be
\Gamma_{\gravitino\to a\axino} 
= 
\frac{\mgravitino^3}{192\pi\MPlanck^2}
=\frac{1}{\tau_{\gravitino}}.
\label{Eq:GammaGravitino}
\ee
Accordingly, gravitinos can be very long-lived.
For example, $\tau_{\gravitino}\simeq 10^{10}\,\seconds$ 
and $10^{5}\,\seconds$
for $\mgravitino\simeq 60~\GeV$ and $6~\TeV$, respectively.
The axions and axinos emitted in
decays of a thermally produced gravitino population 
can thereby contribute substantially to $\Delta\neff$
at late times well after BBN~\cite{Ichikawa:2007jv,Hasenkamp:2011em,Graf:2012hb}.
In fact, the time at which 
the smallest observable modes of the CMB reenter the horizon, 
$t=5.2\times10^{10}\,\seconds$, imposes an upper limit
on $\tau_{\gravitino}$ 
because of the non-observation of a significant release 
of extra radiation thereafter~\cite{Fischler:2010xz}.
The corresponding mass limit is $\mgravitino\gtrsim 35~\GeV$.
For $\msaxion=\mgravitino$, this limit implies
that saxions decay before the onset of BBN
even when small $x$ values of are considered.
With~\eqref{Eq:LifetimeSaxion} being valid 
to a very good approximation in this section also,
one finds $\tau_\saxion\lesssim 0.4~\seconds$ for 
$\fax = 10^{12}\,\GeV$ and $x\gtrsim 0.01$.

While the mass hierarchy $\msaxion\gg\maxino$ 
and the Lagrangian~\eqref{Eq:axion_saxion} now allow
for the additional $\saxion\to\axino\axino$ decay channel, 
the corresponding decay width
\be
\Gamma_{\saxion\to\axino\axino} 
= 
\frac{x^2 \msaxion\maxino^{2}}{\pi\fax^2}
\left[1-\left(\frac{2\maxino}{\msaxion}\right)^{2}\right]
\label{Eq:SaxionAxinoAxino}
\ee
is suppressed by a factor 
of at least $32\maxino^{2}/\msaxion^{2}$
with respect to $\Gamma_{\saxion\to a a}$ 
given in~\eqref{Eq:SaxionAxiAxi}
and thereby negligible for the considered mass splittings.
The saxion lifetime and the relevant branching ratios 
are thus again described
by~\eqref{Eq:LifetimeSaxion}, \eqref{Eq:BRsaxionTwoAxions}, 
and~\eqref{Eq:BRsaxionTwoGluons}, respectively.

As in the previous section,
we encounter the two types of decays of non-relativistic particles.
However, in the scenarios in the previous section,
$\Delta\neff$ originates basically from saxion decays only
and entropy production at two very different times is possible. 
Now there are two possibly significant sources of extra radiation, 
saxion decays and gravitino decays, 
which proceed at very different times, 
whereas entropy can be produced in saxion decays only.
In the following we thus calculate
\be
\Delta\neff(T) 
= 
\frac{120}{7\pi^2 T_\nu^4}\, \rhodr(T),
\label{Eq:DeltaNeffDR}
\ee
where the energy density of dark radiation $\rhodr$  
includes contributions 
of axions from thermal processes in the early universe,
of axions from decays of thermal saxions, 
and 
of axions and axinos from decays of thermally produced gravitinos.
As in the previous section, the possibility of entropy production
in saxion decays is taken into account and decays are treated
beyond the sudden decay approximation. 
Thereby, we update and generalize existing results 
presented in Refs.~\cite{Hasenkamp:2011em,Graf:2012hb}.
For a qualitative understanding of our numerical results,
we again refer to the expressions obtained 
in Appendix~\ref{Sec:AnalyticApprox}.

The following Boltzmann equations describe the time evolution
of the energy densities of gravitinos, saxions, and dark radiation
\begin{align}
\dot{\rho}_{\gravitino} + 3H\rho_{\gravitino} 
&= -\Gamma_{\gravitino} \rho_{\gravitino}, 
\label{Eq:gravEvol2}
\\
\dot{\rho}_{\saxion} + 3H\rho_{\saxion} &= -\Gamma_{\saxion} \rho_{\saxion}, 
\label{Eq:saxEvol2}
\\
\dot{\rho}_{\dr} + 4H\rhodr &= \BR(\saxion\to aa)\Gamma_{\saxion} \rho_{\saxion} + \Gamma_{\gravitino} \rho_{\gravitino},
\label{Eq:drEvol2}
\end{align}
in the epoch well after the one 
in which thermal processes involving EWIPs were efficient 
and when gravitinos and saxions from such processes are non-relativistic.
Here the time evolution of the entropy $S$ and the scale factor $R$
are described respectively by
\be
S^{1/3}\dot{S} = R^4 \left( \frac{2\pi^2}{45} \gstarS \right)^{1/3}  
\left[ 1 - \BR(\saxion\to aa)\right]\Gamma_{\saxion} \rho_{\saxion}.
\label{Eq:EntropyEvol2}
\ee
and the Friedmann equation
\be
H^2 
\simeq 
\frac{8\pi}{3\mPlanck^2}
(\rhodr+\rho_{\saxion}+\rho_{\gravitino}+\rho_\text{rad}),
\label{Eq:Friedmann2}
\ee
with $\rho_\text{rad}$ as given in~\eqref{Eq:RadiationfromEntropy}. 

We solve the closed set of differential 
equations~\eqref{Eq:gravEvol2}--\eqref{Eq:Friedmann2}
numerically.
As in the previous section, 
we start at $t_i=1.6\times10^{-13}\,\seconds$
corresponding to $T_i=1~\TeV$ with $R(t_i)=1~\GeV^{-1}$.
However, the considered end of the evolution is now
set to a much later time of $t_f=10^{12}\,\seconds$ 
corresponding to $T_f\simeq 1~\eV$. 
The initial values of the energy densities are given by
\begin{align}
\rho_{\gravitino}(t_i) 
&= 
\mgravitino \YgravitinoTP s(T_i), 
\\
\rho_{\saxion}(t_i) 
&= 
\msaxion \YsaxeqTP s(T_i), 
\\
\rhodr(t_i) 
&= 
\langle p_{\axion,i}^\text{th}\rangle\YaxeqTP s(T_i), 
\label{Eq:YiDR}
\end{align}
and of the entropy by $S(t_i) = s(T_i)R(t_i)^3$.
Entropy production is quantified 
by the dilution factor $\Delta$
given as in~\eqref{Eq:Delta}.
Note that the contribution of the energy density of axinos
from thermal processes in the early universe can be  
neglected in~\eqref{Eq:Friedmann2} at the considered times.
Also in~\eqref{Eq:YiDR},
this population is neglected,
which contributes at most  
$(\Delta\neff)_{\axino}^{\mathrm{eq/TP}}\sim 0.017$~\cite{Freitas:2011fx},
as mentioned above.
The Boltzmann equation for cold dark matter axions
and the associated contribution in~\eqref{Eq:Friedmann2}
are not mentioned above.
In fact, including this population explicitly
leads to at most a 1-2\% effect in $\Delta\neff$  
and only in settings with 
$\tau_{\gravitino}\gtrsim 10^{10}\,\seconds$.
As in the saxion treatment in the previous section, 
saxions and gravitinos are described as non-relativistic species
throughout the time interval $[t_i,\,t_f]$ although 
saxions and gravitinos, 
e.g., with $m_{\saxion,\gravitino}=100~\GeV$ will be relativistic
at an initial temperature of $T_{i}=1~\TeV$.
This simplified treatment is justified since the contributions
of saxions and gravitinos to the right-hand side 
of the Friedmann equation~\eqref{Eq:Friedmann2}
become relevant only when they are non-relativistic.

\begin{figure*}[t]
\begin{center}
\includegraphics[width=.455\textwidth,clip=true]{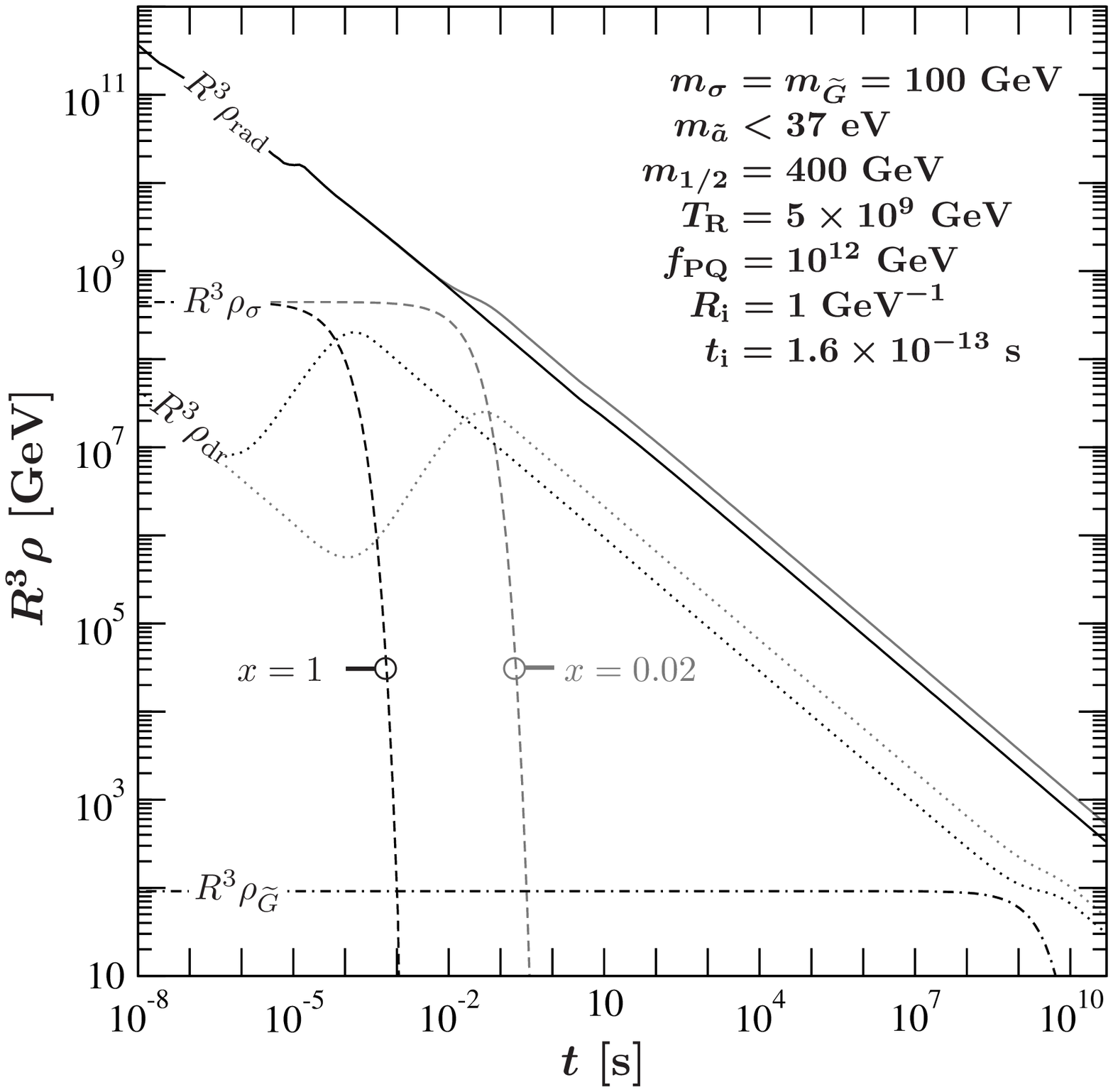}
\hskip 0.75cm
\includegraphics[width=.46\textwidth,clip=true]{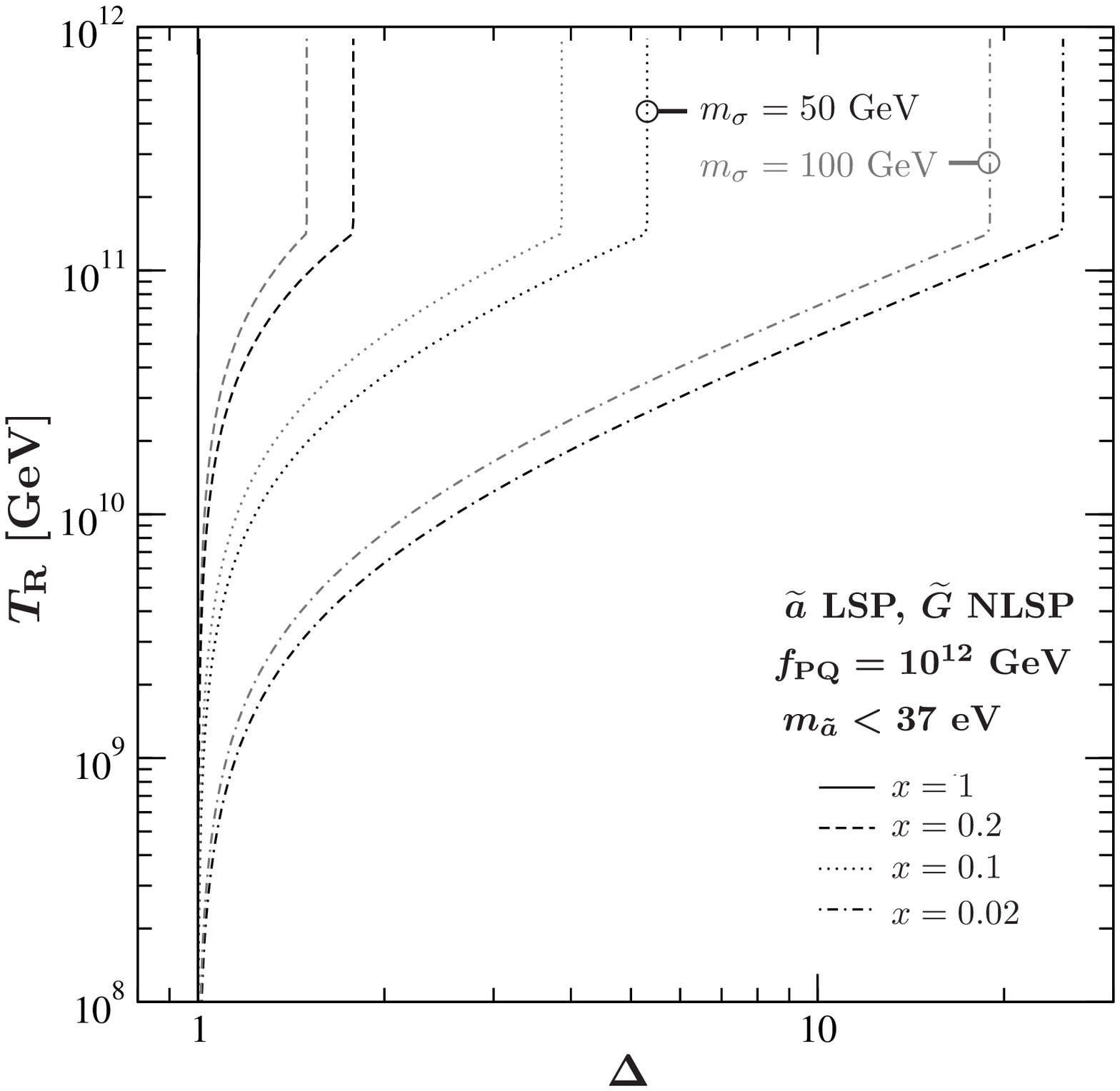} 
\vskip -0.5cm
\makebox[.46\textwidth][l]{\textbf{(a)}}\hfill
\makebox[.46\textwidth][l]{\textbf{(b)}}\hfill
\vskip 0.5cm
\caption{\small  
{(a)}~Time evolution of the energy per comoving volume, $R^{3}\rho$, 
of saxions (dashed), gravitinos (dash-dotted),
dark radiation in the form of axions and axinos (dotted)
and other radiation (solid).
Here $\msaxion=\mgravitino=100~\GeV$, $\maxino\lesssim 37~\eV$,
$m_{1/2}=400~\GeV$, 
$\TR=5\times10^{9}\,\GeV$, and $\fax=10^{12}\,\GeV$. 
The initial value of the scale factor is set to $R_i=1~\GeV^{-1}$ 
at an initial temperature of $T_{i}=1~\TeV$ corresponding to
an initial time of $t_{i}=1.6\times 10^{-13}\,\seconds$.
Black (gray) lines refer to the case with $x=1$ ($0.02$).
{(b)}~The dilution factor $\Delta$  
as a function of the reheating temperature $\TR$
for $x=1$, 0.2, 0.1, and 0.02 shown 
by the solid, dashed, dotted, and dash-dotted lines, respectively,
in axino LSP scenarios with the gravitino NLSP.
Black (gray) lines are obtained with 
$\msaxion=50~(100)~\GeV$,
whereas all other parameter are as in panel~(a).
} 
\label{Fig:aDMEvoDelta}
\end{center}
\end{figure*}

Figure~\ref{Fig:aDMEvoDelta}(a) presents the results of 
the numerical integration for 
$\msaxion=\mgravitino=100~\GeV$, 
$\TR=5\times10^{9}\,\GeV$, 
$\fax=10^{12}\,\GeV$, and
universal gaugino masses at the GUT scale of
$m_{1/2} = 400~\GeV$,
which is compatible with $\mgluino=1~\TeV$
at collider energies.
The time evolution of $R^3\rho$ is shown 
for saxions (dashed), gravitinos (dash-dotted), 
dark radiation (dotted), and other radiation (solid),
where black and gray lines refer to $x=1$ and $0.02$, respectively.
The evolution of entropy $S$ is not shown.
For $x=1$, it is simply a horizontal line and $\Delta=1$.
In the case with $x=0.02$, 
it shows an increase by a factor of $\Delta\simeq 1.5$
when the saxion decay occurs. 
The latter dilution factor can be inferred also 
from the difference of the two solid curves. 
Here one can see that the energy density of the universe 
can be dominated by non-relativistic saxions
just before/during their decay, 
which indicates an early intermediate matter-dominated epoch.
In fact, such an epoch can be even more pronounced
towards larger $\TR$ and/or smaller $\msaxion$
and thereby lead to significantly larger $\Delta$ values, 
as illustrated in Fig.~\ref{Fig:aDMEvoDelta}(b).

In Fig.~\ref{Fig:aDMEvoDelta}(b) 
the $\TR$ dependence of the dilution factor $\Delta$
is shown for $x=1$, 0.2, 0.1, and 0.02 
by the solid, dashed, dotted, and dash-dotted curves, respectively.
Black (gray) lines refer to $\msaxion=\msaxion=50~(100)~\GeV$, 
whereas all other parameters are as in panel~(a). 
The $\TR$ dependence results from the one of $\YsaxeqTP$, 
which explains the kinks at $\TR=\TD^{\saxion}$
encountered already in the previous section.
Again there is an increase of $\Delta$ towards small $x$
due to larger values of $\tau_{\saxion}$ and $\BR(\saxion\to gg)$.
For $x\lesssim 0.1$ and towards large $\TR\gtrsim\TD^{\saxion}$,
$\Delta$ can now be much larger 
than in the previous section because here $\fax=10^{12}\,\GeV$.
The latter implies a larger saxion lifetime~\eqref{Eq:LifetimeSaxion}, 
whereas $\Ysaxeq$ is independent of $\fax$;
see also approximations~\eqref{Eq:DeltaSaxionSmall} 
and~\eqref{Eq:DeltaSaxionLarge} in Appendix~\ref{Sec:AnalyticApprox}.

Let us now turn to $\Delta\neff$. 
The dotted line in Fig.~\ref{Fig:aDMEvoDelta}(a)
illustrates that there are two sizable contributions
at very different times, as advertised above:
$\Delta\neff^{\saxion\to a a}$ residing in axions 
from decays of thermal saxions and
$\Delta\neff^{\gravitino\to a\axino}$
residing in axions and axions 
from decays of thermally produced gravitinos.
For $x=1$ ($0.02$), 
there is an early contribution 
of $\Delta\neff^{\saxion\to aa}\simeq 0.29$ ($0.4$)
and an additional late contribution 
of $\Delta\neff^{\gravitino\to a\axino}=0.39$ ($0.25$)
leading to a sum of 
$\Delta\neff^{\saxion\to aa}+\Delta\neff^{\gravitino\to a\axino}=0.68$ 
($0.65$).
These values are compatible with the $2\sigma$ upper limit 
of $\Delta\neff< 0.79$ ($0.95$)
derived from the Planck+WP+highL(+$\Hzero$)+BAO data set 
quoted in Table~\ref{Tab:Neffconstrains}.

Prior to the announcement of the Planck results,
we found it tempting to suggest 
the substantial difference between 
the posterior maxima of $\Delta\neff\sim 0.8$ from BBN studies 
and the mean of $\Delta\neff\sim 1.8$ 
from pre-Planck precision cosmology as a first indication
towards the realization of the considered axion CDM scenario 
in nature~\cite{Graf:2012hb}; cf.\ Table~\ref{Tab:Neffconstrains}.
The Planck results now disfavor such a substantial difference.
Nevertheless, a small difference remains viable and
the considered scenarios remain attractive
with the axion condensate and thermal leptogenesis
providing natural explanations 
of CDM for $\fax=10^{12}\,\GeV$ and
of the baryon asymmetry for $\TR\gtrsim 10^{9}\,\GeV$,
as already emphasized in Refs.~\cite{Brandenburg:2004du,Hasenkamp:2011em}.

\begin{figure*}[t]
\begin{center}
\includegraphics[width=.46\textwidth,clip=true]{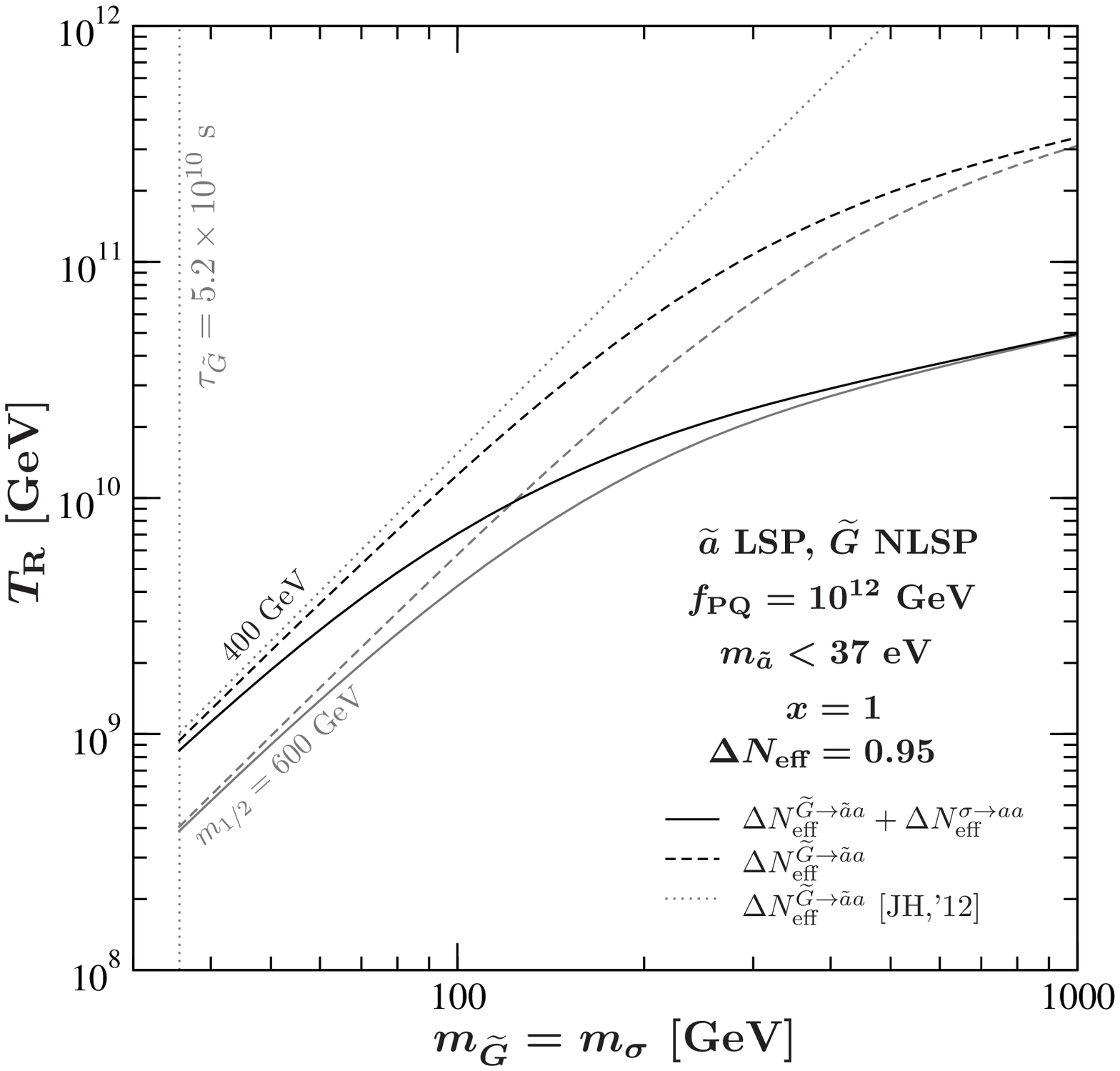}
\hskip 0.75cm
\includegraphics[width=.46\textwidth,clip=true]{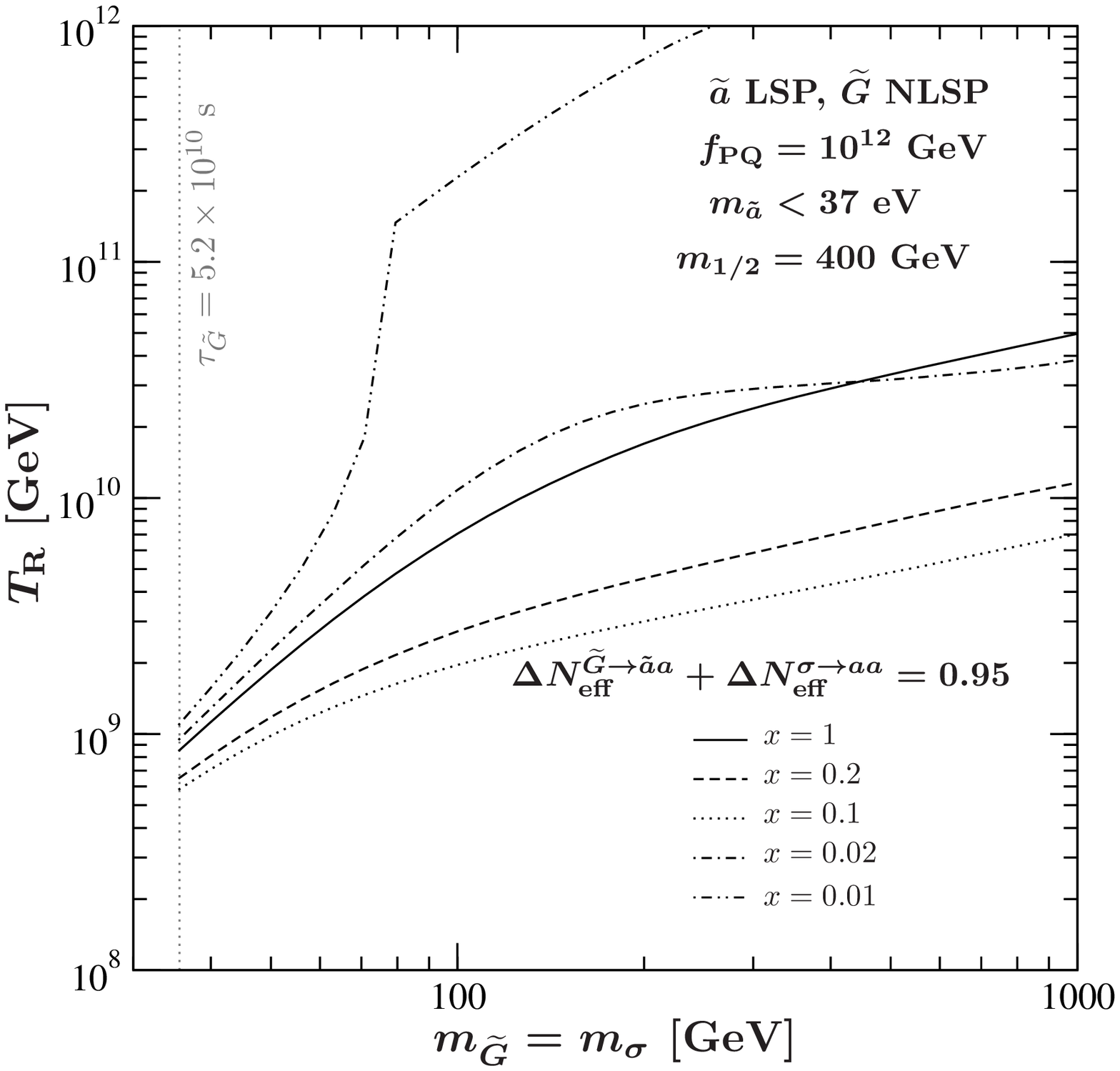} 
\vskip -0.5cm
\makebox[.46\textwidth][l]{\textbf{(a)}}\hfill
\makebox[.46\textwidth][l]{\textbf{(b)}}\hfill
\vskip 0.5cm
\caption{\small 
Contours of $\Delta\neff$ provided by 
axions from decays of thermal saxions, 
$\Delta\neff^{\saxion\to aa}$, 
and by axions and axinos from decays of thermally produced gravitinos,
$\Delta\neff^{\gravitino\to a\axino}$,  
in the $\mgravitino$--$\TR$ parameter plane
in axino LSP scenarios with the gravitino NLSP,
where $\msaxion=\mgravitino$, $\maxino\lesssim 37~\eV$, 
and $\fax=10^{12}\,\GeV$.
In panel~{(a)}, $x=1$ and black (gray) contours
refer to $\monetwo=400~(600)~\GeV$.
Here we show solid contours of 
$\Delta\neff^{\saxion\to aa}+\Delta\neff^{\gravitino\to a\axino}=0.95$
and dashed contours of
$\Delta\neff^{\gravitino\to a\axino}=0.95$.
The diagonal dotted line indicates the latter as well
but as obtained with the $\Delta\neff^{\gravitino\to a\axino}$ estimate 
from Ref.~\cite{Hasenkamp:2011em}.
In panel~{(b)} 
we show solid, dashed, dotted, dash-dotted, and dash-double-dotted contours of 
$\Delta\neff^{\saxion\to aa}+\Delta\neff^{\gravitino\to a\axino}=0.95$ 
for $x=1$, 0.2, 0.1, 0.02, and 0.01, respectively,
and $\monetwo=400~\GeV$.
The regions above these contours are disfavored at the $2\sigma$ level
by the Planck+WP+highL+$\Hzero$+BAO data set~\cite{Ade:2013zuv}; 
cf.\ Table~\ref{Tab:Neffconstrains}.
The vertical dotted line indicates the lower limit 
on $\mgravitino$ from the requirement 
$\tau_{\gravitino}\lesssim 5.2\times10^{10}\,\seconds$ in both panels.} 
\label{Fig:aDMExclude}
\end{center}
\end{figure*}

\begin{figure*}[t]
\begin{center}
\includegraphics[width=.46\textwidth,clip=true]{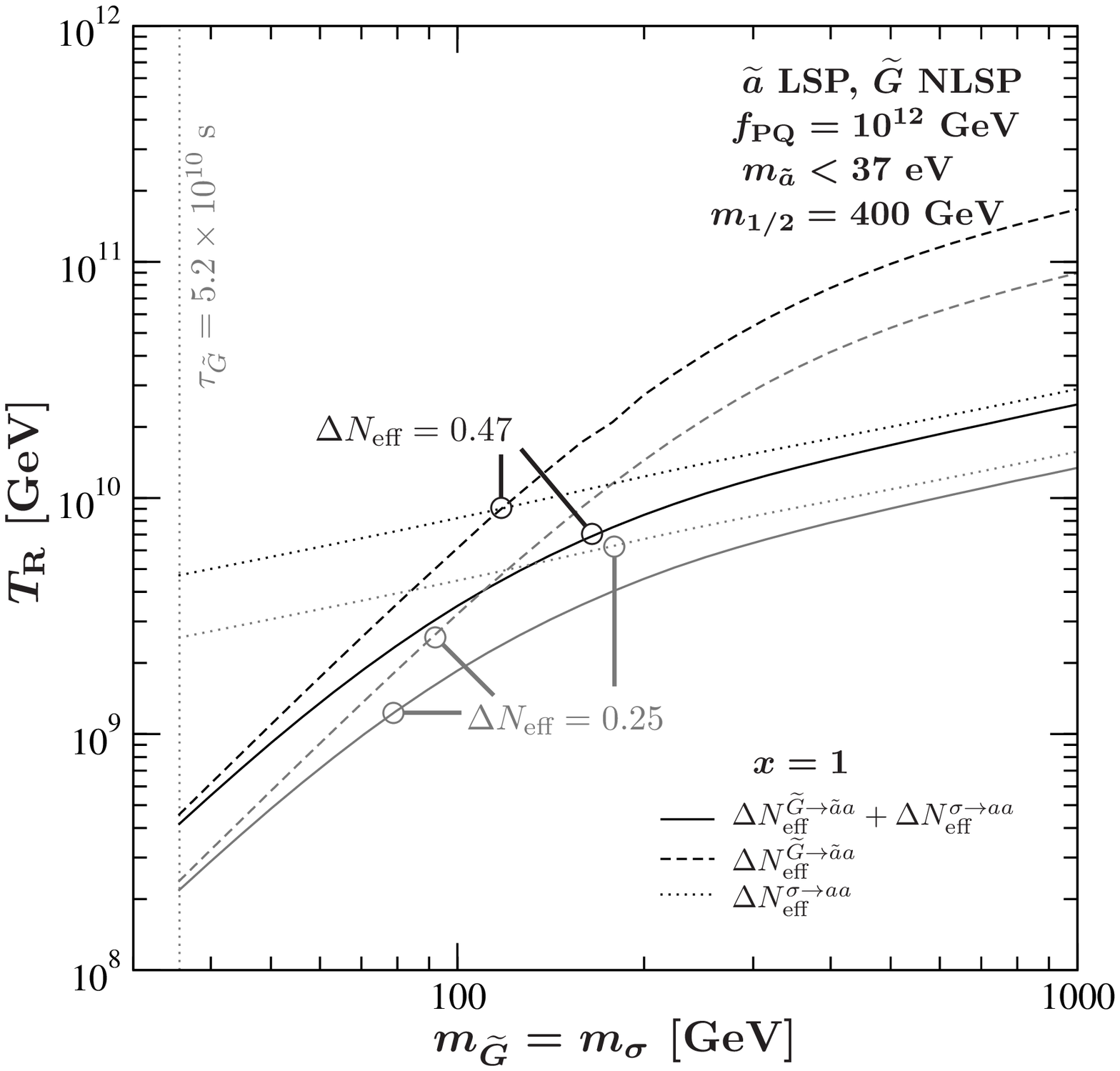} 
\hskip 0.75cm
\includegraphics[width=.46\textwidth,clip=true]{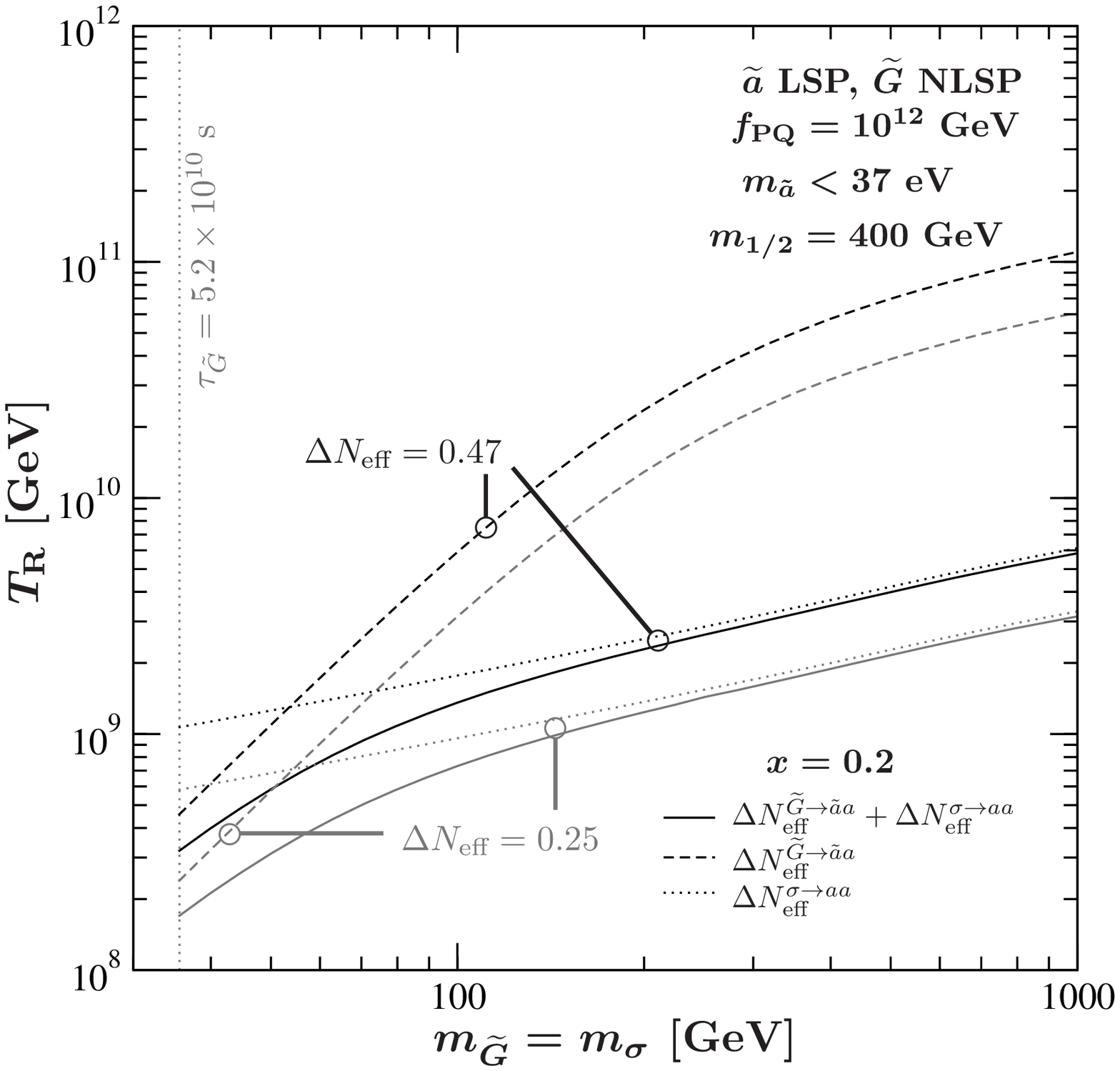} 
\vskip -0.5cm
\makebox[.46\textwidth][l]{\textbf{(a)}}\hfill
\makebox[.46\textwidth][l]{\textbf{(b)}}\hfill
\vskip 0.75cm
\includegraphics[width=.46\textwidth,clip=true]{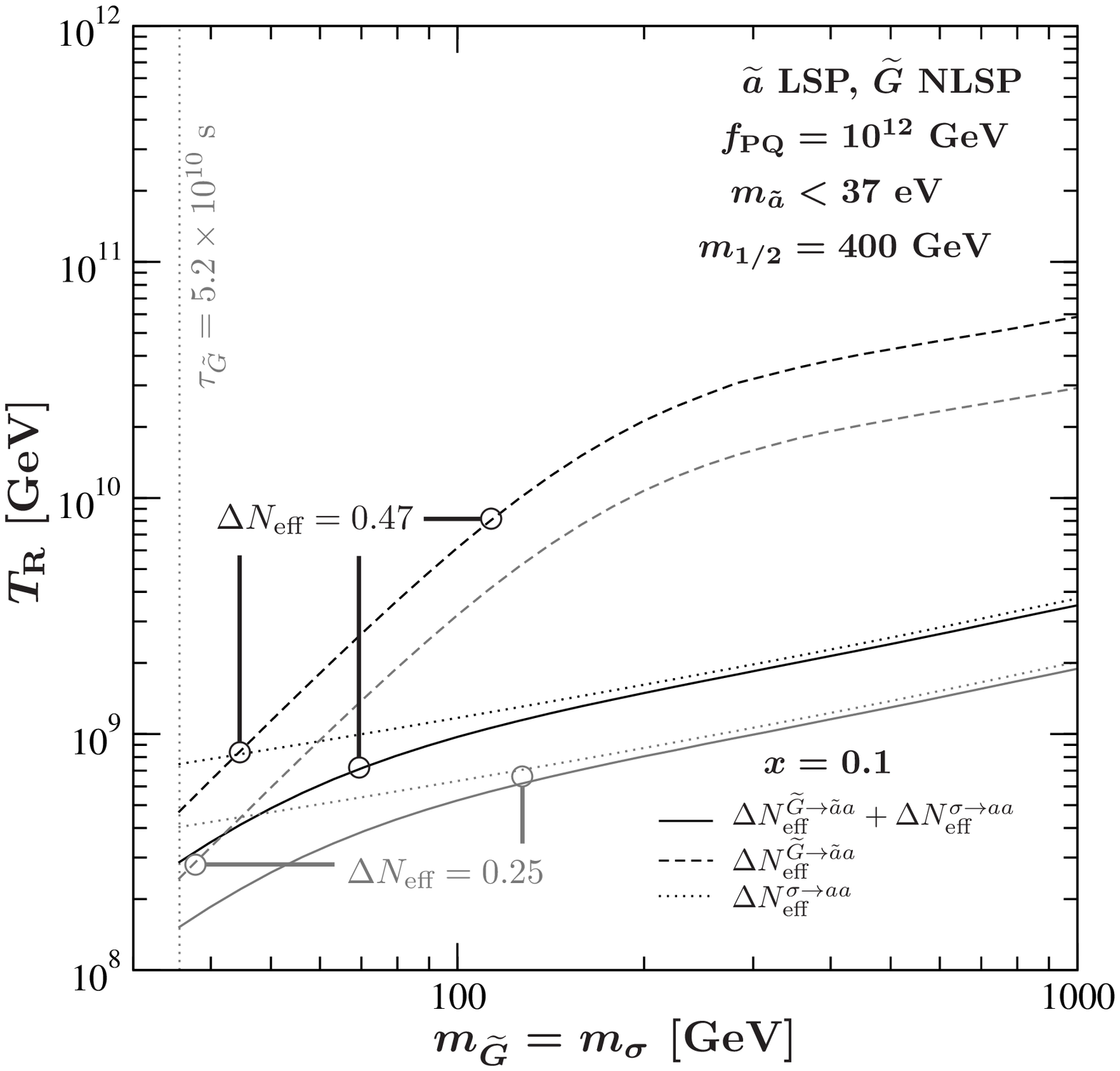} 
\hskip 0.75cm
\includegraphics[width=.46\textwidth,clip=true]{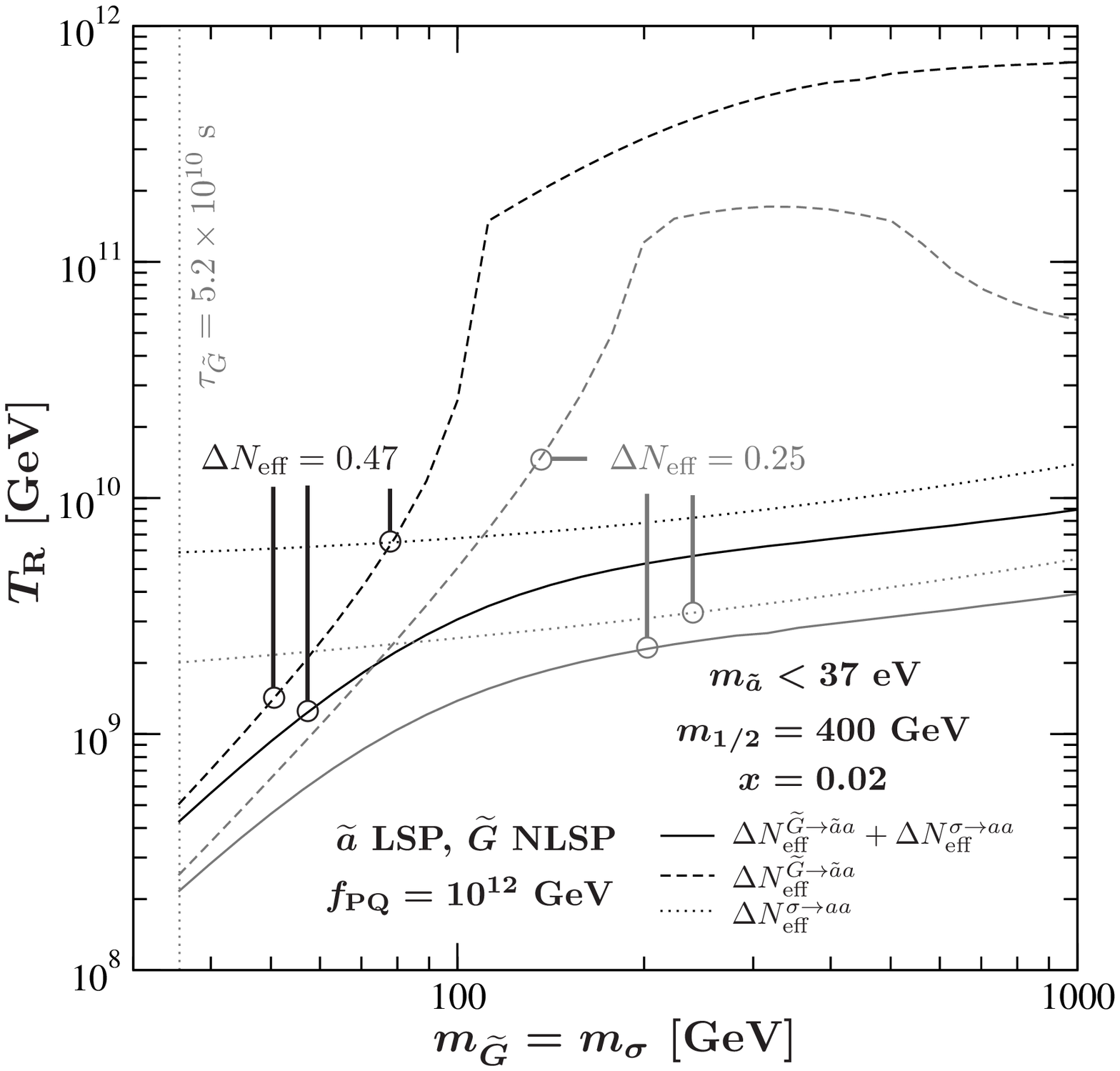} 
\vskip -0.5cm
\makebox[.46\textwidth][l]{\textbf{(c)}}\hfill
\makebox[.46\textwidth][l]{\textbf{(d)}}\hfill
\vskip 0.5cm
\caption{\small 
Contours of $\Delta\neff$ provided by 
axions from decays of thermal saxions, 
$\Delta\neff^{\saxion\to aa}$ (dotted),
by axions and axinos from decays of thermally produced gravitinos,
$\Delta\neff^{\gravitino\to a\axino}$ (dashed),  
and by the sum of both,
$\Delta\neff^{\saxion\to aa}+\Delta\neff^{\gravitino\to a\axino}$ (solid),
in the $\mgravitino$--$\TR$ parameter plane
in axino LSP scenarios with the gravitino NLSP,
$\msaxion=\mgravitino$, $\maxino\lesssim 37~\eV$,
$\fax=10^{12}\,\GeV$, and $m_{1/2}=400~\GeV$.
Black (gray) lines indicate $\Delta\neff=0.47$ ($0.25$)
and thereby the mean inferred 
from the Planck+WP+highL+$\Hzero$+BAO
(Planck+WP+highL+BAO) data set~\cite{Ade:2013zuv}. 
In each of the four panels, a different $x$ value is considered:
(a)~$x=1$, (b)~$0.2$, (c)~$0.1$, and (d)~$0.02$.
The vertical dotted line is as in Fig.~\ref{Fig:aDMExclude}.
}
\label{Fig:aCDMExplainx}
\end{center}
\end{figure*}

In the following we systematically explore $\Delta\neff$ contributions 
in settings with $\fax=10^{12}\,\GeV$ and large $\TR$.
In addition to the latter two features mentioned above,
the saxion energy density residing in coherent saxion oscillations 
with $\sigma_i \sim \fax$ is negligible 
with respect to the one from thermal processes
in that parameter region~\cite{Kawasaki:2007mk}.
Our results are presented 
in Figs.~\ref{Fig:aDMExclude} and~\ref{Fig:aCDMExplainx}
in the $\mgravitino$--$\TR$ parameter plane
for $\msaxion=\mgravitino$, $\maxino\lesssim 37~\eV$,
and universal gaugino masses at the GUT scale 
of $\monetwo=400~\GeV$ or $600~\GeV$.
The latter is compatible with $\mgluino=1.5~\TeV$
at collider energies.
The region with $\tau_{\gravitino}<5.2\times10^{10}\,\seconds$
is not considered and indicated by a vertical gray dotted line
at $\mgravitino\simeq 35~\GeV$.

%------------------------------------------------------------
%------------------------------------------------------------
%
%
%------------------------------------------------------------
%------------------------------------------------------------

In Fig.~\ref{Fig:aDMExclude}(a) the solid black (gray) lines show
$\Delta\neff^{\saxion\to aa}+\Delta\neff^{\gravitino\to a\axino}=0.95$ 
and the dashed lines $\Delta\neff^{\gravitino\to a\axino}=0.95$
for $\monetwo=400~(600)~\GeV$ and $x=1$.
To allow for a comparison, 
the diagonal dotted line indicates
$\Delta\neff^{\gravitino\to a\axino}=0.95$
as obtained for $\mgravitino=1~\TeV$
with the existing result of Ref.~\cite{Hasenkamp:2011em}
based on the sudden decay approximation.
The difference with respect to the corresponding dashed line
is due to the sudden decay approximation, 
which overestimates $\Delta\neff$ by about $13\%$, 
and the omissions of electroweak and spin-3/2 contributions 
in the gravitino yield $\YgravitinoTP$ 
used in Ref.~\cite{Hasenkamp:2011em}.
Including the electroweak contributions increases 
$\YgravitinoTP$ by about $20\%$ at $\mgravitino\sim 35~\GeV$,
while the importance of the spin-3/2 components
becomes much more pronounced towards larger $\mgravitino$.
Comparing the respective dashed and solid lines,
we find that $\Delta\neff^{\saxion\to aa}$ contributions
lead to an additional sizable $\Delta\neff$ increase.
In fact, for $\msaxion\gtrsim 100~\GeV$, 
they tighten the upper limit on $\TR$ imposed by the $2\sigma$ upper limit
$\Delta\neff<0.95$
derived from the Planck+WP+highL+$\Hzero$+BAO data set~\cite{Ade:2013zuv}
by up to almost one order of magnitude.
For further comparison, 
we refer to Fig.~6 in Ref.~\cite{Graf:2012hb}, 
where $\Delta\neff^{\saxion\to aa}$ only
is presented as obtained
in the sudden decay approximation.
Also~\eqref{Eq:DeltaNapprox2} and~\eqref{Eq:DeltaNapprox3}
in Appendix~\ref{Sec:AnalyticApprox} of this work 
are approximate analytical expressions respectively for
$\Delta\neff^{\saxion\to aa}$
and $\Delta\neff^{\gravitino\to a\axino}$
that are based on the sudden decay approximation.

Figure~\ref{Fig:aDMExclude}(a) demonstrates 
how the $\Delta\neff$ contours will move
if LHC experiments point to $\mgluino\gtrsim 1.5~\TeV$ 
and thereby to $\monetwo\gtrsim 600~\GeV$. 
These changes are governed fully by 
$\Delta\neff^{\gravitino\to a\axino}$
whereas $\Delta\neff^{\saxion\to aa}$ is not affected.
At this point, we should stress that a collider measurement
of the LOSP mass $\mLOSP$ will limit $\mgravitino$ from above.
While the chosen $\monetwo$ values can imply an $\mLOSP$
value that is well below $1~\TeV$, we refrain from presenting
such an upper limit for $\mgravitino$ 
since it will depend strongly
on other details of an assumed SUSY model as well. 

The $\Delta\neff=0.95$ contours illustrate the impact 
of the results from the Planck satellite mission.
While Planck does not find any statistically significant hints 
for extra radiation, 
the contour $\Delta\neff=0.95$ provides the new upper limit on $\TR$
at the $2\sigma$ level 
as obtained from the Planck+WP+highL+$\Hzero$+BAO data set~\cite{Ade:2013zuv}.
For $x=1$, the viability of $\TR\gtrsim 10^{9}\,\GeV$
will then depend on $\mLOSP$ 
and on other LOSP-related cosmological constraints discussed below.

Let us now turn to the case of $x<1$.
Figure~\ref{Fig:aDMExclude}(b) shows
$\Delta\neff^{\saxion\to aa}+\Delta\neff^{\gravitino\to a\axino}=0.95$ 
contours for $x=1$ (solid), $0.2$ (dashed), 
$0.1$ (dotted), $0.02$ (dash-dotted), and $0.01$ (dash-double-dotted)
where $\monetwo=400~\GeV$.
Corresponding dilution factors $\Delta$ 
have already been shown in Fig.~\ref{Fig:aDMEvoDelta}(b)
and discussed thereafter.
The dilution factor $\Delta$ for $x=0.01$ has not been shown.
It shows a similar behavior but slightly exceeds the one for $x=0.02$,
i.e., it is slightly below $30$ (above $20$)
for $\msaxion=50~\GeV$ ($100~\GeV$) and $\TR\gtrsim 10^{11}\,\GeV$.
The $x$ dependence of $\Delta\neff^{\gravitino\to a\axino}$ 
results fully from the one of $\Delta$ so that this contribution
decreases towards $x\to 0$.
In contrast, for the same reasons as in the previous section, 
$\Delta\neff^{\saxion\to aa}$ increases towards smaller $x$
in the interval $0.1\lesssim x<1$, 
reaches its maximum at $x\sim 0.1$,
and decreases thereafter, i.e., towards smaller $x\lesssim 0.1$. 
The latter behavior transfers to 
$\Delta\neff^{\saxion\to aa}+\Delta\neff^{\gravitino\to a\axino}$,
as can be seen in Fig.~\ref{Fig:aDMExclude}(b).
Here the most restrictive upper $\TR$ limit is found for $x=0.1$
and the most relaxed one for $x=0.01$, where
$\Delta\neff^{\saxion\to aa}
+
\Delta\neff^{\gravitino\to a\axino}
\simeq
\Delta\neff^{\gravitino\to a\axino}$.

In Fig.~\ref{Fig:aCDMExplainx} 
we explore how  
$\Delta\neff^{\saxion\to aa}
+\Delta\neff^{\gravitino\to a\axino}$ (solid)
can emerge as a composition of 
a late $\Delta\neff^{\gravitino\to a\axino}$ (dashed)
and an early $\Delta\neff^{\saxion\to aa}$ (dotted)
for (a)~$x=1$, (b)~$0.2$, (c)~$0.1$ and (d)~$0.02$.
In each panel, we consider $\monetwo=400~\GeV$ 
and show gray and black contours 
of $\Delta\neff=0.25$ and $0.47$, respectively. 
On the one hand, those $\Delta\neff$ values are the corresponding means 
inferred from the
Planck+WP+highL+BAO and the Planck+WP+highL+$\Hzero$+BAO data sets
obtained by the Planck collaboration~\cite{Ade:2013zuv}.
On the other hand, e.g., a total late $\Delta\neff=0.47$ may be composed
of an early $\Delta\neff=0.25$ from saxion decays
and an additional late $\Delta\neff\simeq 0.22$ from gravitino decays.
The parameter points that allow for this composition 
are the ones at which the gray dotted and the
black solid lines intersect.
Accordingly, this composition is possible 
in all four panels, i.e., for $x=1$, $0.2$, $0.1$, and $0.02$.
In light of the BBN uncertainties with respect to an early $\Delta\neff$,
it should be emphasized that different compositions are possible as well,
as discussed at the end of Sect.~\ref{Sec:ExtraRadiation}.

In an assessment of the simultaneous viability of 
thermal leptogenesis and a certain $\Delta\neff$ composition,
the corresponding dilution factor $\Delta$ 
has to be taken into account 
in the same way as in the previous section.
While this factor can now be much larger,  
the current upper $\TR$ limit imposed by $\Delta\neff<0.95$ 
still allows for that simultaneous viability 
even for $x=0.1$ when $\mgravitino\gtrsim 50~\GeV$;
cf.\ Figs.~\ref{Fig:aDMEvoDelta}(b) and~\ref{Fig:aDMExclude}(b).
This $2\sigma$ upper limit from the 
Planck+WP+highL+$\Hzero$+BAO data set~\cite{Ade:2013zuv}
is a somewhat conservative one.
Nevertheless, even with the more restrictive $2\sigma$ upper limit 
from the Planck+WP+highL+BAO data set~\cite{Ade:2013zuv},
$\Delta\neff<0.79$ or with the mean $\Delta\neff=0.25$ or $0.47$,
thermal leptogenesis can remain viable for $x=1$
and also for smaller $x$ provided $\mgravitino$
can be sufficiently large.
Because of the assumed hierarchy in this section,
$\mgravitino<\mLOSP$, a measurement of $\mLOSP$ 
can thus challenge that simultaneous viability,
in particular, for $x\gtrsim 0.1$
and larger $\monetwo$.

In the considered situation with the axino LSP and the gravitino NLSP,
the LOSP is again a long-lived particle. However, in contrast to
the gravitino LSP setting in Sect.~\ref{Sec:GravitinoCDM}, 
it can not only decay into gravitinos, $\LOSP\to\gravitino\X$,
but also into axinos, $\LOSP\to\axino\X$, 
where the relative importance is governed by $\mgravitino$ 
and $\fax$.
For $\fax=10^{12}\,\GeV$ and $\mgravitino\gtrsim 35~\GeV$,
the decay into the axino is the dominating one, i.e.,
$\Gamma(\LOSP\to\axino\X)\gg\Gamma(\LOSP\to\gravitino\X)$.
Thereby, the LOSP lifetime can be significantly shorter
than in the previous section so that 
the (C)BBN constraints related to a late decaying LOSP
described at the end of that section can be evaded.
In fact, the charged slepton LOSP is now a viable possibility,
which is particularly attractive since it could appear 
as a quasistable charged massive particle in collider experiments.
For example, if the LOSP is the lightest stau 
with $m_{\tilde{\tau}_1}\gtrsim 300~\GeV$, 
there is indeed no limit on the gravitino mass 
other than $\mgravitino<m_{\tilde{\tau}_1}$ 
for $\fax\lesssim 5\times 10^{12}\,\GeV$ 
and already with $\Delta=1$~\cite{Freitas:2011fx}.
Late time entropy production in saxion decays 
with $x\ll 1$ can dilute $\YLOSP$ as described by~\eqref{Eq:YLOSPoverDelta}
with a sizable $\Delta>1$ and thereby imply even more relaxed constraints.
The bino-like neutralino LOSP situation
was considered in Ref.~\cite{Baer:2010gr}
and found to be viable for $\fax\sim 10^{12}\,\GeV$ as well.
This work accounted for entropy production in saxion decays also
but did not address the production of extra radiation.
Similarly, the sneutrino LOSP situation 
is expected to be viable in the considered settings.

Very light axinos with $\maxino\lesssim 37~\eV$
emitted in LOSP decays 
contribute only negligible amounts 
to the density parameter and to extra radiation $\Delta\neff$.
This can be seen by evaluating
\be
\OmegaAxino^{\LOSP\to\axino\X}h^{2}
\simeq
\sqrt{p_{\axino,0}^{2}+\maxino^{2}}\,
Y_{\LOSP} s(T_0)h^2/\rho_c,
\label{Eq:OmegaAxinoLOSPDecay}
\ee
with the present momentum of these axinos $p_{\axino,0}$
as obtained in the sudden decay approximation.
For the $\Delta\neff^{\LOSP\to\axino\X}$ contribution,
this is shown explicitly for the stau LOSP case
in Sect.~4.3 of Ref.~\cite{Freitas:2011fx},
which can easily
be generalized to other LOSP candidates.
Axions and axinos from gravitino decays
are still relativistic today
for the considered values 
of $\maxino$ and $\mgravitino$.
Accordingly, their contribution to the density parameter
can be expressed in terms of $\Delta\neff^{\gravitino\to a\axino}$:
\begin{equation}
\Omega_{a}^{\gravitino\to a\axino}h^{2}
+
\OmegaAxino^{\gravitino\to a\axino}h^{2}
=5.7\times 10^{-6}\,\Delta\neff^{\gravitino\to a\axino}.
\label{Eq:OmegaAxinoAxionGravitinoDecay}
\end{equation}
This clarifies that the shown constraints
will neither be tightened by $\Delta\neff^{\LOSP\to\axino\X}$ 
nor by contributions to the density parameter
described by~\eqref{Eq:OmegaAxinoLOSPDecay}
and~\eqref{Eq:OmegaAxinoAxionGravitinoDecay}.

The density parameter of axinos from thermal processes 
in the early universe is given by
\begin{equation}
\OmegaaxinoeqTP h^2 
\simeq
\sqrt{\langle p_{\axino,0}^\text{th}\rangle^2+\maxino^2}\,\,
\YaxinoeqTP\,\,
s(T_0) h^2/\rho_c ,
\label{Eq:OmegaAxinoThermal}
\end{equation}
where the average momentum of thermal axinos today is given by
$\langle p_{\axino,0}^\text{th}\rangle=3.151\,T_{\axino,0}$
with the present axion temperature of
$T_{\axion,0}=[g_{*S}(T_0)/228.75]^{1/3}\,T_0\simeq 0.06~\meV$.
Expression~\eqref{Eq:OmegaAxinoThermal} relies on the
fact that not only thermal relic but also thermally produced axinos
show basically a thermal spectrum.
When comparing $\langle p_{\axino,0}^\text{th}\rangle$ 
with the axino mass,  
one finds that this axino population can still be 
relativistic today but only when $\maxino<0.2~\meV$.
A similar comparison shows that such a thermal axino population
with, e.g., $\maxino<4~\eV$ was still relativistic at
$t=5.2\times 10^{10}\,\seconds$. 
As mentioned at the beginning of this section,
this axino population is dark radiation
when relativistic.
When non-relativistic, the energy density 
of this axino population
and its contribution to the density parameter 
are governed by $\maxino$.
This allows for quantifying the HDM constraint 
as $\maxino\lesssim 37~\eV$~\cite{Freitas:2011fx}.
Interestingly, the considered axion CDM hypothesis 
will continue to be probed by the 
direct axion search experiment ADMX
exactly in the region around 
$\fax\sim 10^{12}\,\GeV$~\cite{Carosi:2007uc}.
A discovery of axions in this search 
could therefore point towards the realization 
of one of the settings considered in this section.
Further support in favor of those settings 
(and against the ones considered 
in Sect.~\ref{Sec:GravitinoCDM})
would be the discovery of a long-lived charged slepton LOSP
at the LHC.
It could even be that future cosmological analyses 
find hints on the time of the release of extra radiation. 
Such a release may manifest itself in the perturbation spectrum
so that precision cosmology might help to assess 
the lifetime of the gravitino 
whose late decays produce dark radiation
at times before $5.2\times10^{10}\,\seconds$. 
Another strong hint for the scenarios considered here
would be the confirmation of extra radiation prior to BBN
and a significant difference between that amount
with respect to the one at much later times.
In addition to new astrophysical data sets 
from improved direct measurements 
of the Hubble constant $\Hzero$,
this would require advances in BBN-related studies.
In particular, this calls for new high quality spectra
from extragalactic HII regions that should 
allow for a significantly more precise determination 
of $\Delta\neff$ prior to BBN~\cite{Aver:2011bw}.

%______________________________________________
\section{Conclusion}
\label{Sec:Conclusion}
%______________________________________________

We have explored two scenarios 
of hadronic axion models~\cite{Kim:1979if,Shifman:1979if}
in R-parity conserving SUSY settings:
(i)~a gravitino LSP scenario with 
a heavy axino at the TeV scale, 
and (ii)~a scenario with 
a light axino LSP at the eV scale
and the gravitino NLSP.
Both scenarios are found to allow for consistent explanations
of extra radiation and CDM
and for a high reheating temperature $\TR$
of up to about $10^9\,\GeV$
or $10^{11}\,\GeV$, respectively.
Testable cases have been outlined 
that may still allow for the high $\TR$ values 
required by successful thermal leptogenesis
with hierarchical heavy Majorana neutrinos~\cite{Buchmuller:2005eh}.

In the gravitino LSP scenario, CDM resides dominantly in gravitinos
from thermal production and from decays of thermal axions 
and $\Delta\neff$ is explained by thermal saxions 
which decay into axion pairs prior to BBN.
We have shown that up to $\Delta\neff\simeq 0.8$ 
can arise naturally
for $\fax\simeq 10^{10}\,\GeV$ 
and $\TR\simeq 10^7\,\GeV$.
This finding requires that the gluino mass $\mgluino$ is close 
to the current experimental limit of about $1~\TeV$.
For a larger $\mgluino=1.25~\TeV$, we have demonstrated that 
smaller values of $\Delta\neff\simeq 0.5$ remain viable 
for $10^{10}\,\GeV\lesssim\fax\lesssim 10^{11}\,\GeV$ 
and $10^7\,\GeV\lesssim\TR\lesssim 10^9\,\GeV$.
Viability of larger $\Delta\neff$ (i.e., above $0.8$ or $0.5$) 
is found to require a more suppressed saxion-axion coupling, $x\ll 1$,
with a maximum $\Delta\neff$ occurring for $x\sim 0.1$.
There we have shown that the $2\sigma$ limits of $\Delta\neff<0.79$ 
or $0.95$ obtained by the Planck collaboration~\cite{Ade:2013zuv}
translate into new upper limits on $\TR$, which can be the most restrictive ones.

For compatibility of the presented gravitino LSP case 
with cosmological constraints,
the axino must be heavy, $\maxino\gtrsim 2~\TeV$,
so that it decays prior to the decoupling of the LOSP
from the thermal bath.
For the high $\TR$ values considered, 
such a heavy axino can still be produced very efficiently 
in thermal processes in the early Universe. 
Primordial axinos can thereby contribute significantly 
to the total energy density 
just before decaying dominantly into gluinos and gluons.
Calculating the associated entropy production, 
we obtain dilution factors of up to 
$\Delta^{\axino\to g\gluino}\sim 2$ 
that affect the abundances of gravitinos, saxions, 
and axions produced in thermal processes
well before axinos dominate the energy density.
Since also a baryon asymmetry generated prior to that epoch 
is diluted by the same factor, about twice of the observed
value is needed prior to that dilution.
Within the framework of thermal leptogenesis,
this implies that the usually required 
$\TR\sim 10^9\,\GeV$~\cite{Buchmuller:2005eh} 
now has to be basically twice 
as large~\cite{Pradler:2006hh,Hasenkamp:2010if}.
For $x=1$, we find this to be viable for $\fax=10^{11}\,\GeV$, 
$\mgluino\simeq 1~\TeV$, and $\maxino\simeq 6~\TeV$
when $\Delta\neff\lesssim 0.5$.
Towards small $x\ll 1$, the decay of thermal saxions into gluons
can lead to an additional sizable dilution factor of up to
$\Delta^{\saxion\to gg}\sim 3$.
This can dilute even the yield of the LOSP after decoupling
from the thermal plasma and prior to decay and thereby
weaken BBN constraints related to late decaying 
LOSP~\cite{Buchmuller:2006tt,Pradler:2006hh,Hasenkamp:2010if}.
Moreover, for $0.1\lesssim x\ll 1$, $\fax\sim 10^{10}\,\GeV$, 
and $\maxino\simeq 2~\TeV$,
we have found that a significant part 
of the parameter space will allow for the simultaneous
viability of thermal leptogenesis, a sizable $\Delta\neff$
provided by axions from decays of thermal saxions,
and $\OmegaDM$ residing almost fully in thermally produced gravitinos.
Towards $\fax\sim 10^{11}\,\GeV$, small $\mgravitino\lesssim 100~\GeV$,
and large $\maxino\simeq 6~\TeV$,  
gravitinos from decays of thermal axinos 
are found to become an increasingly important component of $\OmegaDM$,
which tightens associated $\TR$ limits considerably.

In the scenario with the light axino LSP and the gravitino NLSP,
CDM resides in axions from the misalignment mechanism,
which provides naturally $\OmegaAxionMIS\simeq\OmegaDM$ 
for $\fax\simeq 10^{12}\,\GeV$.
Remarkably, 
the ongoing direct axion CDM search by ADMX~\cite{Carosi:2007uc}
is sensitive in exactly that $\fax$ range and may find
signals supporting this CDM explanation in the near future.
We have demonstrated that there are now
two sources for a possibly substantial 
$\Delta\neff$ that work at very different times: 
thermal saxions that decay into axion pairs prior to BBN
and thermally produced gravitinos that decay into
axions and axinos well after BBN 
and before $5.2\times 10^{10}\,\seconds$. 
Accordingly, within this scenario, 
we find different possibilities to explain,
e.g., the means of $\Delta\neff=0.25$ or $0.47$ 
obtained recently by the Planck collaboration~\cite{Ade:2013zuv}.
For $\Delta\neff\simeq 0.47$, 
one natural explanation will be the composition
with an early $\Delta\neff^{\saxion\to a a}\simeq 0.25$ 
residing in axions from saxion decays
and an additional late $\Delta\neff^{\gravitino\to a\axino}\simeq 0.22$
residing in axions and axinos from gravitino decays.
However, without more precise BBN limits for $\Delta\neff$,
which may indeed be difficult to obtain in light 
of the systematic uncertainties~\cite{Aver:2010wq},
there remains a significant uncertainty with respect to 
the amount of an early $\Delta\neff$.
Accordingly, e.g., $\Delta\neff\simeq 0.47$ 
can result equally well 
either dominantly from late $\gravitino\to a\axino$ decays
for $\mgravitino\ll100~\GeV$
or dominantly from $\saxion\to a a$ decays prior to BBN
for $\mgravitino\gg 100~\GeV$.
In fact,
also the amount of the late $\Delta\neff$ comes 
with uncertainties
that call for new direct $\Hzero$ measurements.

Our refinements with respect to Refs.~\cite{Hasenkamp:2011em,Graf:2012hb}
have been found to have the following effects.
By treating decays beyond the sudden-decay approximation,
the resulting $\Delta\neff$ values decrease by about 10\%.
Moreover, with the gravitino yield that
accounts for the gravitino-spin-3/2 components
and for electroweak processes,
previously neglected contributions to 
$\Delta\neff^{\gravitino\to a\axino}$ 
are included which become sizable for $\mgravitino\gtrsim 100~\GeV$.
Together with the contributions from $\saxion\to a a$ decays,
this results in significantly larger $\Delta\neff$ values 
in that region.
In turn, our new upper bounds on $\TR$ for $\mgravitino=\Order(100~\GeV)$
are substantially more restrictive than previously expected for $x=1$
in the axino LSP case with the gravitino NLSP.
Towards a more suppressed saxion-axion coupling with $x\sim 0.1$,
we find even larger $\Delta\neff$ values 
that further tighten the $\TR$ limits significantly.
Even smaller values of $x\ll 0.1$ have been found to come
with significant dilution factors of up to
$\Delta^{\saxion\to gg}\sim 30$.
Those can reduce $\Delta\neff$ 
and in turn relax the upper bounds on $\TR$
but have to be included in an assessment of the
viability of thermal leptogenesis.

We have discussed ways in which 
the different explanations of a potentially sizable $\Delta\neff$ 
will be narrowed by ongoing SUSY searches at the LHC.
Particularly important will be new limits 
on $\mgluino$ or measurements thereof.
The gluino mass governs the thermally produced gravitino yield
and thereby limits that relate to this quantity.
Upper limits on $\TR$ 
thus become more restrictive for larger $\mgluino$.
In fact, for $\mgluino\gtrsim 1.1~\TeV$, 
we find that a $\Delta\neff\sim 0.8$
explanation by axions from thermal saxions 
becomes incompatible with 
the then too restrictive $\OmegaGravitino\leq\OmegaDM$ constraint
in the gravitino LSP case.
In the alternative axino LSP case, on the other hand,
it is the $\Delta\neff$-imposed limit that becomes more 
restrictive towards large $\mgluino$ in the small $\mgravitino$ region
where the decay $\gravitino\to a\axino$
contributes significantly to $\Delta\neff$.

Other relevant LHC findings will be a discovery 
of the lightest sparticle within the MSSM (i.e., the LOSP), 
its identification, and a measurement of its mass.
In both of the considered cases, 
the LOSP mass limits $\mgravitino$ from above.
Moreover, the LOSP is expected to be long-lived 
so that additional restrictive cosmological constraints
can occur depending on the nature of the LOSP.
For example, for a long-lived charged slepton LOSP,
which has to be heavier than 
about 300~GeV~\cite{Chatrchyan:2012sp,ATLAS-CONF-2012-075},
CBBN constraints can disfavor 
the presented gravitino LSP case~\cite{Steffen:2006wx,Pospelov:2008ta}.
Remarkably, such an LOSP is found to be compatible 
with the light axino LSP scenario with the 
gravitino NLSP~\cite{Freitas:2011fx}.
The discovery of such an LOSP could thus become 
an important additional hint in favor of the latter scenario.
In the gravitino LSP scenario,
BBN constraints associated with hadronic energy injection 
disfavor the possibilities of a neutralino LOSP~\cite{Feng:2004mt}
or a colored LOSP~\cite{DiazCruz:2007fc} as well.
Nevertheless, that scenario is found to be viable 
with a sneutrino LOSP~\cite{Kanzaki:2006hm,Kawasaki:2008qe}.
One will then face the challenge
to identify a long-lived sneutrino as the 
LOSP~\cite{Covi:2007xj,Ellis:2008as,Katz:2009qx,Figy:2010hu},
which will be a much more difficult task than the identification
of a long-lived charged slepton LOSP.

In summary, we find the presented scenarios appealing 
from the cosmological point of view 
and intriguing with respect to their testability.
In light of those features, 
it will be interesting
to see ways in which model building can allow for the suggested
mass spectra and the large splittings between
the axino mass and the masses of the saxion and the gravitino.
With upcoming new results from 
the direct axion dark matter search experiment ADMX and the LHC,
it will be exciting to see further hints for or against
the viability of the considered scenarios soon.

%______________________________________________
\begin{appendix}
%______________________________________________

%______________________________________________
\section{Thermal Axino Production}
\label{Sec:AxinoTP}
%______________________________________________

Let us present some of the details 
of the calculation that lead to our update 
of thermally produced axino yield~\eqref{Eq:AxinoYieldTP}. 
To obtain a finite result in a gauge-invarant treatment,
we rely on systematic field theoretical methods such as 
HTL resummation~\cite{Braaten:1989mz} and the Braaten--Yuan 
prescription~\cite{Braaten:1991dd}
exactly as applied in Ref.~\cite{Brandenburg:2004du}.
However, 
we now include the quartic axino-squark-antisquark-gluino interaction
described by the second term in the third line of~\eqref{Eq:eff_lag},
which was not considered in Ref.~\cite{Brandenburg:2004du}
as pointed out in Ref.~\cite{Strumia:2010aa}.

\begin{figure}[t]
\begin{description}
\item [Process H]
$\tilde{q}_i+\gluino^a\rightarrow\tilde{q}_j+\axino $  
\begin{center}
\includegraphics[width=0.28\textwidth]{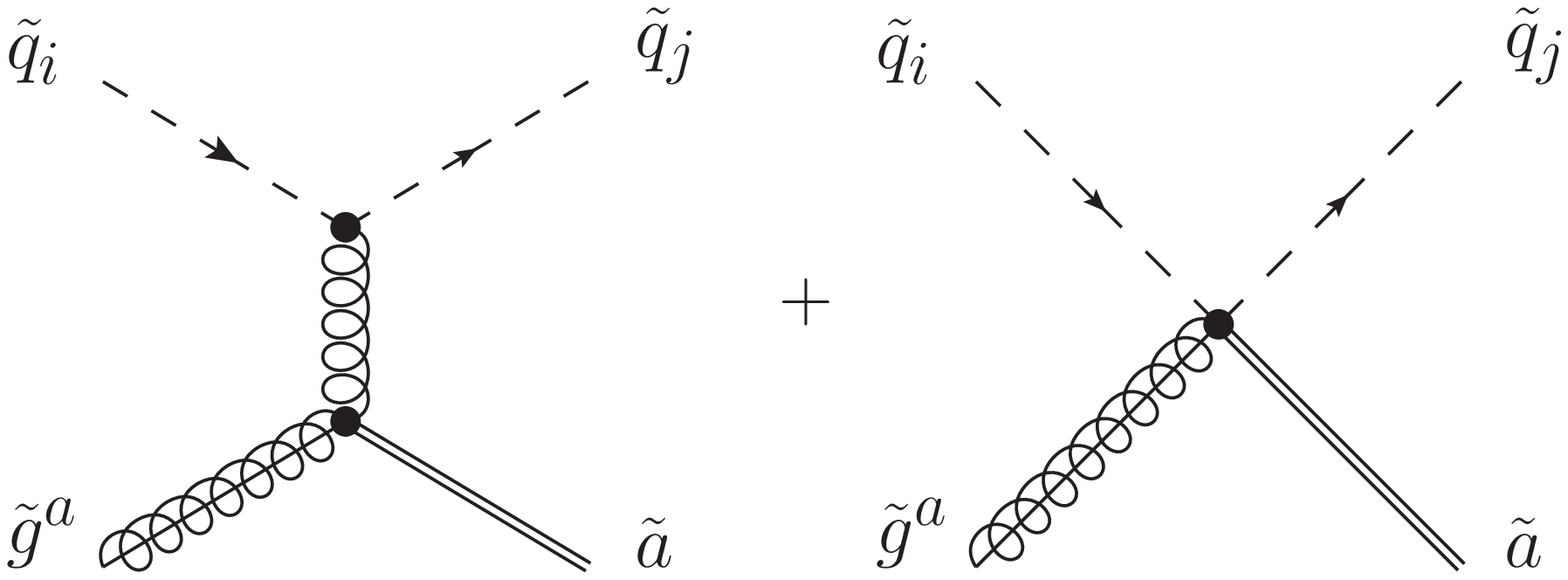}
\end{center}
\item [Process J] 
$\tilde{q}_i+\bar{\tilde{q}}_j\rightarrow\gluino^a+\axino$ 
(Crossing of H)
\end{description}
\caption{
The $2\to2$ processes of axino production 
affected by including the 
quartic axino-squark-antisquark-gluino vertex
described by the second term 
in the third line of~\eqref{Eq:eff_lag}.
Process H is also possible with antisquarks, 
replacing $\tilde{q}_{i,j}$ by $\bar{\tilde{q}}_{i,j}$.
}
\label{Fig:Axinoprod}
\end{figure}

Following~\cite{Brandenburg:2004du} 
and the methods referred to therein closely,
we split the thermal production rate
into a soft part, 
which involves soft gluons 
with momentum transfer of order $g_{s}T$,
and a hard part, in which no soft gluon exchanges occur. 
The soft part is not affected by the additional vertex.
In the hard part, this vertex contributes
additional Feynman diagrams in the process
$\tilde{q}_i+\gluino^a\to\tilde{q}_j+\axino$ 
and its crossing
$\tilde{q}_i+\bar{\tilde{q}}_j\to\gluino^a+\axino$ 
labeled respectively as H and J in~\cite{Brandenburg:2004du}.
Figure~\ref{Fig:Axinoprod} shows the completed set 
of Feynman diagrams for process H
and lists process J as its crossing.
The corresponding squared matrix elements $|M_k|^2$ 
for a single chirality and with sums over initial and final spins read
\begin{align}
|M_\text{H}|^2/\frac{g_s^6}{128\pi^4\fax^2} 
&= 
-2\left(t + 2s + 2\frac{s^2}{t} \right) |T^a_{ji}|^2, 
\\
|M_\text{J}|^2/\frac{g_s^6}{128\pi^4\fax^2} 
&= 
2\left(s + 2t + 2\frac{t^2}{s} \right) |T^a_{ji}|^2
\end{align}
and replace the respective entries 
in Table~1 of Ref.~\cite{Brandenburg:2004du}.
Other entries in that table are not affected. 
The Mandelstam variables are given by $s=(P_{1}+P_{2})^{2}$
and $t=(P_{1}-P_{3})^{2}$, where the particle four-momenta
$P_{i}$ refer to the particles in the order 
in which they are written down above and in Fig.~\ref{Fig:Axinoprod}.
%
%______________________________________________
%______________________________________________

Grouping the processes into different classes 
(depending on the number of external bosons and fermions)
and weighting the matrix elements 
with their respective multiplicities and statistical factors, 
we find that the sums of the corresponding squared matrix
elements $|M_\text{BBF}|^2$ and $|M_\text{BFB}|^2$
given in~(3.7) and~(3.8) of Ref.~\cite{Brandenburg:2004du} 
change to:
\begin{align}
\frac{|M_\text{BBF}|^2}
{\left[\frac{g_{s}^6(N_c^2-1)}{64\pi^4\fax^2}\right]}
\!&= \!
\left(s+2t+\frac{2t^2}{s}\right)\!\left(N_c+n_f\right)\!
+4s n_f,
\\
\frac{|M_\text{BFB}|^2}
{\left[\frac{g_{s}^6(N_c^2-1)}{32\pi^4\fax^2}\right]}
\!&= \!
\left(-t-2s-\frac{2s^2}{t}\right)\!\left(N_c+n_f\right)\!
-4t n_f,
\end{align}
where the notation of the above reference is adopted
with $N_{c}=3$ denoting the number of colors 
and $n_{f}=6$ the number of color triplet 
and anti-triplet chiral multiplets.
Further simplifications lead to
\begin{align}
|M_\text{BBF}|^2 
&\rightarrow 
\frac{g_{s}^6(N_c^2-1)}{32\pi^4\fax^2} 
\left[|M_3|^2\left(N_c+n_f\right)-2|M_2|^2n_f\right], 
\\
|M_\text{BFB}|^2 
&= 
\frac{g_{s}^6(N_c^2-1)}{32\pi^4\fax^2} 
\left[|M_1|^2\left(N_c+n_f\right)-2|M_2|^2n_f\right], 
\end{align}
where $|M_1|^2=-t-2s-(2s^{2}/t)$, $|M_2|^{2}=t$, 
and $|M_3|^{2}=t^{2}/s$.
Using this instead of Eqs.~(3.12) and (3.13)
of Ref.~\cite{Brandenburg:2004du},
our result for the hard part of the thermal production rate
shows a prefactor of $-2n_{f}$ instead of $-3n_{f}/2$
in the fourth line of (E.1) in~\cite{Brandenburg:2004du}
but otherwise agrees with that equation.

By adding the soft and hard parts of the thermal production rate
and by integrating the resulting total thermal production rate
over the energy of the produced axino, 
we arrive at the collision term
\begin{align}
W_{\axino}(T) 
&= 
\frac{(N_c^2-1)}{\fax^2}
\frac{3\zeta(3)g_s^6T^6}{4096\pi^7}
\\
&\quad\times
\left[ 
\ln\left(\frac{1.647\,T^2}{m_g^2} \right)(N_c+n_f) 
+ 0.5781 \,n_f
\right]
\nonumber
\end{align}
with 
$m_g^2=\gS^{2}T^{2}(N_{c}+n_{f})/6$
denoting the squared SUSY thermal gluon mass.
The collision term enters the Boltzmann equation, 
$\dot{n}_{\axino}+3H n_{\axino}=W_{\axino}$,
that describes the time evolution of the axino number density.
Integrating this equation as described in~\cite{Brandenburg:2004du}, 
we get for the thermally produced axino yield
\begin{align}
&\YaxinoTP(T) 
\approx 
\frac{W_{\axino}(\TR)}{s(\TR)H(\TR)} 
\end{align}
and thereby expression~\eqref{Eq:AxinoYieldTP}
given in Sect.~\ref{Sec:HighTR}.
In summary, the constant in the logarithm
in the expression for $\YaxinoTP$
changes from $1.211$ in (E.3) of~\cite{Brandenburg:2004du} 
to $1.271$ in~\eqref{Eq:AxinoYieldTP}
when including the quartic axino-squark-antisquark-gluino vertex.
Accordingly, 
in the $R$-parity conserving axino LSP scenarios
considered in Ref.~\cite{Brandenburg:2004du},
the density parameter of thermally produced CDM axinos
changes to
\begin{align}
\OmegaAxino h^2 
&= 
5.5 g_s^6\ln\left(\frac{1.271}{g_s}\right)
\left(\frac{\maxino}{0.1~\GeV}\right)
\nonumber \\
&\quad\times
\left(\frac{10^{11}\,\GeV}{\fax}\right)^2
\left(\frac{\TR}{10^4\,\GeV}\right).
\end{align}
Nevertheless, the qualitative statements and plots 
of Ref.~\cite{Brandenburg:2004du} 
are only mildly affected by this correction.

%______________________________________________
\section{Approximations for \boldmath$\Delta$ and $\Delta\neff$}
\label{Sec:AnalyticApprox}
%______________________________________________

Here we provide expressions 
that describe approximately the numerical results 
obtained in Sects.~\ref{Sec:GravitinoCDM} and~\ref{Sec:AxionCDM}.
The presented expressions help 
to understand the qualitative behavior of those results
and their dependencies on quantities such as 
$\fax$, $x$, $\TR$, $\msaxion$, $\maxino$, 
$\mgravitino$, and $\mgluino$.

We start with the dilution factor $\Delta$
based on the corresponding considerations 
in Ref.~\cite{Scherrer:1984fd}.
The equation describing the change in entropy 
due to the decay of a single non-relativistic species $\psi$ 
into relativistic particles that rapidly thermalize
reads
\be
S^{1/3}\dot{S} 
= 
R^4 \left(\frac{2\pi^2}{45}\gstarS\right)^{\!\! 1/3}  
\Gamma_{\psi} \rho_{\psi},
\label{Eq:Generalentropy}
\ee
which is the basis for~\eqref{Eq:EntropyEvol1} 
and~\eqref{Eq:EntropyEvol2} in the main text.
Here $\rho_{\psi}$ and $\Gamma_{\psi}$ are the energy density 
and the total decay width of $\psi$, respectively, and $\psi$
is assumed to decay fully into rapidly thermalizing particles.
By integrating~\eqref{Eq:Generalentropy},
one arrives at~\cite{Scherrer:1984fd}
\begin{equation}
\left(\frac{S}{S_i}\right)^{\!\! 4/3} \!\!\!
=  
1 + \frac{4}{3} \rho_{\psi\,i} R_i^4 \!
\int_{t_i}^{t}\! dt' \!
\left(\frac{2\pi^2}{45}\gstarS\right)^{\!\! 1/3}\!
\left[\frac{R(t')}{R_i}\right] e^{-\Gamma_\psi t'} 
\label{Eq:GeneralDelta}
\end{equation}
where the subscript $i$ refers to the respective quantities
at the initial time $t_i$.
This time $t_i$ can differ from the value 
used in our numerical calculations 
in Sects.~\ref{Sec:GravitinoCDM} and~\ref{Sec:AxionCDM}. 
In fact, the main contribution to the integral 
comes from the time interval around $\tau_\psi=1/\Gamma_\psi$
so that, e.g., $t_i=0.01\tau_\psi$ 
is sufficiently early to obtain a good precision.

To solve the integral in~\eqref{Eq:GeneralDelta}, 
one needs to know the evolution of the scale factor. 
This is described by the Friedmann equation 
and therefore depends on the energy content of the Universe
at the relevant times. 
For the following two limiting cases,
an approximate solution for~$\Delta$,  
the ratio of the entropy before and after the decay,
can be obtained analytically.

When the energy density $\rho_{\psi}$ of 
the non-relativistic particle $\psi$ prior to its decay
dominates the one of the Universe,
matter dominates so that $R\propto t^{2/3}$
and~\cite{Scherrer:1984fd}
\be
\Delta_{\text{large}}
\simeq 
1.83 \,\langle\gstarS\rangle^{1/4}
\frac{m_\psi Y_\psi}{(\Gamma_\psi \mPlanck)^{1/2}},
\label{Eq:DeltaLarge}
\ee
where $\langle\gstarS\rangle$ denotes 
a suitably averaged value of $\gstarS$ 
over the integration interval. 
If $\gstarS$ does not change significantly around $t\sim\tau_\psi$, 
$\langle\gstarS\rangle=\gstarS(\tau_\psi)$ 
gives a reasonable approximation.
The subscript ``large'' in~\eqref{Eq:DeltaLarge}
is used because of the large dilution factor,
$\Delta\gg 1$, encountered in such situations
and to indicate 
the correspondingly limited applicability range of~\eqref{Eq:DeltaLarge}.

When the energy density in radiation dominates
the one of the Universe prior and during the epoch in which $\psi$ decays,
$R\propto t^{1/2}$ and~\cite{Scherrer:1984fd}
\be
\Delta_{\text{small}} 
\simeq 
1
+
1.61\,
\frac{\langle\gstarS\rangle^{1/3}}{\gstarS(t_i)^{1/12}}
\frac{m_\psi Y_\psi}{(\Gamma_\psi \mPlanck)^{1/2}} .
\label{Eq:DeltaSmall}
\ee
Again one obtains a good approximation with
$\langle\gstarS\rangle=\gstarS(\tau_\psi)$ 
if $\gstarS$ is (basically) constant 
in the relevant interval.
The subscript ``small'' in~\eqref{Eq:DeltaSmall}
indicates that its applicability is limited to
settings in which $\Delta$ is not much larger than one.

Let us turn to the case of entropy release from axino decay
considered in Sect.~\ref{Sec:GravitinoCDM}. 
For all of the parameter points examined in this work, 
$\rho_{\axino}<\rhorad$ and thus the corresponding 
dilution factor is $\Delta^{\axino\to g\gluino}=\Order(1)$.
Consequently, we can use~\eqref{Eq:DeltaSmall} 
to approximate the dilution factor 
from entropy release in axino decays. 
Indeed, for axinos from thermal processes 
with $\BR(\axino\to g\gluino)\simeq 1$,
our numerical results 
-- shown e.g.\ by the solid line in Fig.~\ref{Fig:GDMEvoDelta}(b) --
are well approximated by
\begin{align}
& \Delta_{\text{small}}^{\axino\to g\gluino} 
\simeq
1 
+ 
2.3\times10^{-2} 
\left(\frac{2~\TeV}{\maxino}\right)^{\!\!1/2}\!
\left(\frac{\fax}{10^{10}\,\GeV}\right)
\nonumber
\\
& \quad \times
\left(\frac{0.1}{\alphaS}\right)\! 
\left( 1- \frac{\mgluino^2}{\maxino^2} \right)^{\!\!-3/2}\!  
\left( \frac{\YaxinoeqTP}{10^{-3}} \right)\!
\frac{\gstarS(\tau_{\axino})^{1/3}}{\gstarS(0.01\tau_{\axino})^{1/12}},
\label{Eq:DeltaAxinoSmall}
\end{align}
where $t_i=0.01\tau_{\axino}$ is used as suggested above.

Saxions can decay both into inert radiation 
and into relativistic particles that rapidly thermalize
with the respective branching ratios~\eqref{Eq:BRsaxionTwoAxions}
and~\eqref{Eq:BRsaxionTwoGluons} governed by $x$.
For $x\gtrsim 0.1$, $\Delta^{\saxion\to gg}=\Order(1)$.
Accordingly, after accounting for $\BR(\saxion\to gg)$,
\eqref{Eq:DeltaSmall} can be used 
to approximate the dilution factor
due to entropy release in decays
of saxions from thermal processes. 
Our numerical results 
-- shown e.g.\ by the dashed and dotted lines 
in Fig.~\ref{Fig:aDMEvoDelta}(b) --
are indeed well described by
\begin{align}
&\Delta_\text{small}^{\saxion\to gg}
\simeq 
1 
+  
1.03\times10^{-2} 
\left(\frac{100~\GeV}{\msaxion}\right)^{1/2} 
\left(\frac{\fax}{10^{10}\,\GeV}\right) 
\nonumber
\\
&\,\,
\times
\frac{\alpha_s^2}{(\alpha_s^2 + 0.5x^2\pi^2)^{3/2}}\!
\left(\frac{\YsaxeqTP}{10^{-3}}\right)\!
\frac{\gstarS(\tau_{\saxion})^{1/3}}
{\gstarS(0.01\tau_{\saxion})^{1/12}} ,
\label{Eq:DeltaSaxionSmall}
\end{align}
where $t_i=0.01\tau_{\saxion}$.
For $x=0.02$, $\fax=10^{12}\,\GeV$, and $\TR\gtrsim5\times10^{10}\,\GeV$
in the axion CDM scenario, $\Delta^{\saxion\to gg}\gtrsim10$ is possible
and there best described by using \eqref{Eq:DeltaLarge}.
Setting $x=0$ in $\Gamma_{\saxion}$ and $\BR(\saxion\to gg)$,
we then obtain
\bea
\Delta_\text{large}^{\saxion\to gg}
&\simeq 
19 
\left(\frac{100\,\GeV}{\msaxion}\right)^{\! 1/2} 
\left(\frac{\fax}{10^{12}\,\GeV}\right) 
\\
&\quad\times
\left(\frac{0.1}{\alphaS}\right)
\left(\frac{\YsaxeqTP}{10^{-3}}\right) 
\left[\frac{\gstarS(\tau_{\saxion})}{10.75}\right]^{1/4},
\label{Eq:DeltaSaxionLarge}
\eea
which deviates by at most $20\%$ 
from our numerical results for $x\leq 0.02$
-- shown e.g.\ by the dot-dashed lines in Fig.~\ref{Fig:aDMEvoDelta}(b) --
in that parameter region with large $\Delta$.
For an approximate treatment of entropy production 
in saxion decays, see also Ref.~\cite{Hasenkamp:2010if}.

In settings in which two non-relativistic species decay 
at different times
and thereby produce entropy at different times, 
the total dilution factor $\Delta$ is given by the product 
of the individual dilution factors $\Delta_j$. 
This occurs, e.g., for $x<1$ in Sect.~\ref{Sec:GravitinoCDM},
where $\Delta=\Delta^{\axino\to g\gluino}\Delta^{\saxion\to gg}$.
There the product of~\eqref{Eq:DeltaAxinoSmall} 
and~\eqref{Eq:DeltaSaxionSmall} describes approximately
the dashed and dotted curves in Fig.~\ref{Fig:GDMEvoDelta}(b).

To arrive at approximate expressions for $\Delta\neff$,
we work with the sudden decay approximation
as, e.g., in Refs.~\cite{Hasenkamp:2011em,Graf:2012hb}.
The decays that can lead to extra radiation
are thus approximated to proceed exactly 
when cosmic time is equal to the lifetime of the decaying species.
The contribution to $\Delta\neff$ of axions from decays
of saxions from thermal processes is then given by
\bea
\Delta\neff^{\saxion\to aa}(T) 
&= 
\frac{120}{7\pi^2T_\nu^4}
\left[\frac{\gstarS(T)}{\gstarS(T_\saxion)}\right]^{4/3}
\left(\frac{T}{T_\saxion}\right)^4 
\\
&\quad\times
\BR(\saxion\to aa) 
\rho_\saxion^\text{eq/TP}(T_\saxion)/\Delta
\label{Eq:DeltaNapprox1}
\eea
with a temperature at the decay time $t=\tausaxion$ of
\bea
T_{\saxion} 
&\simeq 
10.6~\MeV
\left(x^2 + \frac{2\alpha_s^2}{\pi^2}\right)^{1/2} 
\left(\frac{\msaxion}{1~\GeV}\right)^{3/2} 
\\
&\quad\times
\left(\frac{10^{10}\,\GeV}{\fax}\right) 
\left[ \frac{10.75}{\gstar(T_{\saxion})}\right]^{1/4}.
\eea
By going beyond the sudden decay approximation, one finds 
that $\Delta\neff^{\saxion\to aa}$ is overestimated 
by about $10\%$.
Accounting for this by multipling~\eqref{Eq:DeltaNapprox1} with $0.87$,%
\footnote{A similar factor is found in Appendix~A 
of Ref.~\cite{Hasenkamp:2012ii}
in a comparison of the exponential decay behavior 
with the sudden decay approximation; 
see also Appendix of Ref.~\cite{Scherrer:1987rrPart2}.}
\begin{align}
\Delta\neff^{\saxion\to aa}(T) 
\simeq & \,\,
\frac{0.82}{\Delta} 
\left(\frac{1~\GeV}{\msaxion}\right)^{1/2} 
\left(\frac{\fax}{10^{10}\,\GeV} \right) 
\nonumber
\\
& \!\!\!\!
\times
\frac{x^2}{[x^2 + 2(\alphaS/\pi)^2]^{3/2}}  
\left( \frac{\YsaxeqTP}{10^{-3}} \right) 
\label{Eq:DeltaNapprox2}
\\
& \!\!\!\!
\times
\left(\frac{T}{T_\nu}\right)^4 
\left[\frac{\gstarS(T)}{10.75}\right]^{4/3} 
\frac{\gstar(T_{\saxion})^{1/4}}{\gstarS(T_{\saxion})^{1/3}}.
\nonumber
\end{align}
This equation together with 
$\Delta=\Delta^{\axino\to g\gluino}\Delta^{\saxion\to gg}$,
as given by~\eqref{Eq:DeltaAxinoSmall} and~\eqref{Eq:DeltaSaxionSmall} 
for $x\gtrsim 0.1$,
can allow for a better understanding of
the numerical $\Delta\neff^{\saxion\to aa}$ results
obtained in Sect.~\ref{Sec:GravitinoCDM}.

In Sect.~\ref{Sec:AxionCDM} an additional contribution to $\Delta\neff$ 
in the form of axions and axinos from late decays of gravitinos is considered.
In the sudden decay approximation, 
\begin{equation}
\Delta\neff^{\gravitino\to a\axino}(T) 
= 
\frac{120}{7\pi^2T_\nu^4}
\left(\frac{T}{T_{\gravitino}}\right)^{\! 4} 
\rho_{\gravitino}^{\TP}(T_{\gravitino})/\Delta
\label{Eq:DeltaNapprox4}
\end{equation}
with a temperature at the decay time $t=\tau_{\gravitino}$ 
of~\cite{Olive:1984bi,Asaka:2000ew,Hasenkamp:2011em}
\be
T_{\gravitino} 
= 
24~\eV 
\left(\frac{\mgravitino}{100~\GeV}\right)^{3/2}.
\ee
Indeed, the gravitino decay happens always after BBN so that 
$\gstarS(T)=\gstarS(T_{\gravitino})=3.91$
for the values of $\mgravitino$ considered in Sect.~\ref{Sec:AxionCDM}.
Proceeding as above, we obtain
\be
\Delta\neff^{\gravitino \to a\axino} 
\simeq
\frac{0.42}{\Delta} 
\left(\frac{100~\GeV}{\mgravitino}\right)^{\! 1/2} 
 \left( \frac{Y_{\gravitino}^{\TP}}{10^{-11}} \right)
\label{Eq:DeltaNapprox3}
\ee
after multiplying~\eqref{Eq:DeltaNapprox4} 
by a factor of 0.87 
again to compensate for overestimation 
by the sudden decay approximation. 
Without this correction and for $\Delta = 1$, 
the above estimate agrees with the one given in Ref.~\cite{Hasenkamp:2011em}.
To understand better the numerical results of Sect.~\ref{Sec:AxionCDM}
in which 
$\Delta\neff\simeq\Delta\neff^{\saxion\to aa}
+\Delta\neff^{\gravitino \to a\axino}$,
\eqref{Eq:DeltaNapprox2} and~\eqref{Eq:DeltaNapprox3} can be used
together with $\Delta=\Delta^{\saxion\to gg}$,
as given by~\eqref{Eq:DeltaSaxionSmall} for $x\gtrsim 0.1$
or by~\eqref{Eq:DeltaSaxionLarge} for $x\lesssim 0.02$.

%______________________________________________
%
\section{Description of Changes for \boldmath$\msaxion\neq\mgravitino$}
\label{Sec:MassDiffSaxionGravitino}
%______________________________________________

Throughout the main part of this work, 
we use $\mgravitino=\msaxion$. 
In this Appendix we describe the differences with respect to 
the $\mgravitino=\msaxion$ case that one faces
if those two masses differ.
Note that we limit our discussion below to the regime 
with $\tausaxion\lesssim 1~\seconds$. 
In particular, we do not address 
additional cosmological constraints~\cite{Kawasaki:2007mk} 
appearing towards small $\msaxion$ 
that imply longer lifetimes
and thereby saxion decays during/after BBN.

Towards very large $\msaxion$,
additional decay channels into sparticles may open up
such as the decay $\saxion\to\gluino\gluino$, whose
width can be derived from~\eqref{Eq:eff_lag},
\be
\Gamma_{\saxion\to\gluino\gluino} 
= 
\frac{\alphaS^2\msaxion\mgluino^{2}}{4\pi^{3}\fax^2}
\left[1-\left(\frac{2\mgluino}{\msaxion}\right)^{2}\right].
\label{Eq:SaxionGluinoGluino}
\ee
If this decay occurs after the freeze-out of the LOSP,
each of the produced gluinos will lead to one LSP and
thereby contribute
\be
\Omega_{\LSP}^{\saxion\to\gluino\gluino}h^{2}
=2m_{\LSP}
\BR(\saxion\to\gluino\gluino)
\YsaxeqTP(\TL)s(T_0)h^2/\rho_c ,
\label{Eq:OmegaAxinoSaxionGluinoDecay}
\ee
similarly as discussed 
for axino decays in Sect.~\ref{Sec:GravitinoCDM}.
Indeed, a significant excess over~\eqref{Eq:OmegaCDMLimit} 
in the $\gravitino$ CDM case can again be avoided
when the saxion decays prior to the LOSP decoupling.
Because of the additional decay channels into
axions and gluons, this will be easier to realize
than for the axino decay in Sect.~\ref{Sec:GravitinoCDM}.
In the axion CDM case with the very light axino LSP,
\eqref{Eq:OmegaAxinoSaxionGluinoDecay} can be much smaller
because of a much smaller $m_{\LSP}=\maxino\lesssim 37~\eV$.
Here however such decays could lead to additional contributions
to $\Delta\neff$ in the form of relativistic axions.
If the saxion decays prior to the LOSP decoupling,
again no additional constraints will be expected. 
A more detailed discussion of effects related
to the $\saxion\to\gluino\gluino$ decay is
left for future work. 
In the following description of changes
such effects are assumed to be negligible.

In the $\gravitino$ CDM case considered 
in Sect.~\ref{Sec:GravitinoCDM},
increasing (decreasing) $\msaxion$ relative 
to a fixed value of $\mgravitino$ as indicated on the horizontal axes 
moves the $\Delta\neff$ contours to the left (right)
in Figs.~\ref{Fig:GCDMx1}(a)--(c) and~\ref{Fig:GDMxplot}. 
For the $x=1$ case presented in Figs.~\ref{Fig:GCDMx1}(a)--(c), 
there is practically no change of the $\OmegaGravitino h^{2}$ contour
since the entropy released in saxion decays is negligible. 
For $x<1$, $\Delta$ depends on $\msaxion$ 
as can be seen in Fig.~\ref{Fig:GDMEvoDelta}(b). 
For a fixed $\mgravitino$ value,
increasing (decreasing) $\msaxion$ 
reduces (enhances) the dilution due to saxion decay 
and thus results in more (less) restrictive 
upper limits on $\TR$ imposed by $\OmegaGravitino h^{2}<0.124$, 
i.e., the respective contours will more downwards (upwards)
and show a less (more) pronounced dip.

In the $\axion$ CDM case considered in Sect.~\ref{Sec:AxionCDM},
increasing (decreasing) $\msaxion$ relative 
to a fixed $\mgravitino$ value as indicated on the horizontal axis
moves the dotted $\Delta\neff^{\saxion\to aa}$ contours
in Fig.~\ref{Fig:aCDMExplainx}(a)
to the left (right). 
The solid contours depicting the sum 
of both extra radiation components 
change accordingly 
in Figs.~\ref{Fig:aDMExclude}(a) and~\ref{Fig:aCDMExplainx}(a),
while there is no effect on 
the dashed $\Delta\neff^{\gravitino\to a\axion}$ contours for $x=1$.
For $x<1$, the entropy release from saxion decays can become sizable.
For a fixed $\mgravitino$ value,
increasing (decreasing) $\msaxion$ 
gives a smaller (larger) $\Delta$
and affects the $\Delta\neff^{\gravitino\to a\axino}$ contours 
in a way that is qualitatively comparable to their change 
observed for increasing (decreasing) $x$. 
In turn, the dashed, dotted, and dash-dotted curves 
in Fig.~\ref{Fig:aDMExclude}(b)
that present the sum of both $\Delta\neff$ components
change accordingly 
as well as all $\Delta\neff$ contours shown 
in Figs.~\ref{Fig:aCDMExplainx}(b)--(d).

%______________________________________________
\end{appendix}
%______________________________________________

\medskip

%______________________________________________
\section*{Acknowledgments}
%______________________________________________
%
We are grateful to
Georg Raffelt and Fuminobu Takahashi 
for valuable discussions.
This research was partially supported by the 
Cluster of Excellence ``Origin and Structure of the Universe''
and by the  European Union FP7  ITN
INVISIBLES (Marie Curie Actions, PITN- GA-2011- 289442).


\begin{thebibliography}{10}

\bibitem{Izotov:2010ca}
Y.~I.\  Izotov and T.~X.\  Thuan,
\newblock Astrophys.\  J.\  {\bf 710}, L67 (2010), arXiv:1001.4440.
%%CITATION = 1001.4440;%%

\bibitem{Aver:2010wq}
E.~Aver, K.~A.\  Olive, and E.~D.\  Skillman,
\newblock JCAP {\bf 1005}, 003 (2010), arXiv:1001.5218.
%%CITATION = 1001.5218;%%

\bibitem{Hamann:2010bk}
J.~Hamann, S.~Hannestad, G.~G.\  Raffelt, I.~Tamborra, and Y.~Y.\  Wong,
\newblock Phys.\ Rev.\ Lett.\  {\bf 105}, 181301 (2010), arXiv:1006.5276.
%%CITATION = ARXIV:1006.5276;%%

\bibitem{Graf:2012hb}
P.~Graf and F.~D.\  Steffen,
\newblock JCAP {\bf 1302}, 018 (2013), arXiv:1208.2951.
%%CITATION = ARXIV:1208.2951;%%

\bibitem{Komatsu:2010fb}
WMAP, E.~Komatsu {\em et~al.},
\newblock Astrophys.\  J.\  Suppl.\  {\bf 192}, 18 (2011), arXiv:1001.4538.
%%CITATION = 1001.4538;%%

\bibitem{Hamann:2010pw}
J.~Hamann, S.~Hannestad, J.~Lesgourgues, C.~Rampf, and Y.~Y.~Y.\  Wong,
\newblock JCAP {\bf 1007}, 022 (2010), arXiv:1003.3999.
%%CITATION = 1003.3999;%%

\bibitem{Hinshaw:2012fq}
G.~Hinshaw {\em et~al.},
\newblock (2012), arXiv:1212.5226.
%%CITATION = ARXIV:1212.5226;%%

\bibitem{Ade:2013zuv} 
  P.~A.~R.~Ade {\it et al.}  [Planck Collaboration],
  %``Planck 2013 results. XVI. Cosmological parameters,''
  arXiv:1303.5076.
  %%CITATION = ARXIV:1303.5076;%%

\bibitem{Perotto:2006rj}
L.~Perotto, J.~Lesgourgues, S.~Hannestad, H.~Tu, and Y.~Y.~Y.\  Wong,
\newblock JCAP {\bf 0610}, 013 (2006), arXiv:astro-ph/0606227.
%%CITATION = ASTRO-PH/0606227;%%

\bibitem{Hamann:2007sb}
J.~Hamann, J.~Lesgourgues, and G.~Mangano,
\newblock JCAP {\bf 0803}, 004 (2008), arXiv:0712.2826.
%%CITATION = 0712.2826;%%

\bibitem{Riess:2011yx} 
  A.~G.~Riess
%  , L.~Macri, S.~Casertano, H.~Lampeitl, H.~C.~Ferguson, A.~V.~Filippenko, S.~W.~Jha and W.~Li 
  {\it et al.},
  Astrophys.\ J.\  {\bf 730}, 119 (2011)
  [Erratum-ibid.\  {\bf 732}, 129 (2011)]
  arXiv:1103.2976.
  %%CITATION = ARXIV:1103.2976;%%

\bibitem{Hamann:2011ge}
J.~Hamann, S.~Hannestad, G.~G. Raffelt, and Y.~Y.\  Wong,
\newblock JCAP {\bf 1109}, 034 (2011), arXiv:1108.4136.
%%CITATION = ARXIV:1108.4136;%%

\bibitem{Nakayama:2010vs}
K.~Nakayama, F.~Takahashi, and T.~T.\  Yanagida,
\newblock Phys.\ Lett.\  {\bf B697}, 275 (2011), arXiv:1010.5693.
%%CITATION = ARXIV:1010.5693;%%

\bibitem{Boehm:2012gr}
C.~Boehm, M.~J.\  Dolan, and C.~McCabe,
\newblock (2012), arXiv:1207.0497.
%%CITATION = ARXIV:1207.0497;%%

\bibitem{Pastor:2008ti}
S.~Pastor, T.~Pinto, and G.~G.\  Raffelt,
\newblock Phys.\ Rev.\ Lett.\  {\bf 102}, 241302 (2009), arXiv:0808.3137.
%%CITATION = ARXIV:0808.3137;%%

\bibitem{Mangano:2010ei}
G.~Mangano, G.~Miele, S.~Pastor, O.~Pisanti, and S.~Sarikas,
\newblock JCAP {\bf 1103}, 035 (2011), arXiv:1011.0916.
%%CITATION = ARXIV:1011.0916;%%

\bibitem{Chun:1995hc}
E.~J.\  Chun and A.~Lukas,
\newblock Phys.\  Lett.\  {\bf B357}, 43 (1995), arXiv:hep-ph/9503233.
%%CITATION = HEP-PH/9503233;%%

\bibitem{Chun:2000jr}
E.~J.~Chun, D.~Comelli, and D.~H.~Lyth,
\newblock Phys.\ Rev.\  {\bf D62}, 095013 (2000), arXiv:hep-ph/0008133.
%%CITATION = HEP-PH/0008133;%%

\bibitem{Ichikawa:2007jv}
K.~Ichikawa, M.~Kawasaki, K.~Nakayama, M.~Senami, and F.~Takahashi,
\newblock JCAP {\bf 0705}, 008 (2007), arXiv:hep-ph/0703034.
%%CITATION = HEP-PH/0703034;%%

\bibitem{Kawasaki:2007mk}
M.~Kawasaki, K.~Nakayama, and M.~Senami,
\newblock JCAP {\bf 0803}, 009 (2008), arXiv:0711.3083.
%%CITATION = 0711.3083;%%

\bibitem{Fischler:2010xz}
W.~Fischler and J.~Meyers,
\newblock Phys.\ Rev.\  {\bf D83}, 063520 (2011), arXiv:1011.3501.
%%CITATION = ARXIV:1011.3501;%%

\bibitem{Hasenkamp:2011em}
J.~Hasenkamp,
\newblock Phys.\ Lett.\  {\bf B707}, 121 (2012), arXiv:1107.4319.
%%CITATION = ARXIV:1107.4319;%%

\bibitem{Hooper:2011aj} 
  D.~Hooper, F.~S.~Queiroz and N.~Y.~Gnedin,
\newblock Phys.\ Rev.\ {\bf D85}, 063513 (2012),
  arXiv:1111.6599.
  %%CITATION = ARXIV:1111.6599;%%

\bibitem{Choi:2012zn}
K.~Choi, K.-Y.~Choi, and C.~S.~Shin,
\newblock Phys.\ Rev.\ {\bf D86}, 083529 (2012), arXiv:1208.2496.
%%CITATION = ARXIV:1208.2496;%%

\bibitem{Cicoli:2012aq}
M.~Cicoli, J.~P.~Conlon, and F.~Quevedo,
\newblock Phys.\ Rev.\ {\bf D87}, 043520 (2013), arXiv:1208.3562.
%%CITATION = ARXIV:1208.3562;%%

\bibitem{Higaki:2012ar}
T.~Higaki and F.~Takahashi,
\newblock JHEP {\bf 1211}, 125 (2012), arXiv:1208.3563.
%%CITATION = ARXIV:1208.3563;%%

\bibitem{GonzalezGarcia:2012yq} 
  M.~C.~Gonzalez-Garcia, V.~Niro and J.~Salvado,
\newblock JHEP {\bf 1304}, 052 (2013), arXiv:1212.1472.
  %%CITATION = ARXIV:1212.1472;%%

\bibitem{Hasenkamp:2012ii} 
  J.~Hasenkamp and J.~Kersten,
\newblock  JCAP {\bf 1308}, 024 (2013), arXiv:1212.4160.
%%CITATION = ARXIV:1212.4160;%%

\bibitem{Bae:2013qr}
K.~J.~Bae, H.~Baer, and A.~Lessa,
\newblock JCAP {\bf 1304}, 041 (2013), arXiv:1301.7428.
%%CITATION = ARXIV:1301.7428;%%

\bibitem{Jeong:2013axf}
K.~S.~Jeong and F.~Takahashi,
\newblock   JHEP {\bf 1304}, 121 (2013), arXiv:1302.1486.
%%CITATION = ARXIV:1302.1486;%%

\bibitem{Buchmuller:2005eh}
W.~Buchm{\"u}ller, R.~Peccei, and T.~Yanagida,
\newblock Ann.\ Rev.\ Nucl.\ Part.\ Sci.\  {\bf 55}, 311 (2005), 
arXiv:hep-ph/0502169.
%%CITATION = HEP-PH/0502169;%%

\bibitem{Sikivie:2006ni}
P.~Sikivie,
\newblock Lect.\  Notes Phys.\  {\bf 741}, 19 (2008), arXiv:astro-ph/0610440.
%%CITATION = ASTRO-PH/0610440;%%

\bibitem{Kim:2008hd}
J.~E.~Kim and G.~Carosi,
\newblock Rev.\ Mod.\ Phys.\  {\bf 82}, 557 (2010), arXiv:0807.3125.

\bibitem{Raffelt:2006cw}
G.~G.~Raffelt,
\newblock Lect.\ Notes Phys.\ {\bf 741}, 51 (2008), arXiv:hep-ph/0611350.
%%CITATION = HEP-PH/0611350;%%

\bibitem{Beringer}
Particle Data Group, J.~Beringer {\em et~al.},
\newblock Phys.\  Rev.\  {\bf D86}, 010001 (2012).

\bibitem{Chang:1996ih}
S.~Chang and H.~B.~Kim,
\newblock Phys.\ Rev.\ Lett.\  {\bf 77}, 591 (1996), arXiv:hep-ph/9604222.

\bibitem{Asaka:1998ns}
T.~Asaka and M.~Yamaguchi,
\newblock Phys.\  Lett.\  {\bf B437}, 51 (1998), arXiv:hep-ph/9805449.
%%CITATION = HEP-PH/9805449;%%

\bibitem{Choi:2008zq} 
  K.-Y.~Choi, J.~E.~Kim, H.~M.~Lee, and O.~Seto,
\newblock  Phys.\ Rev.\ {\bf D77}, 123501 (2008),
  arXiv:0801.0491.
  %%CITATION = ARXIV:0801.0491;%%
  %45 citations counted in INSPIRE as of 19 Feb 2013

\bibitem{Hasenkamp:2010if}
J.~Hasenkamp and J.~Kersten,
\newblock Phys.\  Rev.\  {\bf D82}, 115029 (2010), arXiv:1008.1740.
%%CITATION = 1008.1740;%%

\bibitem{Hasenkamp:2011xh}
  J.~Hasenkamp and J.~Kersten,
\newblock Phys.\ Lett.\ {\bf B701} 660 (2011),
  arXiv:1103.6193.
  %%CITATION = ARXIV:1103.6193;%%

\bibitem{Strumia:2010aa}
A.~Strumia,
\newblock JHEP {\bf 06}, 036 (2010), arXiv:1003.5847.
%%CITATION = 1003.5847;%%

\bibitem{Brandenburg:2004du}
A.~Brandenburg and F.~D. Steffen,
\newblock JCAP {\bf 0408}, 008 (2004), arXiv:hep-ph/0405158.
%%CITATION = HEP-PH/0405158;%%

\bibitem{Braaten:1989mz}
E.~Braaten and R.~D.~Pisarski,
\newblock Nucl.\  Phys.\  {\bf B337}, 569 (1990).
%%CITATION = NUPHA,B337,569;%%

\bibitem{Braaten:1991dd}
E.~Braaten and T.~C.~Yuan,
\newblock Phys.\  Rev.\  Lett.\  {\bf 66}, 2183 (1991).
%%CITATION = PRLTA,66,2183;%%

\bibitem{Mangano:2005cc}
G.~Mangano {\em et~al.},
\newblock Nucl.\  Phys.\  {\bf B729}, 221 (2005), arXiv:hep-ph/0506164.
%%CITATION = HEP-PH/0506164;%%

\bibitem{Hamann:2007pi}
J.~Hamann, S.~Hannestad, G.~G. Raffelt, and Y.~Y.~Y.~Wong,
\newblock JCAP {\bf 0708}, 021 (2007), arXiv:0705.0440.
%%CITATION = 0705.0440;%%

\bibitem{Reid:2009nq}
B.~A. Reid, L.~Verde, R.~Jimenez, and O.~Mena,
\newblock JCAP {\bf 1001}, 003 (2010), arXiv:0910.0008.
%%CITATION = 0910.0008;%%

\bibitem{GonzalezGarcia:2010un}
M.~C.~Gonzalez-Garcia, M.~Maltoni, and J.~Salvado,
\newblock JHEP {\bf 08}, 117 (2010), arXiv:1006.3795.
%%CITATION = 1006.3795;%%

\bibitem{MNR:MNR13921}
M.~Pettini, B.~J.~Zych, M.~T.~Murphy, A.~Lewis, and C.~C.~Steidel,
\newblock Monthly Notices of the Royal Astronomical Society {\bf 391}, 1499
  (2008).

\bibitem{Asaka:2006bv}
T.~Asaka, S.~Nakamura, and M.~Yamaguchi,
\newblock Phys.\ Rev.\  {\bf D74}, 023520 (2006), arXiv:hep-ph/0604132.
%%CITATION = HEP-PH/0604132;%%

\bibitem{Endo:2007sz}
M.~Endo, F.~Takahashi, and T.~Yanagida,
\newblock Phys.\ Rev.\  {\bf D76}, 083509 (2007), arXiv:0706.0986.
%%CITATION = ARXIV:0706.0986;%%

\bibitem{Kim:1979if}
J.~E.~Kim,
\newblock Phys.\  Rev.\  Lett.\  {\bf 43}, 103 (1979).
%%CITATION = PRLTA,43,103;%%

\bibitem{Shifman:1979if}
M.~A.~Shifman, A.~I.~Vainshtein, and V.~I.~Zakharov,
\newblock Nucl.\  Phys.\  {\bf B166}, 493 (1980).
%%CITATION = NUPHA,B166,493;%%

\bibitem{Choi:2011yf}
K.-Y.~Choi, L.~Covi, J.~E.~Kim, and L.~Roszkowski,
\newblock JHEP {\bf 1204}, 106 (2012), arXiv:1108.2282.
%%CITATION = ARXIV:1108.2282;%%

\bibitem{Dreiner:2008tw}
H.~K.~Dreiner, H.~E.~Haber, and S.~P.~Martin,
\newblock Phys.\ Rept.\  {\bf 494}, 1 (2010), arXiv:0812.1594.
%%CITATION = ARXIV:0812.1594;%%

\bibitem{Drees:2004jm}
M.~Drees, R.~Godbole, and P.~Roy,
{\it Theory and Phenomenology of Sparticles}
\newblock (2004),
Hackensack, USA: Wold Scientific 555p.

\bibitem{Bae:2011jb}
K.~J.~Bae, K.~Choi, and S.~H.~Im,
\newblock JHEP {\bf 1108}, 065 (2011), arXiv:1106.2452.
%%CITATION = ARXIV:1106.2452;%%

\bibitem{Bolz:2000fu}
M.~Bolz, A.~Brandenburg, and W.~Buchm{\"u}ller,
\newblock Nucl.\  Phys.\  {\bf B606}, 518 (2001)
[{\it Erratum ibid.} {\bf B790} (2008) 336],
arXiv:hep-ph/0012052.
%%CITATION = HEP-PH/0012052;%%

\bibitem{Pradler:2006qh} 
  J.~Pradler and F.~D.~Steffen,
\newblock  Phys.\ Rev.\ {\bf D75}, 023509 (2007), arXiv:hep-ph/0608344.
  %%CITATION = HEP-PH/0608344;%%

\bibitem{Pradler:2006hh}
J.~Pradler and F.~D.~Steffen,
\newblock Phys.\ Lett.\ {\bf B648}, 224 (2007), arXiv:hep-ph/0612291.
%%CITATION = HEP-PH/0612291;%%

\bibitem{Porod:2003um}
W.~Porod,
\newblock Comput.\ Phys.\ Commun.\  {\bf 153}, 275 (2003), arXiv:hep-ph/0301101.
%%CITATION = HEP-PH/0301101;%%

\bibitem{Porod:2011nf}
W.~Porod and F.~Staub,
\newblock Comput.\ Phys.\ Commun.\  {\bf 183}, 2458 (2012), arXiv:1104.1573.
%%CITATION = ARXIV:1104.1573;%%

\bibitem{ATLAS:2012us}
ATLAS Collaboration, G.~Aad {\em et~al.},
\newblock (2012), arXiv:1212.6149.
%%CITATION = ARXIV:1212.6149;%%

\bibitem{Graf:2010tv} 
  P.~Graf and F.~D.~Steffen,
\newblock  Phys.\ Rev.\ {\bf D83}, 075011 (2011), arXiv:1008.4528.
  %%CITATION = ARXIV:1008.4528;%%

\bibitem{Moroi:1993mb} 
  T.~Moroi, H.~Murayama, and M.~Yamaguchi,
\newblock Phys.\  Lett.\  {\bf B303}, 289 (1993).
  %%CITATION = PHLTA,B303,289;%%

\bibitem{Bolz:1998ek}  
M.~Bolz, W.~Buchm{\"u}ller, and M.~Pl{\"u}macher,
Phys.\  Lett.\  {\bf B443}, 209 (1998),
arXiv: hep-ph/9809381.
  %%CITATION = HEP-PH/9809381;%%

\bibitem{Covi:2001nw}
  L.~Covi, H.-B.~Kim, J.~E.~Kim, and L.~Roszkowski,
\newblock JHEP {\bf 0105}, 033 (2001),
arXiv: hep-ph/0101009.
  %%CITATION = HEP-PH/0101009;%%

\bibitem{Rychkov:2007uq}
V.~S. Rychkov and A.~Strumia,
\newblock Phys.\ Rev.\  {\bf D75}, 075011 (2007), arXiv:hep-ph/0701104.
%%CITATION = HEP-PH/0701104;%%

\bibitem{Buchmuller:2006tt}
W.~Buchm{\"u}ller, K.~Hamaguchi, M.~Ibe, and T.~T.~Yanagida,
\newblock Phys.\  Lett.\  {\bf B643}, 124 (2006),
arXiv: hep-ph/0605164.
%%CITATION = HEP-PH/0605164;%%

\bibitem{Buchmuller:2002rq}
W.~Buchm{\"u}ller, P.~Di~Bari, and M.~Pl{\"u}macher,
\newblock Nucl.\ Phys.\  {\bf B643}, 367 (2002), arXiv:hep-ph/0205349.
%%CITATION = HEP-PH/0205349;%%

\bibitem{Buchmuller:2002jk}
W.~Buchm{\"u}ller, P.~Di~Bari, and M.~Pl{\"u}macher,
\newblock Phys.\ Lett.\  {\bf B547}, 128 (2002), arXiv:hep-ph/0209301.
%%CITATION = HEP-PH/0209301;%%

\bibitem{Chun:1993vz} 
E.~J.~Chun, H.~B.~Kim and J.~E.~Kim,
\newblock Phys.\ Rev.\ Lett.\  {\bf 72}, 1956 (1994),
arXiv: hep-ph/9305208.
  %%CITATION = HEP-PH/9305208;%%

\bibitem{Kim:1994ub} 
  H.~B.~Kim and J.~E.~Kim,
\newblock Nucl.\ Phys.\  {\bf B433}, 421 (1995),
arXiv: hep-ph/9405385.
  %%CITATION = HEP-PH/9405385;%%

\bibitem{Scherrer:1987rrPart2}
R.~J.~Scherrer and M.~S.~Turner,
\newblock Astrophys.\ J.\  {\bf 331}, 33 (1988).
%%CITATION = ASJOA,331,19;%%

\bibitem{Scherrer:1987rrPart1}
R.~J.~Scherrer and M.~S.~Turner,
\newblock Astrophys.\ J.\  {\bf 331}, 19 (1988).
%%CITATION = ASJOA,331,19;%%

\bibitem{Beltran:2006sq}
M.~Beltran, J.~Garcia-Bellido, and J.~Lesgourgues,
\newblock Phys.\  Rev.\  {\bf D75}, 103507 (2007), arXiv:hep-ph/0606107.
%%CITATION = HEP-PH/0606107;%%

\bibitem{Turner:1985si}
M.~S.~Turner,
\newblock Phys.\ Rev.\  {\bf D33}, 889 (1986).
%%CITATION = PHRVA,D33,889;%%

\bibitem{Lyth:1991ub}
D.~Lyth,
\newblock Phys.\ Rev.\  {\bf D45}, 3394 (1992).
%%CITATION = PHRVA,D45,3394;%%

\bibitem{Visinelli:2009zm}
L.~Visinelli and P.~Gondolo,
\newblock Phys.\ Rev.\  {\bf D80}, 035024 (2009), arXiv:0903.4377.
%%CITATION = ARXIV:0903.4377;%%

\bibitem{Feng:2004mt}
J.~L.~Feng, S.~Su, and F.~Takayama,
\newblock Phys.\ Rev.\  {\bf D70}, 075019 (2004), arXiv:hep-ph/0404231.
%%CITATION = HEP-PH/0404231;%%

\bibitem{Kawasaki:2008qe}
M.~Kawasaki, K.~Kohri, T.~Moroi, and A.~Yotsuyanagi,
\newblock Phys.\ Rev.\  {\bf D78}, 065011 (2008), arXiv:0804.3745.
%%CITATION = ARXIV:0804.3745;%%

\bibitem{Asaka:2000zh}
T.~Asaka, K.~Hamaguchi, and K.~Suzuki,
\newblock Phys.\  Lett.\  {\bf B490}, 136 (2000), arXiv:hep-ph/0005136.
%%CITATION = HEP-PH/0005136;%%

\bibitem{Freitas:2011fx}
A.~Freitas, F.~D.~Steffen, N.~Tajuddin, and D.~Wyler,
\newblock JHEP {\bf 1106}, 036 (2011), arXiv:1105.1113.

\bibitem{Steffen:2006hw} 
  F.~D.~Steffen,
\newblock  JCAP {\bf 0609}, 001 (2006), arXiv:hep-ph/0605306.
  %%CITATION = HEP-PH/0605306;%%

\bibitem{Kanzaki:2006hm} 
  T.~Kanzaki, M.~Kawasaki, K.~Kohri, and T.~Moroi,
\newblock Phys.\ Rev.\  {\bf D75}, 025011 (2007), arXiv:hep-ph/0609246.
  %%CITATION = HEP-PH/0609246;%%

\bibitem{Pospelov:2006sc} 
  M.~Pospelov,
\newblock Phys.\ Rev.\ Lett.\  {\bf 98}, 231301 (2007), arXiv:hep-ph/0605215.
  %%CITATION = HEP-PH/0605215;%%

\bibitem{Steffen:2006wx} 
  F.~D.~Steffen,
\newblock  AIP Conf.\ Proc.\  {\bf 903}, 595 (2007), arXiv:hep-ph/0611027.
  %%CITATION = HEP-PH/0611027;%%
  
\bibitem{Pospelov:2008ta}
M.~Pospelov, J.~Pradler, and F.~D.~Steffen,
\newblock JCAP {\bf 0811}, 020 (2008), arXiv:0807.4287.
%%CITATION = ARXIV:0807.4287;%%

\bibitem{Chatrchyan:2012sp}
CMS Collaboration, S.~Chatrchyan {\em et~al.},
\newblock Phys.\ Lett.\  {\bf B713}, 408 (2012), arXiv:1205.0272.
%%CITATION = ARXIV:1205.0272;%%

\bibitem{ATLAS-CONF-2012-075}
CERN Report No.\ ATLAS-CONF-2012-075, 2012 (unpublished).

\bibitem{Ratz:2008qh} 
M.~Ratz, K.~Schmidt-Hoberg and M.~W.~Winkler,
\newblock JCAP {\bf 0810}, 026 (2008), arXiv:0808.0829.
  %%CITATION = ARXIV:0808.0829;%%
  
\bibitem{Pradler:2008qc} 
  J.~Pradler and F.~D.~Steffen,
\newblock  Nucl.\ Phys.\ {\bf B809}, 318 (2009), arXiv:0808.2462.
  %%CITATION = ARXIV:0808.2462;%%

\bibitem{Lindert:2011td} 
  J.~M.~Lindert, F.~D.~Steffen and M.~K.~Trenkel,
\newblock  JHEP {\bf 1108}, 151 (2011), arXiv:1106.4005.
  %%CITATION = ARXIV:1106.4005;%%

\bibitem{Covi:2007xj}
L.~Covi and S.~Kraml,
\newblock JHEP {\bf 0708}, 015 (2007), arXiv:hep-ph/0703130.
%%CITATION = HEP-PH/0703130;%%

\bibitem{Ellis:2008as}
J.~R.~Ellis, K.~A. Olive, and Y.~Santoso,
\newblock JHEP {\bf 0810}, 005 (2008), arXiv:0807.3736.
%%CITATION = ARXIV:0807.3736;%%

\bibitem{Katz:2009qx}
A.~Katz and B.~Tweedie,
\newblock Phys.\ Rev.\  {\bf D81}, 035012 (2010), arXiv:0911.4132.
%%CITATION = ARXIV:0911.4132;%%

\bibitem{Figy:2010hu}
T.~Figy, K.~Rolbiecki, and Y.~Santoso,
\newblock Phys.\ Rev.\  {\bf D82}, 075016 (2010), arXiv:1005.5136.
%%CITATION = ARXIV:1005.5136;%%

\bibitem{Olive:1984bi}
K.~A.~Olive, D.~N.~Schramm, and M.~Srednicki,
\newblock Nucl.\ Phys.\  {\bf B255}, 495 (1985).
%%CITATION = NUPHA,B255,495;%%

\bibitem{Asaka:2000ew}
T.~Asaka and T.~Yanagida,
\newblock Phys.Lett. {\bf B494}, 297 (2000), arXiv:hep-ph/0006211.
%%CITATION = HEP-PH/0006211;%%

\bibitem{Baer:2010gr}
H.~Baer, S.~Kraml, A.~Lessa, and S.~Sekmen,
\newblock JCAP {\bf 1104}, 039 (2011), arXiv:1012.3769.

\bibitem{Carosi:2007uc}
G.~Carosi and K.~van Bibber,
\newblock Lect.\ Notes Phys.\  {\bf 741}, 135 (2008), arXiv:hep-ex/0701025.
%%CITATION = HEP-EX/0701025;%%

\bibitem{Aver:2011bw}
E.~Aver, K.~A.~Olive, and E.~D.~Skillman,
\newblock JCAP {\bf 1204}, 004 (2012), arXiv:1112.3713.
%%CITATION = ARXIV:1112.3713;%%

\bibitem{DiazCruz:2007fc} 
  J.~L.~Diaz-Cruz, J.~R.~Ellis, K.~A.~Olive, and Y.~Santoso,
\newblock  JHEP {\bf 0705}, 003 (2007), arXiv:hep-ph/0701229.
  %%CITATION = HEP-PH/0701229;%%

\bibitem{Scherrer:1984fd}
R.~J.~Scherrer and M.~S.~Turner,
\newblock Phys.\ Rev.\  {\bf D31}, 681 (1985).
%%CITATION = PHRVA,D31,681;%%

\end{thebibliography}
\end{document}